\documentclass[aps,prd,preprintnumbers,nofootinbib]{revtex4}
\usepackage{epsfig}

\def\CA{{\cal A}}

\def\CL{{\cal L}}
\def\CO{{\cal O}}

\def\CV{{\cal V}}
\def\CU{{\cal U}}
\def\CM{{\cal M}}

\def\sfno#1#2{{(\gamma_{#1}\otimes\xi_{#2})}}
\def\semitimes{\mathrel>\joinrel\mathrel\triangleleft}

\def\bar{\overline}

\def\tilde{\widetilde}

\def\Tr{\textrm{Tr}}
\def\Str{\textrm{Str}}

\def\etal{{\it et al.}}

\def\spose#1{\hbox to 0pt{#1\hss}}
\def\ltapprox{\mathrel{\spose{\lower 3pt\hbox{$\mathchar"218$}}
 \raise 2.0pt\hbox{$\mathchar"13C$}}}
\def\gtapprox{\mathrel{\spose{\lower 3pt\hbox{$\mathchar"218$}}
 \raise 2.0pt\hbox{$\mathchar"13E$}}}
\def\inapprox{\mathrel{\spose{\lower 3pt\hbox{$\mathchar"218$}}
 \raise 2.0pt\hbox{$\mathchar"232$}}}

\preprint{UW/PT 04-15}
\begin{document}

\title{Staggered Chiral Perturbation Theory at Next-to-Leading Order}

\author{Stephen R. Sharpe}
\email{sharpe@phys.washington.edu}
\affiliation{Physics Department, University of Washington, Seattle, WA 98195-1560}
\author{Ruth S. Van de Water}
\email{ruthv@u.washington.edu}
\affiliation{Physics Department, University of Washington, Seattle, WA 98195-1560}

\date{\today}

\begin{abstract}
We study taste and Euclidean rotational symmetry  violation for staggered fermions at nonzero lattice spacing using staggered chiral perturbation theory.  We extend the staggered chiral Lagrangian to $\CO(a^2 p^2)$, $\CO(a^4)$ and $\CO(a^2 m)$, the orders necessary for a full next-to-leading order calculation of pseudo-Goldstone boson  masses and decay constants including analytic terms.  We then calculate a number of $SO(4)$ taste-breaking quantities, which involve only a small subset of these NLO operators.  We predict relationships between $SO(4)$ taste-breaking splittings in masses, pseudoscalar decay constants, and dispersion relations.  
We also find predictions for a few quantities that are not $SO(4)$ breaking.
All these results hold also for theories in which the fourth-root of the fermionic determinant is taken to reduce the number of quark tastes; testing them will therefore provide evidence for or against the validity of this trick.
\end{abstract}

\maketitle

\section{Introduction}
\label{sec:intro}

Simulations of lattice QCD using staggered fermions~\cite{Susskind} are 
able, at present, to reach significantly smaller dynamical quark masses
than those using other fermion discretizations~\cite{BerlinPanel,Daviesetal}.
This computational advantage is due in part to the 
axial $U(1)$ symmetry remaining at nonzero lattice spacing,
which protects the quark mass from additive renormalization. 
The primary disadvantage of staggered fermions is that each continuum flavor
comes in four tastes. In the continuum limit,
the tastes become degenerate and one can formally remove the extra tastes
by taking the fourth-root of the quark determinant 
(the ``$\sqrt[4]{\mbox{Det}}$ trick'').
At non-zero lattice spacing, however, the $SU(4)$ taste symmetry is
broken by discretization errors of $\CO(a^2)$, 
where $a$ is the lattice spacing. These errors
are numerically significant in present simulations~\cite{MILCspectrum}.
This has two related implications.
First, since one must, in practice, take the fourth-root of the determinant 
before taking the continuum limit,
 the resulting underlying quark action is likely non-local.
It is thus not guaranteed to lie in the same universality class as QCD,
even if the non-local part is, in some sense, of $\CO(a^2)$ 
and thus small.\footnote{%
For summaries of this and related problems with
the $\sqrt[4]{\mbox{Det}}$ trick see Refs.~\cite{Jansenetal,Kennedy}.}
Second, assuming that the continuum limit is correct,
it is necessary to perform a combined continuum and
chiral extrapolation taking into account taste-breaking effects
of $\CO(a^2)$~\cite{MILCf}. 

The first issue is clearly more fundamental, but also more difficult
to address. To date,
support for the use of the $\sqrt[4]{\mbox{Det}}$ trick comes from
an accumulation of numerical or indirect evidence: 
the accurate agreement of lattice and experimental results
for all ``gold-plated'' quantities that have been calculated~\cite{Daviesetal};
the success of fits of the light pseudoscalar meson masses
and decay constants to the predictions of chiral perturbation theory
including the $\CO(a^2)$ taste-breaking errors~\cite{MILCspectrum};
the near four-fold degeneracy of eigenvalues of the staggered Dirac operator
and its ability to correctly measure the topology of the gauge 
configurations~\cite{Follanaetal,Durretal}; and the results of
analytic studies of the time continuum limit~\cite{Adams1,Adams2}.
None of these tests purports to be conclusive.
Nevertheless, in the absence of a more theoretical approach, it is important
to have further numerical tests. This is one of the motivations for
the present work.

Our major focus, however, is on the further development of 
chiral perturbation theory for staggered fermions
including discretization errors. This ``staggered chiral perturbation
theory" (or S$\chi$PT for short) is the theoretical tool which determines
the functional forms needed to do the combined continuum and chiral
extrapolations incorporating taste violations.
The development of S$\chi$PT for staggered fermions \emph{without} the
$\sqrt[4]{\mbox{Det}}$ trick, i.e. with four tastes per flavor, is
a standard application of the methods of effective field theory~\cite{LS,BA1}. 
In order to apply S$\chi$PT to simulations done with the $\sqrt[4]{\mbox{Det}}$
trick, however, one must alter the $\chi$PT Feynman rules by hand
(or alternatively use a variant of the replica trick~\cite{ABreplica}),
so there is no underlying ``staggered chiral Lagrangian"
which can reproduce the Feynman rules~\cite{BA1}. 
Nevertheless, if the resulting expressions 
describe the outcomes of numerical simulations done using the $\sqrt[4]{\mbox{Det}}$
trick, one gains confidence that the long distance physics of
the staggered simulations reproduces that of QCD in the continuum limit. 
Indeed, the expressions in S$\chi$PT go over to those of standard QCD $\chi$PT in the
continuum limit, and give an explicit example of how the effect of the
$\sqrt[4]{\mbox{Det}}$ trick can vanish smoothly.

We consider here only S$\chi$PT applied to the properties
of pseudo-Goldstone bosons (PGBs).\footnote{%
There is a potential ambiguity in this terminology. We use the term 
PGB to refer to \emph{all} pseudoscalar mesons (and fermions, once
we consider partially quenched theories) whose masses vanish
in the combined chiral and continuum limits. This includes the
pion whose mass vanishes in the chiral limit
even at non-zero lattice spacing, which we refer to as
the ``lattice Goldstone" pion, as well as all the other
light pseudoscalars with different tastes.}
S$\chi$PT is a combined expansion in powers of the
quark masses and lattice spacing, 
in which the usual power counting is $m\sim a^2$.
The leading order (LO) correction to continuum $\chi$PT is
the $\CO(a^2)$ potential, which was determined for a single
flavor of staggered fermion in Ref.~\cite{LS}.
This potential retains an $SO(4)$ subgroup of the continuum $SU(4)$
taste symmetry. Since this subgroup is larger than the discrete
lattice symmetry group, the masses of the PGBs in certain lattice
irreps are predicted, at LO, to be degenerate.
Reference~\cite{BA1} generalized the potential to multiple staggered flavors, and showed that the degeneracies remain.
These predictions of S$\chi$PT work very well;
both in old quenched results~\cite{JLQCD1, JLQCD2, OT}
 and in recent dynamical simulations
(using the $\sqrt[4]{\mbox{Det}}$ trick)~\cite{MILCspectrum}.
This success gives one confidence in the utility of
S$\chi$PT.

It is well known from continuum $\chi$PT that practical
applications to the properties of the physical PGB mesons requires
one to work at least to next-to-leading order (NLO).
The same is true of chiral extrapolations of lattice 
data~\cite{MITpanel}. Thus for S$\chi$PT to be a practical
tool for guiding extrapolations, it must be extended beyond LO.
Important first steps were taken in Refs.~\cite{BA1} and~\cite{BA2},
where the one-loop contributions to the mass and decay
constant of the lattice Goldstone boson (that with taste $\xi_5$) 
were calculated. These forms, and in particular the
taste-breaking built into them, have proved essential to describe
the numerical results~\cite{MILCf}. 
Conversely, the success of the fits provides more evidence
for the applicability of S$\chi$PT, and the lack of problems introduced
by the $\sqrt[4]{\mbox{Det}}$ trick.

In this paper we continue the extension of S$\chi$PT to NLO by
determining all the operators proportional to $a^2 p^2$, $a^2 m$
and $a^4$, including source terms for left- and right-handed
currents and for scalar and pseudoscalar densities.
These operators give the analytic NLO contributions that
incorporate the effects of discretization. 
Each is multiplied by a different, unknown low energy coefficient
(LEC). The full NLO result is obtained by adding 
their contributions to those from one-loop
diagrams and from the continuum NLO operators.

In continuum $\chi$PT there are of order ten NLO operators,
and it is possible to consider enough physical quantities to
determine their coefficients and make predictions.
In S$\chi$PT we find that there are an order of magnitude
more NLO operators. While, in principle, there is a corresponding
increase in the number of physical quantities that can be calculated
(e.g. scattering amplitudes with many choices of external tastes),
in practice it will be difficult to carry through a NLO program
along the lines of that in the continuum. 

Nevertheless, our enumeration of operators is not a purely academic
exercise, for there is one set of quantities for which very few
operators contribute, and a number of new predictions can be made.
These are quantities which are non-vanishing only because
the $SO(4)$ taste symmetry is broken.
Since this symmetry is preserved by the LO Lagrangian, it is also
preserved by the one-loop contributions, as these involve LO vertices.
Thus the only source of $SO(4)$ breaking is the NLO operators.
Furthermore, we find that for pion masses and pseudoscalar decay constants
(i.e. the vacuum to pion matrix element of the pseudoscalar density),
the only contribution to $SO(4)$ breaking comes from the operators
proportional to $a^2 p^2$, and that there are few enough of these that
we can make several predictions. These are valid only at NLO, 
being violated by higher order terms.  Comparison of these S$\chi$PT predictions with lattice data numerically tests the $\sqrt[4]{\mbox{Det}}$ trick.

\medskip

Discussions of staggered fermions can quickly become rather
technical, so in an attempt to keep the main text of this paper accessible,
we relegate most of the details of operator enumeration to an appendix. 
The precise organization is as follows.
In Sections \ref{sec:Review} and
\ref{sec:SChPT} we review the construction of the continuum quark-level effective
staggered action and leading order staggered chiral perturbation theory,
respectively.  In Section~\ref{sec:newops} we discuss the symmetry-breaking pattern exhibited by the $\CO(a^2 p^2)$, $\CO(a^4)$, and $\CO(a^2 m)$
operators, and outline the general consequences for various PGB properties.  Our specific predictions for relations between $SO(4)$ breaking in masses, dispersion relations and pseudoscalar decay constants are presented
and explained in Section \ref{sec:decay}. We conclude in Section~\ref{sec:conc}.  We determine the NLO staggered operators in Appendix~\ref{app:ops}, and collect them in the tables of Appendix~\ref{app:tables}.
          
\section{Continuum effective Lagrangian for staggered quarks}\label{sec:Review}

The construction of the chiral Lagrangian including discretization
errors proceeds in two steps~\cite{ShSi}. 
One first determines the continuum effective Lagrangian for
quarks with momenta much smaller than the lattice cut-off $1/a$. 
The dominant power-law
dependence on the lattice spacing is thereby made explicit.
In the second step
one maps the continuum quark-level Lagrangian onto an effective
chiral Lagrangian.
In this section we briefly review the first step for staggered fermions,
including in the effective Lagrangian the
leading discretization effects, which are proportional to $a^2$.
This Lagrangian was determined for a single flavor in Ref.~\cite{LS},
and generalized to multiple flavors in Ref.~\cite{BA1}.
While we obtain no new results here,
we attempt to clarify aspects of the discussion in these papers.
%We do not discuss the impact of the
%$\sqrt[4]{\mbox{Det}}$ trick in this section.

\medskip

The construction of the effective quark Lagrangian follows the
standard method developed by Symanzik~\cite{Symanzik}, in which  
one determines all local continuum operators of dimension up to
and including six that are invariant under the lattice symmetry group.
Given the complexity of the symmetry group for staggered 
fermions~\cite{GS},
a simple way to perform this enumeration is to first write down 
allowed \emph{lattice} operators of dimension up to six, and then
match them onto continuum operators.
The first step utilizes all of the lattice symmetries, including
translations. It is discussed in detail in Appendix A of Ref.~\cite{LS}.
The second step uses only transformations that map the $2^4$ hypercube
onto itself, which are related to spin and taste transformations
in the continuum~\cite{Verstegen}.
As is well known, there are no dimension 5 operators~\cite{Sharpe2,Luo2}.
At dimension 6 one finds purely gluonic operators,
fermion bilinears and four-fermion operators.
These operators were enumerated in Refs.~\cite{Luo} and~\cite{LS}.

Neither the
gluonic operators nor the fermion bilinears break the continuum taste
symmetry because they do not contain any taste matrices.
Since our major focus is taste symmetry breaking, we do not reproduce
these operators here. They do, however, break the continuum $SO(4)$
rotation symmetry down to its hypercubic subgroup, $SW_{4}$,
and we discuss this issue briefly in Appendix~\ref{app:ops}.

$SU(4)$ taste symmetry is broken by the four-fermion operators. We first
recall the results for a single flavor and then generalize to multiple
flavors. We write quark bilinears using the direct-product (spin
$\otimes$ taste) notation of Refs.~\cite{DS, SP, PS}.\footnote{In the
standard staggered convention, the spin matrices, $\gamma_S$, are
Euclidean gamma matrices, while the taste matrices, $\xi_T$, are
complex-conjugated Euclidean gamma matrices.}  
Each four fermion operator in the effective Lagrangian
turns out to be a product of two bilinears with the same spin and
taste. Thus one can use a compact notation to represent them.
Following Refs.~\cite{SP,LS} we write the pair of spin matrices in the
two bilinears as
\begin{eqnarray}
&&S \equiv 1\otimes 1\,, \quad P \equiv \gamma_5\otimes
\gamma_5\,,\quad V \equiv \sum_\mu \gamma_\mu \otimes \gamma_\mu
\,,\quad A \equiv \sum_\mu \gamma_{\mu5} \otimes \gamma_{5\mu}\,,\quad
T \equiv \sum_{\mu<\nu} \gamma_{\mu\nu} \otimes \gamma_{\nu\mu} \,,
\label{eqn:LSnotation}
\end{eqnarray}
using the definitions $\gamma_{\mu\nu}\equiv \gamma_\mu \gamma_\nu$
and $\gamma_{\mu5}\equiv \gamma_\mu \gamma_5$.
The pair of taste matrices are denoted 
in the same manner with $\gamma_\mu \to \xi_\mu$.  
The four fermion operator is then denoted by the
spin and taste matrices that it contains. For example
\begin{equation}
[A\times T] \equiv
\sum_{\mu}\sum_{\nu<\rho}
\bar{Q} \sfno{\mu5}{\nu\rho} Q
\,\bar{Q} \sfno{5\mu}{\rho\nu} Q
\,,
\label{eq:AT}
\end{equation}
where $Q,\bar{Q}$ are Dirac fields with spinor and taste indices.
We stress that $Q,\bar{Q}$ are at this stage continuum fields, 
and that the two $Q$'s and two $\bar{Q}$'s in this operator 
reside at the same position in Euclidean space.

The fields $Q,\bar{Q}$ also have color indices, which can be contracted
in (\ref{eq:AT}) either within each bilinear or between them.
These two color contractions are related by Fierz transformations.
In Ref.~\cite{LS}, these transformations were used to select
a basis in which all operators had the same type of color contraction.
Here we proceed differently, following Ref.~\cite{BA1}. 
We keep both color contractions,
and use the notation such as $[A\times T]$ to implicitly refer to both.
This is a sensible notation because, when we map these operators into
the chiral Lagrangian, the choice of color contraction is irrelevant:
the two operators map onto the same chiral operators.
Using this notation allows us to shorten the list of operators,
since by Fierz transformations we can always bring the operators
into a form in which both bilinears have  ``odd distance''
(see Ref.~\cite{LS} for an explanation of this terminology).
For example, the operator $\{[S\times S] - [P\times P]\}$,
which consists of two bilinears both of distance zero, can be
transformed into a linear combination of operators with odd-distance bilinears.

We now list the resulting four-fermion operators. They can
be divided into two types based on how badly they break taste symmetry.  
Twelve have their spin and taste indices contracted separately, 
as in the example of (\ref{eq:AT}),
and are therefore rotationally-invariant:
\begin{eqnarray}
\CL_6^{\rm FF(A)} &\sim&
\;\; [S\times A] + [S\times V] + [A\times S] + [V\times S] + [P\times V] + [P\times A] + [V\times P] + [A\times P] +\nonumber \\
&& \;\; [T\times V] + [T\times A] + [V\times T] + [A\times T] 
%+ \left\{[S\times S]-[P\times P]\right\} +
%\left\{[S\times P]-[P\times S]\right\} +\nonumber \\
%&& \left\{[S\times T]-[P\times T]\right\} +
%\left\{[T\times S]-[T\times P]\right\} + 
%\left\{[V\times V]-[A\times A]\right\} + 
%\left\{[V\times A]-[A\times V]\right\} 
\,.
\label{eq:L6FFA}
\end{eqnarray}
Each operator has a unique coefficient, 
which we do not show, that is proportional to $a^2$,
but also depends on $g(a)^2$ and ln($a$).  
Because the indices on the taste matrices are contracted between
bilinears, these operators preserve an $SO(4)$ subgroup of the full
$SU(4)$ taste symmetry.  The four remaining four fermion operators have
spin and taste indices contracted together:
\begin{eqnarray}
\CL_6^{\rm FF(B)} &\sim&
\;\;\;\;[T_\mu\times V_\mu] + [T_\mu\times A_\mu] + [V_\mu\times T_\mu] +
\,[A_\mu\times T_\mu] 
%+ \nonumber \\
%&&\left\{([V_\mu\times V_\mu]-[A_\mu\times A_\mu]) -\frac{1}{4}
%([V\times V]-[A\times A])\right\} + \nonumber \\
%&&\left\{([V_\mu\times A_\mu]-[A_\mu\times V_\mu]) -\frac{1}{4}
%([V\times A]-[A\times V])\right\} 
\,.
\label{eq:L6FFB}
\end{eqnarray}
[The precise meaning of this notation can be seen from 
the example in (\ref{eq:VTdef}).]
They are therefore only invariant under certain combined spin and
taste rotations, and break the full spin-taste symmetry down to the
discrete subgroup respected by the lattice theory,
$\Gamma_4 \semitimes SW_4$~\cite{KS}.
In other words, these ``FF(B)'' operators 
break the continuum symmetry maximally.

One advantage of this basis is that the $U(1)_A$ symmetry of
the lattice theory (which is exact in the chiral limit) is manifest.
Under this symmetry
\begin{equation}
Q \longrightarrow \exp(i\alpha[\gamma_5\otimes \xi_5])\, Q\,,\quad
\bar{Q} \longrightarrow \bar{Q}\, \exp(i\alpha[\gamma_5\otimes \xi_5]) \,,
\label{eq:U1}
\end{equation}
and each of the operators in (\ref{eq:L6FFA}) and (\ref{eq:L6FFB})
is separately invariant.
Another advantage is that there are fewer operators to consider;
Ref.~\cite{LS} had an additional six $FF(A)$ and two $FF(B)$ operators.

\medskip

Thus far we have only considered the single flavor theory.
We now recall the generalization to $n>1$ flavors~\cite{BA1}.
The quark field picks up a flavor index, $Q\to Q_i$,
and bilinear operators generalize straightforwardly,
e.g. $m \bar{Q} Q \to \sum_i m_i \bar{Q}_i Q_i$.
The generalization of the four-fermion operators is
more subtle. Since these operators are the result of 
hard gluon exchange, and gluons do not change flavor,
they can be written in a form containing flavor diagonal bilinears, i.e.
\begin{equation}
\sum_{i,j} \bar{Q}_i (\gamma_S \otimes \xi_T) Q_i 
\bar{Q}_j (\gamma_S \otimes \xi_T) Q_j \,.
\label{eq:FFunmixed}
\end{equation}
Of course we could Fierz transform such an operator to bring
it into a ``flavor-mixed'' form, but it is advantageous
not to do so. This is because, in the unmixed basis (\ref{eq:FFunmixed}),
each bilinear must have odd distance in order to be invariant under
the lattice axial symmetry, as discussed in the following paragraph.
Thus the spin-taste structure of the $FF(A)$ and $FF(B)$ four-fermion operators 
remains as in (\ref{eq:L6FFA}) and (\ref{eq:L6FFB}), respectively,
but the flavor structure is now given by 
(\ref{eq:FFunmixed}).
As for the single flavor case, 
there are two possible color contractions of each operator, 
but we do not show these explicitly as they play no role
when we map the operators into the chiral Lagrangian.

We emphasize that the result that only odd-distance bilinears
appear in the flavor unmixed basis follows from symmetries alone.
The discussion of gluon exchanges is given only to provide
motivation for considering the flavor unmixed basis,
but is not strictly necessary.
Since the four-fermion operators remain in the chiral limit,
they must be invariant under the symmetries of the lattice
theory in that limit. These include 
$U(n)$ axial transformations, which act as in (\ref{eq:U1}),
except that $\alpha$ is now an Hermitian $n\times n$ flavor matrix,
and $Q$ a flavor vector.\footnote{%
These transformations do not form a group for $n>1$, but this is
not important for our considerations.}
In the lattice theory, this corresponds to rotating even and odd sites
by opposite $U(n)$ transformations.
It is easy to check that the operators in (\ref{eq:L6FFA}) and (\ref{eq:L6FFB}) 
are invariant under $U(n)$ transformations if they
are composed of flavor diagonal bilinears.
On the other hand, compound operators with spin-taste structure such as 
$[S\times S] - [P\times P]$
are not invariant for more than one flavor.
We stress that it is not sufficient to consider only $U(1)^n$ transformations
(in which each flavor is rotated independently),
since this does not rule out compound operators of the form 
\begin{equation}
\sum_{i}\left[
 \bar{Q}_i (\gamma_I \otimes \xi_I) Q_i \bar{Q}_i (\gamma_I \otimes \xi_I) Q_i
- 
 \bar{Q}_i (\gamma_5 \otimes \xi_5) Q_i \bar{Q}_i (\gamma_5 \otimes \xi_5) Q_i
\right]
 \,.
\end{equation}
These are only ruled out by their lack of invariance under general $U(n)$ transformations.

\medskip

Finally, we comment briefly on operators of yet higher dimension.
Since we aim to determine operators in S$\chi$PT proportional to $a^4$,
one might have thought that we need to enumerate all dimension 8 operators
in the continuum effective theory (dimension 7 operators are
forbidden by the axial symmetry). Fortunately this is not the case.
The naive argument for this is that the dimension 6 operators break
the continuum symmetries down to the lattice subgroup, and so dimension
8 operators break no further symmetries. Since it is the
symmetries that are used to map to the chiral Lagrangian, the
dimension 8 operators are not needed. This argument is, however,
incomplete. What matters in addition is the number of taste matrices
with correlated indices in the quark-level operators.
If there are more such taste matrices, then one can build different
operators in the chiral Lagrangian.
We explain this point in Appendix~\ref{app:ops}, where in
Section~\ref{app:singleFFB} we show that certain operators in the chiral
Lagrangian do not arise until $\CO(a^8)$, 
since they require four tensor taste matrices with fully correlated
indices, and this does not occur unless one has an eight-quark operator.
Now, the dimension 6 operators have at most two taste matrices,
as exemplified by $[A\times T]$ in (\ref{eq:AT}).
Moving to dimension 8 adds further derivatives or gluon fields,
but no further bilinears, and thus no more taste matrices.
For this reason we do not need to enumerate the dimension 8 operators.

\section{Chiral Lagrangian for staggered quarks at Leading Order}\label{sec:SChPT}

In this section we review the mapping of the continuum quark effective
Lagrangian into the effective chiral Lagrangian, working at leading
order in our joint our expansion about the chiral and continuum limits.
This was done for a single staggered flavor in Ref.~\cite{LS},
and generalized to multiple flavors in Refs.~\cite{Bernard,BA1}.

If there are $n$ flavors of staggered fermions, then,
in the combined chiral-continuum limit, the theory possesses  
an $SU(4n)_L \times SU(4n)_R$ chiral symmetry.
We assume that, as in QCD, this breaks spontaneously down to
$SU(4n)_V$.\footnote{%
We do not need to worry about a possible change in the dynamics
for large numbers of flavors, e.g. the loss of asymptotic freedom,
because practical applications always involve three or less light 
\emph{dynamical} flavors, for the which the chiral symmetry breaking
pattern of QCD should apply. The reduction from $4n$ quarks to
2-3 sea quarks is accomplished by partial quenching, as explained
at the end of this section. The theoretical status of symmetry
breaking in partially quenched theories has been discussed in
Ref.~\cite{SS1}.}
This symmetry breaking pattern leads to $16n^2-1$ pseudo-Goldstone bosons.  
We can collect these in the usual way into an $SU(4n)$ matrix:
\begin{equation}
\Sigma=\exp(i\Phi / f) \,,
\end{equation}
where $\Phi$ is a traceless $4n \times 4n$ matrix:
\begin{eqnarray}\label{eq:Phi}
	\Phi = \left( \begin{array}{cccc} U & \pi^+ & K^+ & \cdots \\*
	\pi^- & D & K^0 & \cdots \\* K^- & \bar{K^0} & S & \cdots \\*
	\vdots & \vdots & \vdots & \ddots \end{array} \right)\,,
\end{eqnarray}
with $4 \times 4$ submatrices:
\begin{equation}
U = \sum_{a=1}^{16} U_a T_a \,,
\end{equation}
and so forth.  The pion decay constant, $f$, is normalized such that
$f_\pi \approx 132\,$MeV.  We choose to express the $SU(4)$ generators
in the following Hermitian basis:
\begin{equation}\label{eq:T_a}
	T_a = \{ \xi_5, i\xi_{\mu 5}, i\xi_{\mu\nu}, \xi_{\mu},
	\xi_I\}\,,
\end{equation}
where $\xi_I$ is just the $4 \times 4$ identity matrix.  It is
important to retain the taste singlet meson, $U_I \propto tr(U)$,
because with $n$ flavors (and thus $4n$ tastes), only the overall
$SU(4n)$ singlet decouples.  Under
an $SU(4n)_L \times SU(4n)_R$ chiral symmetry transformation, $\Sigma$
transforms as
\begin{equation}
\Sigma \rightarrow  L \Sigma R^{\dagger}\,,
\label{eq:sigtrans}
\end{equation}
where $L \in SU(4n)_L$ and $R \in SU(4n)_R$.

We use the following power-counting scheme when determining the
staggered chiral Lagrangian:
\begin{equation}
p^2/\Lambda_{QCD}^2 \approx m/\Lambda_{QCD} \approx a^2
\Lambda_{QCD}^2 \,,
\label{eqn:PC}\end{equation}
which is consistent with parameters of current simulations.
The lowest order Lagrangian, which is of $\CO(p^2,m,a^2)$, is 
\begin{equation}\label{eq:LOLagrangian}
	\CL_\chi = \frac{f^2}{8} \Tr(\partial_{\mu}\Sigma
	\partial_{\mu}\Sigma^{\dagger}) - \frac{1}{4}\mu f^2 \Tr(\CM
	\Sigma + \CM \Sigma^{\dagger}) 
% + \frac{2m_0^2}{3}(U_I + D_I + S_I + \cdots)^2 
        + a^2\CV\,,
\end{equation}
where $\CM$ is the quark mass matrix:
\begin{eqnarray}
	\CM = \left( \begin{array}{cccc} m_u I & 0 &0 & \cdots \\* 0 &
	m_d I & 0 & \cdots \\* 0 & 0 & m_s I & \cdots\\* \vdots &
	\vdots & \vdots & \ddots \end{array} \right)\,.
\end{eqnarray}
and $\CV$ is the taste-breaking potential resulting from the
four-fermion operators in the quark effective action.
Note that $\Tr$ is a full $4n \times 4n$ trace in both flavor and
taste space.  The dimensionful constant, $\mu$, is of
$\CO(\Lambda_{QCD})$, and is defined such that the 
PGB mass is given at LO by
\begin{equation}
	\left(m_{\pi}^2\right)_{\rm LO}=2\mu \frac{m_i + m_j}{2} + a^2 \Delta_F \,.
\label{eqn:LOmass}\end{equation}
The labels $i$ and $j$ indicate the flavors of quarks in the PGB,
here assumed to be different, 
while the splitting, $\Delta_F$, depends on the taste.

\medskip

The mapping of the four-fermion operators enumerated in
the previous section into the mesonic operators in $\CV$ 
is done by treating the taste matrices as spurions.
Having worked out the implications of the symmetry under axial $U(n)$ 
transformations at the quark level, 
no further subtleties arise in the mapping.
The method of Ref.~\cite{LS}, generalized to multiple flavors~\cite{BA1},
can be used. A key result is that only the $FF(A)$ operators,
i.e. those in (\ref{eq:L6FFA}), contribute to $\CV$, so
the potential has a larger symmetry than the underlying lattice theory.
We give a brief recapitulation of the determination of the form of $\CV$
at the end of Secs.~\ref{app:FFAVA} and \ref{app:FFASP}, as a byproduct
of our extension of the methodology to NLO.
In the notation of Ref.~\cite{BA1}, the result is
\begin{equation}
\CV = \CU +\CU\,'
\end{equation}
where
\begin{eqnarray}
	-\CU \equiv \sum_{k} C_k \CO_k & = & C_1
	\Tr(\xi^{(n)}_5\Sigma\xi^{(n)}_5\Sigma^{\dagger}) \nonumber
	\\* & & +C_3\frac{1}{2} \sum_{\nu}[ \Tr(\xi^{(n)}_{\nu}\Sigma
	\xi^{(n)}_{\nu}\Sigma) + h.c.] \nonumber \\* & &
	+C_4\frac{1}{2} \sum_{\nu}[ \Tr(\xi^{(n)}_{\nu 5}\Sigma
	\xi^{(n)}_{5\nu}\Sigma) + h.c.] \nonumber \\* & & +C_6\
	\sum_{\mu<\nu} \Tr(\xi^{(n)}_{\mu\nu}\Sigma
	\xi^{(n)}_{\nu\mu}\Sigma^{\dagger})
\label{eq:U}\end{eqnarray}
and
\begin{eqnarray}
	-\CU\,' \equiv \sum_{k'} C_{k'} \CO_{k'} & = &
	C_{2V}\frac{1}{4} \sum_{\nu}[ \Tr(\xi^{(n)}_{\nu}\Sigma)
	\Tr(\xi^{(n)}_{\nu}\Sigma) + h.c.] \nonumber \\*
	&&+C_{2A}\frac{1}{4} \sum_{\nu}[ \Tr(\xi^{(n)}_{\nu
	5}\Sigma)\Tr(\xi^{(n)}_{5\nu}\Sigma) + h.c.] \nonumber \\* & &
	+C_{5V}\frac{1}{2} \sum_{\nu}[ \Tr(\xi^{(n)}_{\nu}\Sigma)
	\Tr(\xi^{(n)}_{\nu}\Sigma^{\dagger})]\nonumber \\* & &
	+C_{5A}\frac{1}{2} \sum_{\nu}[ \Tr(\xi^{(n)}_{\nu5}\Sigma)
	\Tr(\xi^{(n)}_{5\nu}\Sigma^{\dagger}) ],
\label{eq:U_prime}\end{eqnarray}
Here $\xi_T^{(n)}$ is the $4n\times 4n$ generalization of the taste-matrix:
\begin{equation}
	\xi_T^{(n)} = \left( \begin{array}{cccc} \xi_T & 0 & 0 &
	\cdots\\* 0 & \xi_T & 0 & \cdots\\* 0 & 0 & \xi_T & \cdots \\*
	\vdots &\vdots & \vdots &\ddots\end{array} \right)\,,
\label{eq:xi_T}\end{equation}
with $\xi_T$ the ordinary $4 \times 4$ taste matrix.
This flavor diagonal form mirrors that of the underlying quark operators,
(\ref{eq:FFunmixed}).

The potential is both rotationally and $SO(4)$ taste symmetric.
Therefore the PGB sector respects a \emph{larger} symmetry than the
lattice theory at $\CO(a^2)$.  Although the 16 PGB masses (for a given
choice of constituent quark flavors) are no longer 
degenerate as in the continuum, they split into five degenerate groups
according to $SO(4)$ representations having tastes $\xi_5, \xi_A, \xi_T, \xi_V$, and
$\xi_I$.  The taste $\xi_5$ PGB is the lattice Goldstone boson and receives no
mass correction from $\CV$ because of the exact axial symmetry.  
Thus $\Delta_F$ in (\ref{eqn:LOmass}) in fact only depends
on the $SO(4)$ representation of the particular taste.

The potential can be simplified for a single flavor using Fierz transformations.
In particular, all double-trace operators can be transformed into 
operators with only single traces. This single trace basis is that used in Ref.~\cite{LS}.
Such transformations, however, are not useful for multiple flavors because they
do not maintain the flavor diagonal form of $\xi_T^{(n)}$.
Indeed, as realized in Ref.~\cite{BA1}, the double trace operators of
$\CU'$ give rise to a previously unnoticed effect, namely the presence
of hairpin diagrams (quark disconnected contractions) for flavor singlet
mesons with non-trivial taste. In the LO potential,
these occur only for vector and axial tastes, but, as we show below,
at NLO they arise for all tastes.

Given the importance of the axial $U(n)$ transformations in restricting the
form of the underlying quark operators, as explained in the previous section,
it is worthwhile discussing these transformations at the mesonic level.
They are simply $SU(4n)_A$ transformations, acting on $\Sigma$ as in (\ref{eq:sigtrans}),
with $L=R^\dagger= \exp( i \alpha \otimes \xi_5)$. Here,
as in the previous section, $\alpha$ is an Hermitian $n\times n$ matrix
acting in flavor space, while $\xi_5$ acts in taste space.  
Using the commutation relations among the taste matrices it is easy to show that
\begin{eqnarray}
L \xi_\mu^{(n)} L &= &\xi_\mu^{(n)}\,, 
\quad  L \xi_{\mu 5}^{(n)} L = \xi_{\mu 5}^{(n)} 
\nonumber \\ 
L \xi_5^{(n)} L^\dagger &= &\xi_5^{(n)}\,,\quad
 L \xi_{\mu\nu}^{(n)} L^\dagger =\xi_{\mu\nu}^{(n)}\,.
\label{eqn:UnA}\end{eqnarray}
It then follows immediately that all the operators in 
the potential $\CV$ are invariant under these transformations.

\medskip

We conclude by generalizing the results to partially quenched (PQ) theories,
since partial quenching is often used in practical simulations.
Following Ref.~\cite{BG}, one introduces valence quark flavors
with ghost partners, each coming here in four tastes.
There are then $N=4 n_{\rm val} + 4 n_{\rm sea}$ quarks
and $M=4 n_{\rm val}$ ghosts in total, 
where $n_{\rm val}$  and $n_{\rm sea}$ are the number of valence
and sea flavors, respectively.
Note that the unquenched sea-quark sector lies within the PQ theory,
so one loses no generality by considering the larger theory~\cite{SS2}.

As noted in Ref.~\cite{BA1}, 
the form of the chiral Lagrangian in the PQ theory is unchanged, 
except that the chiral symmetry group becomes $SU(N|M)_L\times SU(N|M)_R$,
traces are replaced by supertraces, and the PGB matrix is enlarged.
For example, in a theory with two valence quarks ($x$ and $y$) 
and three sea quarks ($u$, $d$ and $s$), 
the PGB and mass matrices become (schematically)
\begin{eqnarray}
	\Phi = \left( \begin{array}{cccccccc} X & P^+ & \cdots &
	\cdots & \cdots & \cdots & \cdots \\* P^- & Y & \cdots &
	\cdots & \cdots & \cdots & \cdots \\* \vdots & \vdots & U &
	\pi^+ & K^+ & \cdots & \cdots \\* \vdots & \vdots & \pi^- & D
	& K^0 & \cdots & \cdots \\* \vdots & \vdots & K^- & \bar{K^0}
	& S & \cdots & \cdots \\* \vdots & \vdots & \vdots & \vdots &
	\vdots & \tilde{X} & \tilde{P}^+ \\* \vdots & \vdots & \vdots
	& \vdots & \vdots & \tilde{P}^- & \tilde{Y} \\* \end{array}
	\right)\,, \;\;\;\; \CM = \left( \begin{array}{cccccccc} m_x I
	& 0 & \cdots & \cdots & \cdots & \cdots & \cdots \\* 0 & m_y I
	& 0 & \cdots & \cdots & \cdots & \cdots \\* \vdots & 0 & m_u I
	& 0 & \cdots & \cdots & \cdots \\* \vdots & \vdots & 0 & m_d I
	& 0 & \cdots & \cdots \\* \vdots & \vdots & \vdots & 0 & m_s I
	& 0 & \cdots \\* \vdots & \vdots & \vdots & \vdots & 0 & m_x I
	& 0 \\* \vdots & \vdots & \vdots & \vdots & \vdots & 0 & m_y I
	\\* \end{array} \right)\,,\end{eqnarray}
where each block is a $4\times4$ taste matrix.
The taste matrices $\xi_T^{(n)}$ are also enlarged to size $(N+M)^2$,
but they maintain their flavor diagonal form, (\ref{eq:xi_T}).
In fact, for the sake of brevity we drop the superscript $(n)$
in the remaining sections.
Finally, for practical calculations in S$\chi$PT, it may be simplest
to enlarge $\Sigma$ to be a $U(N|M)$ matrix, and then remove the
singlet component by adding a mass term to $\CL_\chi$
proportional to $m_0^2 \Str(\Phi)^2$, where $m_0$ is sent to infinity at the end~\cite{SS1}.

\section{Chiral Lagrangian for staggered quarks at NLO}\label{sec:newops}

We have extended the construction reviewed in the previous two sections
to NLO. At this order, the new operators introduced by discretization 
are proportional to $a^2 p^2$, $a^2 m$ and $a^4$. We explain in detail
how we enumerate these operators in Appendix~\ref{app:ops}.
This requires combining appropriate numbers of
taste spurions, mass spurions and Lie derivatives
into independent operators that are invariant under the chiral group,
forming parity invariant combinations, and reducing them
to a minimal set using the equations of motion.
We use the graded group theory method of Ref.~\cite{Me} to
determine the number of linearly-independent operators.
We include sources for left- and right-handed currents and scalar and pseudoscalar
densities, so that we can calculate matrix elements of these operators.
This leads to additional operators of $\CO(a^2 p^2)$ that are proportional
to the sources, and thus contribute only to PGB decay constants.
We refer to these as ``source'' operators.  We collect all of the NLO taste-violating operators in the tables of Appendix~\ref{app:tables}.

\medskip

In this section we focus on the pattern
of symmetry-breaking exhibited by these NLO operators.  Unlike the LO staggered potential, they come from four-fermion operators in both $S_6^{FF(A)}$ and $S_6^{FF(B)}$.  Thus they break the $SO(4)$-taste symmetry down to the lattice symmetry group.  In order to understand the symmetries and symmetry-breaking, we examine the consequences of these operators for physical observables.  We discuss the symmetries respected by these quantities order-by-order in the combined chiral-continuum expansion.  It is useful to keep in mind the general form of this expansion, illustrated by the NLO PGB mass:
\begin{equation}
\left(m_\pi^2\right)_{\rm NLO} = \left(m_\pi^2\right)_{\rm LO} 
+ \left(\delta m_\pi^2\right)_{1-loop}
+ \left(\delta m_\pi^2\right)_{m^2}
+ \left(\delta m_\pi^2\right)_{a^2 m}
+ \left(\delta m_\pi^2\right)_{a^4} 
\,. \label{eqn:NLOform}
\end{equation}
The subscripts indicate the order, in S$\chi$PT, of the terms which
contribute to the physical quantity,
with $m$ indicating both $m$ and $p^2$.
We use a similar notation for the axial and pseudoscalar decay constants, 
$f^A$ and $f^P$.\footnote{We use ``axial decay constant" to refer to the 
the standard decay constant, which comes from the vacuum-to-pion matrix 
element of the axial current, $\langle 0|A_\mu|\pi \rangle \propto f^A$.
  In a slight 
abuse of terminology, ``pseudoscalar decay constant" refers to the decay 
constant from the corresponding matrix element of the \emph{pseudoscalar 
density}, $\langle 0|P|\pi\rangle \propto f^P$.}

The LO pion mass is given in (\ref{eqn:LOmass}), while the LO decay constants are
\begin{equation}
\left(f^A\right)_{\rm LO} = f\,,\qquad
\left(f^P\right)_{\rm LO} = \mu f\,.
\label{eq:fLO}
\end{equation}
The results of the previous section imply that the LO contributions to
these quantities respect the $SO(4)$ taste symmetry, since 
the potential $\CV$ does. This has already been discussed for the PGB mass,
but also holds for the decay constants.
In fact, the LO decay constants are $SU(4)$ symmetric, 
since $\CV$ does not contribute to either one.
The staggered potential also generates interesting predictions for the difference between
the properties of flavor singlet and non-singlet particles.
This difference is predicted to vanish for decay constants of all tastes,
and for the masses of taste $\xi_5$ and $\xi_{\mu\nu}$ pions
(since there are only vector and axial hairpins).\footnote{%
We are phrasing this discussion as if the hairpin vertices iterate to give
only a mass shift for flavor singlets. 
As is well known, however, this is not generally
true in PQ theories because valence quark loops are absent,
and there are factors of $1/4$ added by hand due to the 
$\sqrt[4]{\mbox{Det}}$ trick. There will generally be additional, unphysical
double poles in correlators~\cite{SS2}.
Nevertheless, one can always determine the size of the hairpin vertices
by measuring the residue of the double pole and comparing to
the prediction of PQ S$\chi$PT, as discussed below.}
One of our aims is to determine at what order these predictions fail.

The form of the 1-loop contributions is
\begin{eqnarray}
\left(\delta m_\pi^2\right)_{1-loop} &\sim&
\left[\left(m_\pi^2\right)_{\rm LO} + a^2\right]^2
\mbox{ln}\left(m_\pi^2\right)_{\rm LO}
\label{eqn:1loopmass}\\
\left(\delta f^{A,P}\right)_{1-loop} &\sim&
\left[\left(m_\pi^2\right)_{\rm LO} + a^2\right]
\mbox{ln}\left(m_\pi^2\right)_{\rm LO}
\label{eqn:1loopf}\,.
\end{eqnarray}
These involve only LO vertices, and thus do not 
break the $SO(4)$ taste symmetry.
They do, however, break the $SU(4)$ symmetry of decay constants
down to $SO(4)$.
Note that, unlike in continuum $\chi$PT, the coefficients of the logarithms are not proportional to
the LO mass---there are additional $a^2$ terms from the hairpin vertices and four-pion vertices produced by $\CV$. The exception is in the
lattice Goldstone mass, which is protected by the $U(1)_A$ symmetry.
The one-loop contributions to the lattice Goldstone mass and axial decay constant have
been calculated in Ref.~\cite{BA1,BA2}.

Now we turn to the analytic NLO contributions.
The generic form of these contributions are as follows:
\begin{equation}
\left(\delta m_\pi^2\right)_{m^2} \sim m^2\,,\qquad
\left(\delta m_\pi^2\right)_{a^2 m} \sim a^2 m\,,\qquad
\left(\delta m_\pi^2\right)_{a^4} \sim a^4\,,\qquad
\label{eq:deltampiNLO}
\end{equation}
\begin{equation}
\left(\delta f^{A,P}\right)_{m^2} \sim m\,, \qquad
\left(\delta f^{A,P}\right)_{a^2 m} \sim a^2\,.
\label{eq:deltafNLO}
\end{equation}
\begin{equation}
\left(\CA^{4 \pi}\right)_{m^2} \sim m^2 + m p^2 + p^4\,, \qquad
\left(\CA^{4 \pi}\right)_{a^2 m} \sim a^2 m + a^2 p^2\,, \qquad
\left(\CA^{4 \pi}\right)_{a^4} \sim a^4\,.
\label{eq:deltaANLO}
\end{equation}
Here we begin to include the 4-pion scattering amplitude, $A^{4 \pi}$, as it is the simplest quantity showing the most general pattern of contributions from the NLO operators.  In these expressions we use a schematic notation, leaving out factors of $\Lambda_{\rm QCD}$, but distinguishing between mass and momentum-dependent terms.
Note that there is no $\CO(a^4)$ contribution to $f^{A,P}$ because such terms
in the chiral Lagrangian contain no sources.
In order to determine whether there are relationships between splittings,
one needs to know which types of operators from
Appendix~\ref{app:tables} contribute to which of the corrections
in (\ref{eq:deltampiNLO}-\ref{eq:deltaANLO}).
This information is collected
in Table~\ref{tab:NSpattern} for single supertrace operators,
and in Table~\ref{tab:Spattern} for double supertrace operators.
The former are the only contributions to the masses
and decay constants of flavor non-singlet mesons,
while the latter gives the additional
hairpin contributions for flavor singlets.  The distinction between single and double supertraces is less significant for the scattering amplitude, so we only include it in the first table.

\begin{table}\begin{tabular}{lccccccc}  \hline\hline

\multicolumn{1}{c}{\emph{Operator}} 
\vphantom{$\left(m_\pi^2\right)_{a^2 m}$} 
& Table & $\;SO(4)\;$? &
$\;\left(\delta m_\pi^2\right)_{a^2 m}\;$ &
$\;\left(\delta m_\pi^2\right)_{a^4}\;$ &
$\;\left(\delta f^A\right)_{a^2 m}\;$ &
$\;\left(\delta f^P\right)_{a^2 m}\;$ &
$\;\left(\delta \CA^{4\pi}\right)_{a^2 m,a^4}\;$ 
\\[0.5mm] \hline

$a^2 p^2$ ($FF(A)$) & \ref{tab:FFA_VA},\ref{tab:FFA_SPa2} & Y & 
Y & N & Y & Y & Y\\[0.5mm]

$a^2 m$ ($FF(A)$) & \ref{tab:FFA_VA},\ref{tab:FFA_SPa2} & Y & 
Y & N & N & Y & Y\\[0.5mm]

$a^4$ ($FF(A)$ \& $FF(B)$) & \ref{tab:a4_VA_VA}-\ref{tab:a4_FFB} & Y & 
N & Y & N & N & Y \\[0.5mm]

$a^2$ source ($FF(A)$) & \ref{tab:source_FFA} & Y & 
N & N & Y & N & N\\[0.5mm]
\hline

$a^2 p^2$ ($FF(B)$) & \ref{tab:FFB} & N & 
Y & N & Y & Y & Y \\[0.5mm]

$a^2$ source ($FF(B)$) & \ref{tab:source_FFB} & N & 
N & N & Y & N & N\\[0.5mm]

$a^4$ ($FF(B)$) & \ref{tab:a4_FFB} & N & 
N & N & N & N & Y \\[0.5mm]

\hline\hline
\end{tabular}\caption{Contributions of 
single supertrace NLO S$\chi$PT operators to physical quantities.
The column labeled ``$SO(4)$?'' indicates
whether the contributions are consistent with $SO(4)$ taste symmetry.
The last five columns indicate which quantities receive contributions
from the particular operators. For further explanation see text.
}\label{tab:NSpattern}\end{table}

\begin{table}\begin{tabular}{lcccccc}  \hline\hline

\multicolumn{1}{c}{\emph{Operator}} 
\vphantom{$\left(m_\pi^2\right)_{a^2 m}$} 
& Table & $SO(4)\;$? &
$\;\left(\delta m_\pi^2\right)_{a^2 m}\;$ &
$\;\left(\delta m_\pi^2\right)_{a^4}\;$ &
$\;\left(\delta f^A\right)_{a^2 m}\;$ &
$\;\left(\delta f^P\right)_{a^2 m}\;$ 
\\[0.5mm] \hline

$a^2 p^2$ ($FF(A)$) & \ref{tab:FFA_VA},\ref{tab:FFA_SPa2} & Y & 
Y & N & Y & Y \\[0.5mm]

$a^2 m$ ($FF(A)$) & \ref{tab:FFA_VA},\ref{tab:FFA_SPa2} & Y & 
$\xi_{\mu(5)}$ only & N & N & $\xi_{\mu(5)}$ only \\[0.5mm]

$a^4$ ($FF(A)$ \& $FF(B)$) & \ref{tab:a4_VA_VA}-\ref{tab:a4_FFB} & Y & 
N & Y & N & N \\[0.5mm]

$a^2$ source ($FF(A)$) & \ref{tab:source_FFA} & Y & 
N & N & $\xi_{\mu(5)}$ only & N \\[0.5mm]
\hline

$a^2 p^2$ ($FF(B)$) & \ref{tab:FFB} & N & 
Y & N & Y & Y \\[0.5mm]

$a^2$ source ($FF(B)$) & \ref{tab:source_FFB} & N & 
N & N & $\xi_{\mu(5)}$ only & N \\[0.5mm]

\hline\hline
\end{tabular}\caption{Contributions of NLO S$\chi$PT operators 
with two supertraces to PGB masses and decay constants. 
Notation as in Table~\ref{tab:NSpattern}. $\xi_{\mu(5)}$ indicates
both vector and axial taste.
}\label{tab:Spattern}\end{table}

As indicated in the tables, there is an important distinction between
the contributions from underlying $FF(A)$ and $FF(B)$ four fermion operators: the former
cannot break the taste $SO(4)$ symmetry, while the latter can.
These two types of operator differ primarily in their index structure.
Those resulting from $FF(A)$ operators must have
indices contracted in pairs, while those from $FF(B)$ operators
must contain more than two identical indices. 
This can be seen by comparing the
$\CO(a^2p^2)$ $FF(A)$ operators in Tables~\ref{tab:FFA_VA} and \ref{tab:FFA_SPa2}
to the corresponding $FF(B)$ operators in Table~\ref{tab:FFB},
or the $\CO(a^4)$ $FF(A)$ operators in 
Tables~\ref{tab:a4_VA_VA}-\ref{tab:a4_VA_SP} to the corresponding
$FF(B)$ operators in Table~\ref{tab:a4_FFB}.
Note, however, that the there are no $\CO(a^2 m)$ contributions
from $FF(B)$ operators because there are not enough indices to contract more than two at a time. Thus $\CO(a^2 m)$ operators do not
break taste $SO(4)$.

An interesting feature of the pattern of $SO(4)$ breaking is
that the $\CO(a^4)$ mesonic operators resulting from double insertions of
underlying $FF(B)$ operators (which are listed in Table~\ref{tab:a4_FFB})
do not break the $SO(4)$ symmetry of pion masses,
despite the fact that they break $SO(4)$ in general.
This is because, when calculating tree-level
masses from operators such as 
\begin{equation}
\sum_{\mu} \sum_{\nu \neq \mu} \Str(\xi_{\mu\nu} \Sigma \xi_\mu \Sigma \xi_{\nu \mu} \Sigma^\dagger \xi_\mu \Sigma
^\dagger)\,,
\end{equation}
two of the four $\Sigma$'s in each operator must be replaced
by the identity, so the taste matrices collapse into
a form which is $SO(4)$ symmetric. 
In contrast, one contribution to the four-pion scattering amplitude replaces each $\Sigma$ with a pion field, 
giving an $SO(4)$ breaking contribution to $\CA^{4\pi}$.
Other contributions do not, however, break $SO(4)$. This is why,
in Table~\ref{tab:NSpattern}, operators of type 
``$a^4$ $FF(B)$'' are listed as both conserving and breaking
$SO(4)$, and why the latter contributes only to $\CA^{4\pi}$.

This distinction between two- and four-pion contributions
does not arise for the $SO(4)$ breaking $\CO(a^2 p^2)$ operators.
These violate $SO(4)$ by correlating indices on derivatives
and taste matrices, e.g.
\begin{equation}
\sum_\mu \Str(\Sigma D_\mu \Sigma^\dagger \xi_\mu
\Sigma^\dagger D_\mu \Sigma \xi_\mu)\,,
\end{equation}
where $D_\mu$ is the covariant derivative.
They produce direction-dependent contributions to masses and 
decay constants, as well as to scattering amplitudes.

As is shown in the tables,
the $\CO(a^2 m)$ operators contribute both to pion masses and
to $f^P$ (since $M$ and $M^\dagger$ are sources for $P$),
but not to $f^A$ (since they contain no derivatives).
In particular, each $a^2 m$ contribution
to PGB properties comes with two independent mass dependencies, $m_x + m_y$
(where $x$ and $y$ are the valence flavors) and
$\sum m_{\rm sea}$, The only exception is the mass of the taste $\xi_5$ pion,
which is guaranteed to vanish when $m_x+m_y \to 0$ by the axial symmetry,
and indeed has no $a^2 \sum m_{\rm sea}$ contributions.
Similarly, it is easy to see that it has no contributions proportional
to $a^4$.  Another feature indicated in the tables is that the ``$a^2$ source'' operators contribute only to $f^A$.

Finally, we compare the entries in the single and double supertrace operator tables, keeping in mind that the double supertrace operators are responsible for splittings between flavor singlet and non-singlet PGB properties.  The entries are almost identical. 
The only difference is that some of the ``Y'' entries in the latter
are replaced by ``$\xi_{\mu(5)}$ only''.  This is  because the $a^2 m$ 
and $a^2$ source two supertrace 
terms only contain vector and axial tastes.  (Recall that the same is true of the LO two supertrace operators in $\CU'$.)
In contrast, operators of $\CO(a^2 p^2)$ and $\CO(a^4)$ with two supertraces contain all possible taste matrices.  

\medskip

We now summarize the consequences of the previous
observations, with the goal in mind to generate \emph{predictions} of S$\chi$PT which can be tested on the lattice.  We focus on the masses and decay constants,
as these quantities are straightforward to calculate
in simulations.
\begin{enumerate}
\item
The $SU(4)$ symmetry in the LO decay constants of flavor
non-singlet PGBs is broken
down to $SO(4)$ by the NLO terms resulting from $FF(A)$ 
operators.\footnote{%
We have checked that there are sufficient independent operators
to completely break the symmetry down to $SU(4)$ in this
and all other similar cases that we mention.}
There is no relation between the $SU(4)\to SO(4)$ breaking
in $f^A$ and $f^P$, since the $a^2 p^2$ and $a^2 m$ operators
contribute differently to these two decay constants.
Furthermore, 1-loop contributions to $m_\pi^2$, $f^A$ and $f^P$
will give independent contributions to the splittings.
Thus there are no predictions between splittings at this stage.

\item
The breaking of $SO(4)$ down to $\Gamma_4\semitimes SW_4$
occurs first in the NLO analytic terms, and not in the 1-loop terms.
It arises only from two types of
operator---$a^2 p^2$ and ``$a^2$ source".  In particular, since both $m_\pi$ and $f^P$ only receive $SO(4)$-breaking contributions from $\CO(a^2 p^2)$ operators, there are relations
between the splittings within $SO(4)$ multiplets.  The ``$a^2$ source" operators break all such relations among splittings in $f^A$.
There are also predictions for the rotational symmetry breaking
in the dispersion relations. 
These results hold separately for flavor singlet and non-singlet pions.

\item
The absence of taste $\xi_5$ and $\xi_{\mu\nu}$ hairpins at LO
does not hold at NLO, where there are \emph{hairpin vertices for all tastes}.
We note that the presence of a hairpin vertex for taste $\xi_5$ is
consistent with its Goldstone nature since
the vertex is proportional to $p^2$, and thus $m_\pi^2$ at the pole.

\item
The hairpin contributions to
$f^A$ and $f^P$ are related to each other
for tastes P and T, since there are no two supertrace operators with sources for either the pseudoscalar density or the axial current.  This is true for both the $SO(4)$-conserving and the $SO(4)$-\emph{violating} contributions.
\end{enumerate}
We discuss the various relations and predictions in detail in the
following section. Here we stress 
that none of these relations follows from a symmetry of the
lattice theory---indeed, we have checked that all of them
are broken by higher dimension operators in S$\chi$PT.

\section{NLO relations for PGB masses and decay constants}\label{sec:decay}

In this section we work out the detailed form of the relations
which follow from the particular form of the NLO analytic terms.

\medskip

We first study $SO(4)$-taste and rotational symmetry breaking in the pion dispersion relations.
This arises from the $a^2 p^2$ $FF(B)$ operators,
which are listed in Table~\ref{tab:FFB}.  
Of the 18 such operators, only 8 contribute to $SO(4)$
breaking in single pion properties. The others, such as
\begin{equation}
\sum_\mu \sum_{\nu\ne\mu} \Str(\partial_\mu \Sigma^\dagger \partial_\mu
 \Sigma \Sigma^\dagger \xi_{\mu\nu} \Sigma \xi_{\nu\mu}) + p.c.
\,,
\end{equation}
(with $p.c.$ indicating parity conjugate)
give either vanishing, or taste symmetric, contributions to
two pion properties, because the two pions must be drawn from
the fields with derivatives acting on them.
The contribution of the $SO(4)$ breaking operators
to the chiral Lagrangian is
\begin{eqnarray}
&& a^2  \;\;\sum_\mu\sum_{\nu \ne \mu} \left\{ 
C_2 \Str(\partial_\mu
\Sigma^\dagger \xi_{\mu\nu} \partial_\mu \Sigma \xi_{\nu\mu}) 
+
C_7 \Str( \Sigma \partial_\mu \Sigma^\dagger \xi_{\mu\nu})
\Str( \Sigma^\dagger \partial_\mu \Sigma \xi_{\nu\mu})  \right.\nonumber \\
&&\left. + C_{10} \left[\Str( \Sigma\partial_\mu
\Sigma^\dagger \xi_{\mu\nu} \Sigma \partial_\mu \Sigma^\dagger \xi_{\nu\mu})
+ p.c. \right]
+ C_{13} \left[ \Str( \Sigma \partial_\mu \Sigma^\dagger \xi_{\mu\nu})
\Str( \Sigma \partial_\mu \Sigma^\dagger \xi_{\nu\mu}) + p.c. \right] \right\}
\nonumber \\ 
&&+ a^2 \;\;\sum_\mu \left\{
C_{36V}\Str(\Sigma \partial_\mu \Sigma^\dagger
\xi_\mu \Sigma^\dagger \partial_\mu \Sigma \xi_\mu) 
+ C_{36A}\Str(\Sigma \partial_\mu \Sigma^\dagger \xi_{\mu 5}
\Sigma^\dagger \partial_\mu \Sigma \xi_{5 \mu}) 
 \right. \nonumber \\
&& \left. + C_{41V} \Str(\partial_\mu \Sigma^\dagger\xi_\mu)\Str(\partial_\mu \Sigma \xi_\mu)  
+ C_{41A} \Str(\partial_\mu \Sigma^\dagger\xi_{\mu 5})
\Str(\partial_\mu\Sigma \xi_{5 \mu})  \right\}\,. 
\label{eqn:decay_ops}\end{eqnarray}
The eight operators reduce to six for two pion contributions,
three single supertrace and three double supertrace operators:
\begin{eqnarray}
&& \frac{a^2}{f^2} \;\;\sum_\mu \Big\{
\sum_{\nu\ne\mu} 
\left[ (C_2-2 C_{10}) \Str(\partial_\mu \Phi \xi_{\mu\nu} \partial_\mu \Phi \xi_{\nu\mu})
+ (C_7-2 C_{13}) \Str(\partial_\mu \Phi \xi_{\mu\nu}) \Str(\partial_\mu \Phi \xi_{\mu\nu})
\right] \nonumber \\
&& \qquad + C_{36V}\; \Str(\partial_\mu \Phi \xi_\mu \partial_\mu \Phi \xi_\mu)
+ C_{41V}\; \Str(\partial_\mu \Phi \xi_\mu) \Str(\partial_\mu \Phi \xi_\mu)
\nonumber \\[.5em]
&&  \qquad
+ C_{36A}\; \Str(\partial_{\mu} \Phi \xi_{\mu5} \partial_\mu \Phi \xi_{5\mu})
+ C_{41A}\; \Str(\partial_{\mu} \Phi \xi_{\mu5})\Str(\partial_\mu \Phi \xi_{5\mu})
\Big\}\,.
\label{eq:decay_ops_2pi}
\end{eqnarray}

We consider first the effect of these operators on the properties of
flavor non-singlet mesons, and in particular those composed of a quark
of flavor $a$ and an antiquark of a different flavor $b$, where these flavors
can be either valence or sea quarks. Examples are kaons and charged pions,
both unquenched and partially quenched.\footnote{%
The following relations also hold for flavor neutral mesons,
e.g the $\pi_0$-like states with flavor structure $(aa-bb)$,
if flavors $a$ and $b$ are degenerate.}
For these particles only the
single supertrace operators contribute, and we find the following results
for their dispersion relations:
%The contribution of these operators to the dispersion relations of the PGB
%mesons is straightforward to calculate. We find
%
\begin{eqnarray}
E^2_{I} & = & \vec{p}\;{}^2 + m_I^2 (1 + \delta_I)\,, 
\\
E^2_{5} & = & \vec{p}\;{}^2 + m_5^2 (1 + \delta_5)\,,
\\
E^2_{k} & = & (p_i^2 + p_j^2) + p_k^2 (1 + \delta_k - \delta_4)+
m_\mu^2 (1 + \delta_k) \,, \\
E^2_{4} & = & \vec{p}\;{}^2 (1 + \delta_4-\delta_k) + m_\mu^2 (1 + \delta_4)\,,
\\
E^2_{{k5}} & = & (p_i^2 + p_j^2) + p_k^2 (1 + \delta_{k5} - \delta_{45})
 + m_{\mu5}^2 (1 + \delta_{k5}) \,, 
\\ 
E^2_{{45}} & = & \vec{p}\;{}^2 (1 + \delta_{45} - \delta_{k5})
 + m_{\mu5}^2 (1 + \delta_{45}) \,,
\\
E^2_{{lm}} & = & (p_l^2 + p_m^2) (1 + \delta_{lm} -\delta_{k4}) + p_k^2 +
m_{\mu \nu}^2 (1 + \delta_{lm}) \,,
\\
E^2_{{k4}} & = & (p_l^2 + p_m^2)(1 + \delta_{k4} - \delta_{lm}) + p_k^2 +
m_{\mu \nu}^2 (1 + \delta_{k4}) \,.
\label{eqn:dispersion}\end{eqnarray}
Here $E_F$ is the energy of the taste $F$ PGB determined from the exponential
fall-off of the two-point function along the Euclidean time direction.
The quantities $m_F$ are the masses including all NLO contributions except
that from the $SO(4)$ breaking operators in (\ref{eq:decay_ops_2pi}).
They are thus labeled by indices of the $SO(4)$ irreps: $I$, $5$, $\mu$,
$\mu5$ and $\mu\nu$. The additional contributions from the $SO(4)$ breaking
operators are denoted $\delta_F$.
Here the labels, like those on the energies,
are those of the 8 irreps of the lattice timeslice group,
which distinguishes between spatial and timelike indices~\cite{Golterman}.
This is the maximal splitting among tastes consistent with the lattice
symmetry for states at zero spatial momentum.
The expressions show that, at non-zero momentum, the states fall into
even smaller representations, due to the coupling of the spin and taste
transformations in the lattice symmetry group. These irreps have been
discussed in Ref.~\cite{KS}.

The expressions for the $SO(4)$ breaking mass shifts ($\delta m^2_\pi / m^2_\pi$) are
\begin{eqnarray}    
   	\delta_I & = &  - \frac{8 a^2}{f^2}(C_{36V}+C_{36A})\,,
   	\nonumber \\ 
     \delta_5 & = & \frac{8 a^2}{f^2}(C_{36V}+C_{36A}) \,,
\nonumber \\ 
     \delta_k & = & - \frac{8 a^2}{f^2}( C_2 - 2C_{10}) + \frac{8
   	a^2}{f^2}(C_{36V}-C_{36A}) \,,
\nonumber \\ 
     \delta_4 & = &
   	+ \frac{24 a^2}{f^2}( C_2 - 2C_{10}) - \frac{8
   	a^2}{f^2}(C_{36V}-C_{36A}) \,,
% + \frac{64 a^2}{f^2}C_{47V} \,,
\nonumber \\ 
     \delta_{k5} & = & -\frac{8 a^2}{f^2}( C_2 -
   	2C_{10}) - \frac{8 a^2}{f^2}(C_{36V}-C_{36A}) \,,
\nonumber \\
   	\delta_{45} & = & +\frac{24 a^2}{f^2}( C_2 - 2C_{10})
   	+ \frac{8 a^2}{f^2}(C_{36V}-C_{36A}) \,,
%+ \frac{64 a^2}{f^2}C_{47A}\,,
   	\nonumber \\ 
\delta_{lm} & = &  - \frac{8 a^2}{f^2}(C_{36V}+C_{36A})
   \,,	\nonumber \\ 
\delta_{k4} & = & 
%-\frac{16 a^2}{f^2}(C_7 - 2C_{13}) 
+ \frac{8 a^2}{f^2}(C_{36V}+C_{36A}) \,.
\label{eqn:FFBdecay}\end{eqnarray}
%
%Those for the non-singlets are obtained by dropping the contributions
%of $C_7$, $C_{13}$ and $C_{47}$.
There are three $SO(4)$ breaking splittings, those between
tastes $\xi_{4}$ and $\xi_{i}$, between $\xi_{45}$ and $\xi_{i5}$,
and between $\xi_{ij}$ and $\xi_{k4}$.  The six independent coefficients
 are sufficient to give independent contributions to each of these splittings.
%for both flavor singlets and non-singlets separately.
Thus there are no relations predicted between these mass splittings.

What is predicted, however, are relations between the 
violation of rotation symmetry in the dispersion relations 
and the $SO(4)$-taste breaking in the mass splittings. 
These can be written as
\begin{eqnarray}
\frac{E_k^2 - E_4^2}{m_k^2 - m_4^2} &=& 1 + 
\frac{p_i^2 + p_j^2 + 2 p_k^2}{(m_k^2+m_4^2)/2} \,,
\nonumber\\
\frac{E_{k5}^2 - E_{45}^2}{m_{k5}^2 - m_{45}^2} &=& 1 + 
\frac{p_i^2 + p_j^2 + 2 p_k^2}{(m_{k5}^2+m_{45}^2)/2} \,,
\nonumber\\
\frac{E_{lm}^2 - E_{k4}^2}{m_{lm}^2 - m_{k4}^2} &=& 1 + 
2 \frac{p_l^2 + p_m^2}{(m_{lm}^2+m_{k4}^2)/2} \,.
\label{eq:prediction1}
\end{eqnarray}
%independently for flavor singlets and non-singlets.
In these relations $m_F$ is the full NLO mass. Note that
the relations are trivial when $\vec p = 0$.  The fact that
the precise form of the mass used in the denominator on the 
r.h.s. does not matter at this order allows these expressions 
to be written in terms of quantitities that are directly measurable on the lattice.  
%Thus tests of these predictions should be relatively straightforward.

Note that neither the taste singlet nor the taste Goldstone PGBs 
receive rotational symmetry violating
contributions to their dispersion relations at this order.  
Such terms are prohibited by the fact that these tastes transform 
trivially under rotations.  Therefore they can only
feel the effect of rotational symmetry breaking through operators with four or more derivatives~\cite{BA1},
\begin{equation}
	a^2 \sum_\mu \partial_\mu^2 \pi \partial_\mu^2 \pi\,,
\end{equation} 
which are of $\CO(a^2 p^4)$ and thus contribute only at NNLO.

\medskip
Next we turn to the relations involving the decay constants,
still considering only flavor non-singlet mesons. As explained
in the previous section, the $SO(4)$ breaking operators of $\CO(a^2p^2)$
contribute to both the masses and the decay constants. For the pseudoscalar
decay constants, as for the masses,
the contribution is through wavefunction renormalization.
Since the same terms contribute to both masses and decay constants, there are
three simple relationships between splittings:
\begin{eqnarray}
\Big(\frac{f^P_k - f^P_4}{f^P_k + f^P_4}\Big) & = & 
\frac{1}{2} \Big(\frac{m_k^2 - m_4^2}{m_k^2 + m_4^2}\Big) \,, 
\nonumber \\
\Big(\frac{f^P_{k5} - f^P_{45}}{f^P_{k5} + f^P_{45}}\Big) & = & 
\frac12 \Big(\frac{m_{k5}^2 - m_{45}^2}{m_{k5}^2 + m_{45}^2}\Big)
\,, \nonumber \\
\Big(\frac{f^P_{lm} - f^P_{k4}}{f^P_{lm} + f^P_{k4}}\Big) & = &
\frac12 \Big(\frac{m_{lm}^2 - m_{k4}^2}{m_{lm}^2 + m_{k4}^2}\Big)\,.
\label{eqn:relations}\end{eqnarray}
These are likely to be the most simple predictions to test in practice. 
It is essential for these expressions that the $Z$-factors are $SO(4)$-invariant, and therefore identical for both tastes in the expression.  They can thus be tested using \emph{bare lattice operators}, thereby avoiding $\CO(a^2)$ ambiguities in matching 
lattice and continuum operators that could destroy the relationship. This is not true in general (e.g. for $f^A$) but does hold for the pseudoscalar operators 
$\gamma_5\otimes \xi_F$.  For example, the operators with spin-taste structure $\gamma_5\otimes \xi_i$ are related to $\gamma_5\otimes \xi_4$ by a lattice rotation.

This argument shows why there can be no similar relationships
involving the axial decay constants, since the $Z$-factors in that
case are not $SO(4)$ invariant. For example, $\gamma_{45}\otimes\xi_i$
is a ``2 link" operator while $\gamma_{45}\otimes \xi_4$ is 
a ``4 link" operator, so they are clearly not related by any
lattice symmetries, and thus have different $Z$'s.
In fact, it is not even clear how to unambiguously calculate the quantities
on the l.h.s. of eq.~(\ref{eqn:relations}) for $f^P\to f^A$, as the
$Z$-factors are not universal at $\CO(a^2)$.
Thus it is reassuring that S$\chi$PT predicts no relations involving
$f^A$, because of the ``$a^2$ source" mesonic operators, which contribute
$SO(4)$ breaking contributions to $f^A$ alone.
The only prediction for axial decay constants is a qualitative one:
the splittings between $SO(4)$ irreps should not be different in
magnitude from those within the irreps. In other words, there
should be no pattern of approximate degeneracies, unlike for the
masses which exhibit approximate $SO(4)$ symmetry.
This expectation is consistent with the 
results of Ref.~\cite{JLQCD2}, in which $f^A$ was calculated for
all tastes in the quenched approximation.

\medskip
We now turn to $SO(4)$-breaking predictions for PGBs for which the double supertrace,
or ``hairpin'', operators also contribute. These are particles which have
a flavor singlet component, i.e. $\bar u u+\bar dd$ and $\bar s s$ in the
unquenched sector, and $\bar x x$ or $\bar y y$ (and their ghost companions)
in the valence sector. These states are mixed by hairpin contributions, and because,
in general, they start off before mixing with different masses (due to the different
masses of the quarks), there are no simple predictions for the properties of
the resulting mixed states. The situation is yet more complicated if the
 $\sqrt[4]{\mbox{Det}}$ trick is being used. In fact, for the vector and axial tastes
these complications arise at LO, due to the presence of hairpins, and have
been addressed in detail in Ref.~\cite{BA1}.

The only theories for which the flavor non-singlet predictions  
still hold as written are unquenched theories
with $N\ge 1$ {\em degenerate} flavors.
 In these theories there is a single flavor-singlet state, 
and thus no mixing. This state can have any taste except $\xi_I$, which is,
as noted above, not a PGB. The predictions for the other fifteen tastes 
take exactly the same form as those given above for the flavor non-singlets,
except that there are additional contributions to the $\delta$'s from the
double supertrace operators:
\begin{eqnarray}    
     \delta_4^{\rm singlet} & = & - N \frac{32 a^2}{f^2}C_{41V} \,,
\nonumber \\
   	\delta_{45}^{\rm singlet} & = & - N \frac{32 a^2}{f^2}C_{41A}\,,
   	\nonumber \\ 
\delta_{k4}^{\rm singlet} & = & -N \frac{16 a^2}{f^2}(C_7 - 2C_{13})  \,.
\label{eqn:deltasinglet}\end{eqnarray}
Thus the $SO(4)$ breaking splittings for the flavor singlets are not related
to those of the non-singlets, but the form of the predictions for dispersion
relations, (\ref{eq:prediction1}), and of the relations between mass and decay constant
splittings, (\ref{eqn:relations}), remain unchanged.

The situation is similar if one has $N$ degenerate flavors and uses the
 $\sqrt[4]{\mbox{Det}}$ trick. The only complication is that the disconnected
quark contractions must be multiplied, by hand, by a factor of $1/4$ before
being added to the connected quark contractions. Then the predictions
of (\ref{eq:prediction1}) and (\ref{eqn:relations}) apply, with
the $\delta$'s having the additional contributions given in (\ref{eqn:deltasinglet}),
except that $N\to N/4$.

For theories with non-degenerate sea quarks, such as QCD,
and for PQ theories, there are no simple relations involving masses and
decay constants. Instead, we can obtain simple relations involving the
hairpin vertices by considering disconnected valence correlators.
These have the flavor structure $11 \leftrightarrow 22$, 
where 1 and 2 refer to valence flavors ($x$, $y$, \dots).
This picks out the desired hairpin contractions at the \emph{quark level}. 
To cancel the LO decay constants, and to obtain a quantity with a simple
dependence on Euclidean time, we take the ratio of the disconnected correlator to the connected ($12 \leftrightarrow 21$) correlator. 
This ratio has been studied extensively in the quenched approximation.
In the present context, it has the attractive property that the derived relationships
 hold even if one uses the $\sqrt[4]{\mbox{Det}}$ trick.  
For simplicity, we restrict ourselves to degenerate valence quarks, i.e. $m_1 = m_2$.  

The earlier flavor nonsinglet predictions related splittings within $SO(4)$ irreps in 
masses and pseudoscalar decay constants.  As a natural generalization to  
flavor singlets, we therefore consider $SO(4)$-breaking in disconnected 
correlators of the pseudoscalar density.  For example, for tensor taste we
calculate
\begin{equation}
	\frac{\langle P_{11}(0) P_{22}(\vec p=0,t)\rangle_{\xi_{\ell m}}
       - \langle P_{11}(0) P_{22}(\vec p=0,t)\rangle_{\xi_{k4}}}
         {\langle P_{11}(0) P_{22}(\vec p=0,t)\rangle_{\xi_{\mu\nu}}}\,,
\label{eqn:SO4_PP_T}\end{equation}  
where $P$ indicates a pseudoscalar operator with the flavor 
indicated by the subscript.  The taste of the operators
are shown by the subscript to the expectation value.
Note that, in the denominator, any choice of tensor taste can be used,
as the difference between the correlators for timeslice irreps 
$\xi_{\ell m}$ and $\xi_{k 4}$ is subleading.      

\begin{figure}
	\epsfxsize=1.9in \epsffile{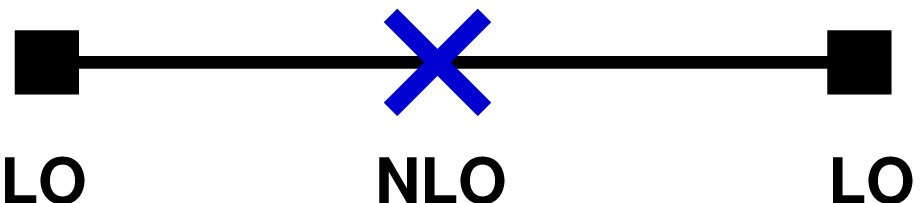}
\caption{NLO contribution to the $\langle PP\rangle$ flavor-disconnected correlator.  
The two black squares represent insertions of the pseudoscalar density, 
while the cross represents the hairpin vertex.  For tensor taste, 
this vertex comes from both $\CO(a^2 p^2)$ and $\CO(a^4)$ two supertrace operators,
but only two of the  $\CO(a^2 p^2)$ operators lead to $SO(4)$ breaking.  
The analogous diagram, with pseudoscalar sources changed to axial currents, 
contributes to the $\langle AA \rangle$ flavor-disconnected correlator.}\label{fig:CorrVertex}
\end{figure}

We describe the calculation of (\ref{eqn:SO4_PP_T}) in detail, 
as the remaining predictions in this section all rely on 
some generalization of this method.  
At NLO, the numerator is due to one diagram, 
shown in Figure~\ref{fig:CorrVertex}, in which two LO insertions of the flavor singlet pseudoscalar density are contracted with a NLO hairpin vertex.  
While the full NLO tensor taste hairpin vertex comes from both 
$\CO(a^2 p^2)$ and $\CO(a^4)$ operators, only two operators, $C_7$ and $C_{13}$ in (\ref{eqn:decay_ops}), produce $SO(4)$-breaking.\footnote{Recall that $\CO(a^4)$ operators are $SO(4)$-invariant for 2-PGB properties, and only two-supertrace operators produce hairpin vertices.}  
Thus, before Fourier transformation, the numerator of (\ref{eqn:SO4_PP_T}) is
\begin{eqnarray}
	\langle P_{11}(0) P_{22}(p)\rangle_{\xi_{\ell m}}
 - \langle P_{11}(0) P_{22}(p)\rangle_{\xi_{k 4}}
 &=&  16 a^2 \mu^2 p^2 \;( C_{7} - 2 C_{13} )\; 
\bigg(\frac{1}{p^2 + m^2_{\mu \nu}}\bigg)^2\,,
\\
& = &  16 a^2 \mu^2\;( C_{7} - 2 C_{13} )\; \left\{  \frac{1}{p^2 + m^2_{\mu \nu}}
 - m^2_{\mu \nu} \bigg(\frac{1}{p^2 + m^2_{\mu \nu}}\bigg)^2 \right\}
\,.
\end{eqnarray}
The mass $m_{\mu\nu}$ is that of the non-singlet taste tensor meson with
flavor $12$. In this equation it is the LO mass, which is the same for all
components of the $SO(4)$ irrep. At NLO accuracy, it can, however, be replaced
by the NLO (or all orders) mass of any member of the irrep.  
In the second line we have shown explicitly the single and double-pole contributions.
In an unquenched theory, the former would give rise to a correction to the
pseudoscalar decay constant, the latter to the PGB mass. The S$\chi$PT prediction
is that they have a common coefficient since they result from the same operators.

To denominator of the ratio (\ref{eqn:SO4_PP_T}) is just the LO contribution to the connected $\langle PP\rangle$ correlator. Before Fourier transformation it is
\begin{equation}
	\langle P_{11}(0) P_{22}(p) \rangle_{\xi_{\mu\nu}}
\;=\; \mu^2 f^2 \frac{1}{p^2 + m^2_{\mu \nu}}\,.
\end{equation}
We now set $\vec p=0$, Fourier transform to Euclidean time, and form the ratio,
yielding:
\begin{eqnarray}
	\frac{\langle P_{11}(0) P_{22}(\vec p=0,t)\rangle_{\xi_{\ell m}}
       - \langle P_{11}(0) P_{22}(\vec p=0,t)\rangle_{\xi_{k4}}}
         {\langle P_{11}(0) P_{22}(\vec p=0,t)\rangle_{\xi_{\mu\nu}}}
 &=& - \delta_{k4}^{\rm singlet}(1 - [1 + m_{\mu\nu} \:|t|\:]/2 ) \nonumber \\
 &=& - \delta_{k4}^{\rm singlet}(1 - m_{\mu\nu} \:|t|\:)/2\,,
\label{eq:hairpinP}
\end{eqnarray}
where $t$ is Euclidean time, and $\delta_{k4}^{\rm singlet}$ is 
given in (\ref{eqn:deltasinglet}) with $N=1$.
On the first line, the $1$ on the r.h.s. is from the single pole in the numerator, 
while the $[1 + m_{\mu\nu} \:|t|\:]/2$ is from the double pole.
We imagine that the mass $m_{\mu\nu}$ is first determined from the 
connected correlator. 
The prediction of S$\chi$PT is then that the constant term and
the term linear in $m_{\mu\nu}|t|$ should have \emph{opposite coefficients}.
That it is possible to separate these two terms has been shown in quenched
simulations~\cite{bardeen}. Note that for this prediction to hold one must
be at large enough Euclidean times that the PGB contributions dominate the
correlators in both the numerator and denominator of the ratio,
but small enough that $|\delta_{k4}^{\rm singlet} m_{\mu\nu} t| \ll 1$.
If the latter condition does not hold, further terms in the iteration of
the NLO hairpin must be included. 

As noted above, the prediction (\ref{eq:hairpinP}) holds irrespective of
whether the $\sqrt[4]{\mbox{Det}}$ trick has been used. This is because the hairpin correlators do not, at this order, access the sea sector of the theory. The only caveat
is that one must ensure that the mass of the tensor taste valence mesons
differs significantly 
from that of any of the tensor taste mesons composed of sea quarks.
If not, mixing of valence and sea sectors, although of higher than NLO,
can be enhanced by the proximity to intermediate poles.

Another feature of the flavor-singlet prediction common with those earlier is that one can
use the bare lattice pseudoscalar densities, since $Z$ factors cancel in
the ratio. One might be concerned that the $Z$ factors in the numerator
and denominator might differ, since the calculation of the former involves
additional quark disconnected diagrams with intermediate gluons.
One can show, however, that the difference between the $Z$ factors 
vanishes in the continuum limit. This is actually true irrespective of
the spin of the bilinear and follows because the operator has non-trivial taste.
Thus the difference between $Z$'s in the numerator and denominator is of
$O(a^2)$, and does not contribute at NLO.

We have not been able to find a prediction as simple as (\ref{eq:hairpinP})
for the vector or axial taste PGBs.  The analogous $SO(4)$ breaking
quantity to consider for, say, vector taste is:
\begin{equation}
	\frac{\langle P_{11}(0) P_{22}(\vec p=0,t)\rangle_{\xi_4} 
       - \langle P_{11}(0) P_{22}(\vec p=0,t)\rangle_{\xi_k}}
         {\langle P_{11}(0) P_{22}(\vec p=0,t)\rangle_{\xi_\mu}}\,.
\label{eqn:SO4_PP_V}\end{equation}  
Using the methods of Refs.~\cite{SS1,BA1} one can determine the
form of this ratio. The calculation is complicated by
the presence of the LO hairpin vertex, which means that there
are not only terms proportional to $\delta_4^{\rm singlet}$,
but also terms proportional to the non-singlet $SO(4)$ breaking
quantity $(\delta_4-\delta_k)$. Furthermore, the fact that the LO hairpins
must be iterated to all orders brings in the sea sector through intermediate propagators,
and the momentum dependence is quite complicated.
Thus we do not give explicit expressions.

\medskip

Finally, we discuss the fourth point of our summary in the previous section.
We can relate the hairpin contributions to $f^P$ and $f^A$ for pseudoscalar and
tensor taste because they only arise from operators of
$\CO(a^2 p^2)$ and $\CO(a^4)$.  In particular, there are no two supertrace $\CO(a^2 m)$ operators, which would contribute to $f^P$, or $\CO(a^2 source)$ operators, 
which would only modify $f^A$;  all NLO hairpin operators contribute to either 
both or neither quantity.  It is a straightforward exercise to calculate the 
contribution of these operators to $f^P$ and $f^A$, and, when we do, we find 
a simple relation for both tastes:
\begin{eqnarray}
	\bigg(\frac{\delta f_5^P}{f^P}\bigg)_{\rm hairpin} = - \bigg(\frac{\delta f_5^A}{f^A}\bigg)_{\rm hairpin}\,, & \qquad & \bigg(\frac{\delta f_{\mu \nu}^P}{f^P}\bigg)_{\rm hairpin} = - \bigg(\frac{\delta f_{\mu \nu}^A}{f^A}\bigg)_{\rm hairpin}\,.
\label{eq:deltafhairpin}
\end{eqnarray}
The contributions of the NLO two supertrace operators are equal and opposite.  
The minus sign comes from the fact that the $\CO(a^2 p^2)$ terms contribute to $f^P$ only through wavefunction renormalization, but to $f^A$ also through the axial sources contained in the covariant derivative.  Note that these relationships do involve both
$SO(4)$ conserving and breaking NLO contributions---in this sense they are
more general than those considered above.

These expressions can only be evaluated as written in unquenched theories with
degenerate quarks. One needs to determine the decay constants of flavor singlet
and non-singlet PGBs, and take the difference---this is $\delta f_{\rm hairpin}$. 
For non-degenerate sea quarks and/or PQ theories, however, this will not work.
As above, one can instead isolate the hairpin contributions using disconnected valence
correlation functions.
We consider the following ratios of pseudoscalar densities and axial currents:
\begin{eqnarray}
	\frac{\langle P_{11}(0) P_{22}(\vec p=0,t)\rangle_{\xi_T}}
           {\langle P_{12}(0) P_{21}(\vec p=0,t)\rangle_{\xi_T}}\,, 
& \qquad &  \frac{\langle A^{4}_{11}(0) A^4_{22}(\vec p=0,t)\rangle_{\xi_T}}
                 {\langle A^4_{12}(0) A^4_{21}(\vec p=0,t)\rangle_{\xi_T}}\,.
\end{eqnarray}
The superscript on the axial currents indicates the Lorentz index, and not taste.
Again we assume equal valence quark masses, $m_1=m_2$,
and that the valence PGBs have masses differing from those of all PGBs composed
of sea quarks. In both these ratios bare operators can be used as $Z$ factors cancel
as above.

First consider taste $\xi_5$. In this case 
$\CO(a^4)$ hairpin vertices are prohibited because of the lattice axial
symmetry, so that the numerators of the ratios only involve two operators:
\begin{equation}
	a^2  \;\;\sum_\mu  \left\{ C_{7 P} \Str(\Sigma \partial_\mu \Sigma^\dagger \xi_5) \Str(\Sigma^\dagger \partial_\mu \Sigma \xi_5)
+ C_{13 P} [\Str(\Sigma \partial_\mu \Sigma^\dagger \xi_5)\Str(\Sigma \partial_\mu \Sigma^\dagger \xi_5) + p.c.]  \right\}
\end{equation}  
The contribution to the pseudoscalar ratio can be worked out as above. Only
the diagram of Fig.~\ref{fig:CorrVertex} contributes to the numerator, and one
finds at NLO
\begin{equation}
\frac{\langle P_{11}(0) P_{22}(\vec p=0,t)\rangle_{\xi_5}}
           {\langle P_{12}(0) P_{21}(\vec p=0,t)\rangle_{\xi_5}}
= - \frac{16 a^2 \mu^2}{f^2} ( C_{7 P} - 2 C_{13 P} ) 
\bigg(\frac{1-m_5 \:|t|\:}{2}\bigg)\,.
\label{eq:PP5ratio}
\end{equation}
In fact, without even relating this to the $\langle AA\rangle$ ratio, we have
another prediction of S$\chi$PT: the constant and $m_5|t|$ terms should have
opposite signs.

\begin{figure}
	\epsfxsize=2.1in \epsffile{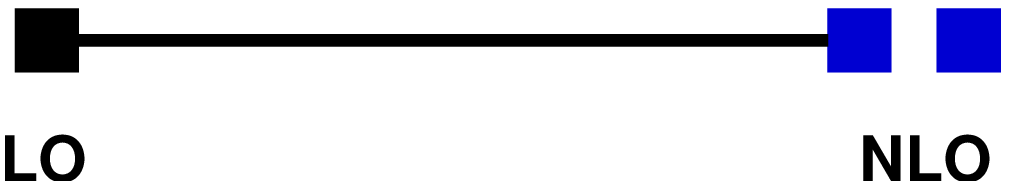}
\caption{NLO contribution to the $\langle AA\rangle$ 
flavor-disconnected correlator from axial current renormalization.  The single black square represents an insertion of the LO axial current, while the 
 double box represents an insertion of the NLO axial current.  It is shown as two boxes because the hairpin vertex is in the current itself.}\label{fig:CorrSource}
\end{figure}

For the axial current correlator, the numerator receives contributions not
only from wave function renormalization (Fig.~\ref{fig:CorrVertex}), but
also through direct renormalization of the axial current (shown in Fig.~\ref{fig:CorrSource}).
The flavor-disconnected axial correlator for taste $\xi_5$ is:
\begin{equation}
{\langle A^4_{11}(0) A^4_{22}(p)\rangle_{\xi_5}}
	= - 16 a^2( C_{7 P} - 2 C_{13 P} ) p_4^2 
\left\{
\frac{p^2}{(p^2 + m_5^2)^2} - 2 \frac{1}{(p^2 + m_5^2)}\right\}\,,
\end{equation}
where the first term is from wavefunction renormalization, and the second is from axial current renormalization.  If we set $\vec p=0$, this can be simplified to: 
\begin{equation}
{\langle A^4_{11}(0) A^4_{22}(\vec p=0,p_4)\rangle_{\xi_5}}
 = 16 a^2( C_{7 P} - 2 C_{13 P} ) \left\{2 - \frac{m_5^4}{(p_4^2 + m_5^2)^2}\right\} \,.
\end{equation}
Thus, in the axial correlator, the
fact that only $\CO(a^2 p^2)$ operators contribute for 
taste $\xi_5$ results in \emph{the absence of a single pole term}.
To test this in practice one Fourier transforms and takes the ratio to
the connected correlator:
\begin{equation}
\frac{\langle A^{4}_{11}(0) A^4_{22}(\vec p=0,t)\rangle_{\xi_5}}
                 {\langle A^4_{12}(0) A^4_{21}(\vec p=0,t)\rangle_{\xi_5}}
= - \frac{16 a^2 \mu^2}{f^2} ( C_{7 P} - 2 C_{13 P} ) \bigg(\frac{-1-m_5 \:|t|\:}{2}\bigg)\,.
\label{eq:AA5ratio}
\end{equation}
The absence of a single pole translates into the predicted $(1+m_5|t|)$ dependence.

Now we can return to the relation between hairpin contributions to
$f^P$ and $f^A$. This shows up in the fact that the overall coefficients 
in the $\langle PP\rangle$ and $\langle AA\rangle$ ratios are the same.
The prediction in (\ref{eq:deltafhairpin}) becomes the result that the constant terms have
opposite signs. On the other hand, the linear terms must have the same sign
as they correpond to a mass shift which does not depend on the external operators.
One way to test the prediction of opposite constant terms is to add the
two ratios:
\begin{equation}
\frac{\langle P_{11}(0) P_{22}(\vec p=0,t)\rangle_{\xi_5}}
           {\langle P_{12}(0) P_{21}(\vec p=0,t)\rangle_{\xi_5}}
+
\frac{\langle A^{4}_{11}(0) A^4_{22}(\vec p=0,t)\rangle_{\xi_5}}
                 {\langle A^4_{12}(0) A^4_{21}(\vec p=0,t)\rangle_{\xi_5}}
\propto m_5 \:|t|
\,.
\label{eqn:5_predict}\end{equation}
The S$\chi$PT prediction at NLO is then that there is no constant term
in this quantity.

Because the tensor taste PGB is not a lattice Goldstone pion, 
it has $\CO(a^4)$ hairpin vertices in addition to those of $\CO(a^2 p^2)$. 
Because of this, neither of the predictions (\ref{eq:PP5ratio}) or 
(\ref{eq:AA5ratio}) apply---the
constant and linear terms in $|t|$ within each individual ratio are not related.
Since the $\CO(a^4)$ hairpins lead only to mass renormalization, one expects,
however, for the constant terms in the $\langle PP \rangle$ and $\langle AA\rangle$
ratios to be opposite, as for taste $\xi_5$. This turns out to be the case.
Thus we find
\begin{equation}
\frac{\langle P_{11}(0) P_{22}(\vec p=0,t)\rangle_{\xi_{\mu\nu}}}
           {\langle P_{12}(0) P_{21}(\vec p=0,t)\rangle_{\xi_{\mu\nu}}}
+
\frac{\langle A^{4}_{11}(0) A^4_{22}(\vec p=0,t)\rangle_{\xi_{\mu\nu}}}
                 {\langle A^4_{12}(0) A^4_{21}(\vec p=0,t)\rangle_{\xi_{\mu\nu}}}
\propto m_{\mu\nu} \:|t|
\,.
\label{eqn:T_predict}\end{equation}
The absence of a constant term in this quantity is really two independent predictions,
one for taste $\xi_{4 k}$ and another for $\xi_{\ell m}$.

Since the predictions (\ref{eq:PP5ratio}), (\ref{eq:AA5ratio}) and (\ref{eqn:5_predict}) 
do not require $SO(4)$ breaking, there can be and in fact are
one-loop contributions to these ratios involving LO vertices. Note that,
as discussed in Ref.~\cite{SS1}, quark disconnected correlators receive
contributions at one-loop from single supertrace vertices. We have evaluated
this contributions, and find that they are consistent with all three predictions.
Thus these are full NLO predictions of S$\chi$PT.

\section{Conclusion}\label{sec:conc}

Staggered fermion simulations are currently able to reach much lower dynamical quark masses than other fermion discretizations.  However, taste-symmetry breaking is numerically significant at current lattice spacings.  Thus combined chiral and continuum extrapolations incorporating discretization errors are crucial for correct extrapolation of physical quantities.  Moreover, the $\sqrt[4]{\mbox{Det}}$ trick, which must be in practice be used before taking the continuum limit, may or may not change the universality class of the theory.  Thus it is unclear whether one is actually studying QCD.  We have addressed both of these issues in this paper.  

By enumerating all of the NLO operators in the staggered chiral Lagrangian, 
including source terms,
we allow a full NLO calculation, including analytic terms, of PGB properties.  Because $SO(4)$-taste symmetry breaking first enters at this order, these operators correctly reflect the true symmetry group of the underlying lattice action.  They are therefore necessary, combined with one-loop contributions, 
for accurate extrapolations and precise determinations of physical quantities.

As we noted in the Introduction, the large number of operators that appear at
NLO implies that one must calculate a correspondingly large number of physical
quantities in order to obtain predictions. Thus, for most new
quantities one calculates (e.g. the decay constants of PGBs of different tastes),
the NLO analytic terms simply give to an independent contribution proportional to
$a^2$. The most important NLO terms are then the one-loop contributions, which
lead to non-analytic dependence on the quark mass and lattice spacing.

The exceptions to this observation are quantities which are only non-vanishing
because of $SO(4)$ breaking. These only recieve contributions from the NLO
operators we have enumerated, and, in fact, from only a handful of these.
There are no loop contributions at NLO.
This allows us to make a number of testable predictions 
relating $SO(4)$-breaking splitting in PGB masses, decay constants, and dispersion relations. These predictions can be tested by calculating only two-point
correlators of unrenormalized lattice operators. For the hairpin operators
we suggest a method similar to that used successfully in quenched simulations.
Thus we hope that it will be practical to carry out these tests.

We also find a few predictions for quantities that do not involve $SO(4)$
breaking, but reflect the structure of S$\chi$PT. These are NLO analogues of 
the absence of the taste tensor hairpin at LO. 
We stress that these predictions, like those above, are {\em not} consequences
of lattice symmetries, and indeed are violated at NNLO or higher order 
in the chiral-continuum expansion.
In a similar vein, we note that not all operators in the
effective chiral theory which are consistent with
the lattice symmetries and power counting actually appear at a given order.
One must enumerate quark-level operators and then match these onto
mesonic operators, and this leads, at NLO, to restrictions on the contraction
of indices.

In our view, the most important application of the predictions we have given
is to test the applicability of the effective chiral theory when one
uses the $\sqrt[4]{\mbox{Det}}$ trick, and by so doing to
test the assumption that this trick does not modify
the continuum limit. Most of our predictions are unchanged in form when
one implements the $\sqrt[4]{\mbox{Det}}$ trick by hand in the effective theory.
This is both good and bad. It is good because it makes the predictions somewhat
more theoretically robust; it is bad because one would like to calculate quantities
which directly check the factors of $1/4$ put in by hand.
In any case, the key point is that if S$\chi$PT fails to describe results
from numerical simulations with small quark masses,
this would shed serious doubt on the correctness of the $\sqrt[4]{\mbox{Det}}$ trick.
Testing our predictions will thus provide further empirical evidence either for or
against the validity of this trick.

\section*{Acknowledgments}
We thank Oliver B\"ar and Claude Bernard for useful discussions.  This work was supported in part by the US Department of Energy through grant DE-FG02-96ER40956.
R.V. thanks the theory group at Fermilab for hospitality while some of this work
was undertaken.

\newpage

\appendix

\section{Determination of ${\cal O}(a^2 p^2, a^2m, a^4)$ operators}\label{app:ops}

Here we determine all operators in staggered chiral perturbation theory
which are NLO in our power counting
and arise from discretization errors. We do not discuss the 
remaining NLO operators, which are 
of ${\cal O}(p^4, p^2 m, m^2)$, since they are unchanged from 
continuum chiral perturbation theory.

We first discuss the
chiral operators of ${\cal O}(a^2 p^2, a^2 m)$.
These  arise
from a single insertion of the quark-level ${\cal O}(a^2)$ operators.
Then we turn to the ${\cal O}(a^4)$ chiral operators, which require
either two insertions of quark-level ${\cal O}(a^2)$ operators, or a single
insertion of an ${\cal O}(a^4)$ operator.
We close this subsection with a brief note on operators which do not arise until
higher than NLO in our expansion.
Finally, we consider the $\CO(a^2 p^2)$ operators involving covariant
derivatives acting on taste spurions, which lead to additional
contributions to the vector and axial currents.

\subsection{Single insertion of gluonic operators and fermion bilinears}
\label{app:bilin}

Most of the dimension six gluonic operators and fermion bilinears
do not break continuum symmetries,
and can only generate corrections proportional to
terms in the continuum chiral Lagrangian. Thus at NLO they 
only give rise to
the two operators in the LO chiral Lagrangian,
\begin{equation}
\label{eq:LOtimesasq}
	 \Str(\partial_\mu \Sigma \partial_\mu
	\Sigma^\dagger)\ \ \mathrm{and} \ \ 
         \Str( M^\dagger \Sigma +  M \Sigma^\dagger)
	 \,,
\end{equation}
multiplied by independent coefficients of size $a^2$.
In other words, they generate independent $\CO(a^2)$ corrections to 
the parameters $f$ and $\mu$.  As we will see, many additional contributions 
to the two operators in (\ref{eq:LOtimesasq}) arise from 
the four-fermion operators discussed in the following sections.
However, these, too, can simply be absorbed into the unknown coefficients multiplying
each operator.

Here we use mass spurions which transform in
the same manner as the $\Sigma$ field under a chiral symmetry
transformation:
\begin{equation}
	M \rightarrow L M R^\dagger\,, \;\;\;\;\; M^\dagger
	\rightarrow R M^\dagger L^\dagger\,.
\end{equation}
These spurions also function as sources for the scalar
and pseudoscalar densities, with $M= s + i p$, and $s$ and $p$
general Hermitian fields. In the absence of external fields
we set $s\to \CM$ and $p\to 0$.

The remaining two gluonic and bilinear operators violate Euclidean rotation symmetry,
\begin{equation}
\sum_\mu \Tr \left(D_\mu F_{\mu\nu} D_\mu F_{\mu\nu}\right), 
\;\;\;\;\; \sum_\mu \bar Q (\gamma_\mu\otimes1) D_\mu^3 Q\,,
\end{equation}
although they are taste symmetric.
They can give rise to rotationally non-invariant mesonic operators,
but this requires four derivatives~\cite{LS}, e.g.
\begin{equation}
	\sum_{\mu} \Str(\partial_\mu \Sigma \partial_\mu
	\Sigma^\dagger \partial_\mu \Sigma \partial_\mu
	\Sigma^\dagger) \,.
\end{equation}
These operators are of $\CO(a^2p^4)$, and thus NNLO, which is one higher order than we consider here.

\subsection{Single insertions of operators from $S_6^{FF(A)}$}
\label{app:FFA}

It is important to remember that the four fermion operators in
$S_6^{FF(A)}$ are invariant under rotations and $SO(4)$ taste
transformations, so they must map onto mesonic operators that also
respect these symmetries.  For discussion purposes, we divide them
according to how they transform under $SU(4N|4M)_L \times
SU(4N|4M)_R$ chiral rotations.

\subsubsection{Operators with spin structure $V$ or $A$}
\label{app:FFAVA}

Here we consider mesonic operators arising from single insertions of
four-fermion operators of the form $[V\times F]$ and $[A\times F]$,
where $F=S,P,T$ indicates the taste of the bilinears.  
The chiral structure of these operators is
\begin{equation}
\CO_F= \pm \sum_\mu \left(\bar {Q_R} (\gamma_\mu\otimes F_R) Q_R \pm 
\bar{Q_L} (\gamma_\mu\otimes F_L) Q_L \right)^2 \,,
\label{eq:OFform}
\end{equation}
where and the upper and lower signs correspond to spins V and A,
respectively, and $F_{R,L}$ are Hermitian taste matrices.  To
determine the corresponding mesonic operators, 
we promote the taste matrices to spurion fields,
with transformations chosen
so that $\CO_F$ is invariant under chiral transformations:
\begin{equation}
F_L\to L F_L L^\dagger, \;\;\;\;\; F_R \to R F_R R^\dagger \,.
\end{equation}
At the end we will set $F_L=F_R=F$, where $F$ is the specific taste
matrix appearing in the operator.

We first construct the resulting $\CO(a^2 p^2)$ operators.
These must be chiral singlets, respect parity
($F_L\leftrightarrow F_R$ and $\Sigma\leftrightarrow \Sigma^\dagger$),
be quadratic in $F$ (odd powers of
$F$ are forbidden by the $F \rightarrow -F$ symmetry and quartic terms
are higher order in $a$), and contain two derivatives.  $S_6^{FF(A)}$ is
rotationally invariant, so the indices of the derivatives must be
contracted with each other, and cannot be correlated with any of the
indices associated with the taste matrices.  It is easiest to
construct linearly-independent, chiral singlet operators using
elements with the same chiral transformation properties.  We use
left-handed building blocks, the choices for which are:
\begin{eqnarray}
	F_L & \rightarrow &L (F_L) L^\dagger \nonumber \\ \Sigma F_R
	\Sigma^\dagger & \rightarrow & L (\Sigma F_R \Sigma^\dagger)
	L^\dagger \nonumber \\ \Sigma \partial_\mu \Sigma^\dagger &
	\rightarrow & L (\Sigma \partial_\mu \Sigma^\dagger) L^\dagger
	\,. \label{eqn:basis1}
\end{eqnarray}

\begin{figure}
	\epsfysize=0.65in \vspace{0in}\hspace{-.0in}\rotatebox{0}{
	\leavevmode\epsffile{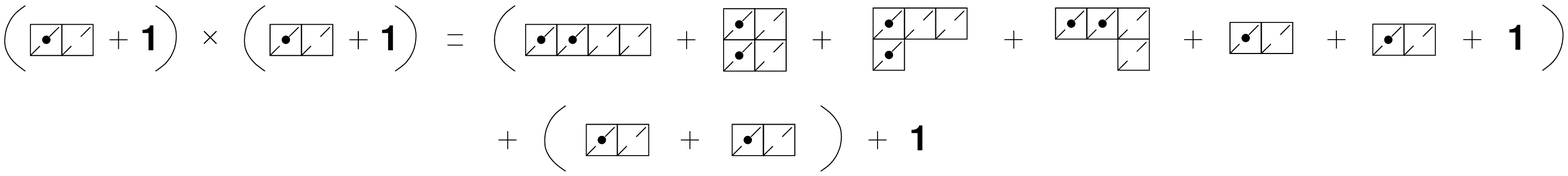}} \caption{The product of two
	bifundamentals in $SU(4N | 4M)$.  The result also applies for
	$SU(N)$ for $N > 3$ (with the dashed lines removed).}
	\vspace{-0in} \label{fig:bigGYT}
\end{figure}

\begin{figure}
	\epsfysize=0.40in \vspace{0in}\hspace{-.0in}\rotatebox{0}
	{\leavevmode\epsffile{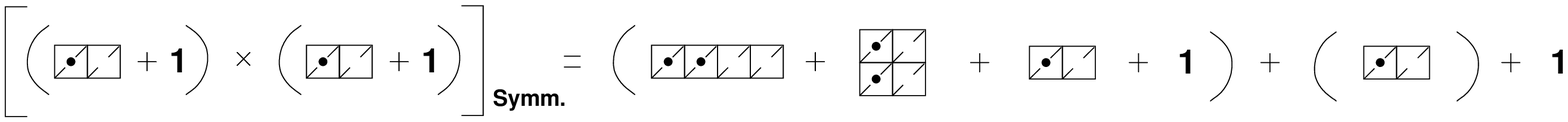}} \caption{The symmetric
	product of two bifundamentals in $SU(4N | 4M)$.  The result
	also applies to $SU(N)$ for $N > 3$ (with the dashed lines
	removed).}  \vspace{-0in} \label{fig:symGYT}
\end{figure}

\begin{figure}
	\epsfysize=0.38in
	\vspace{0in}\hspace{-.0in}\rotatebox{0}{\leavevmode\epsffile{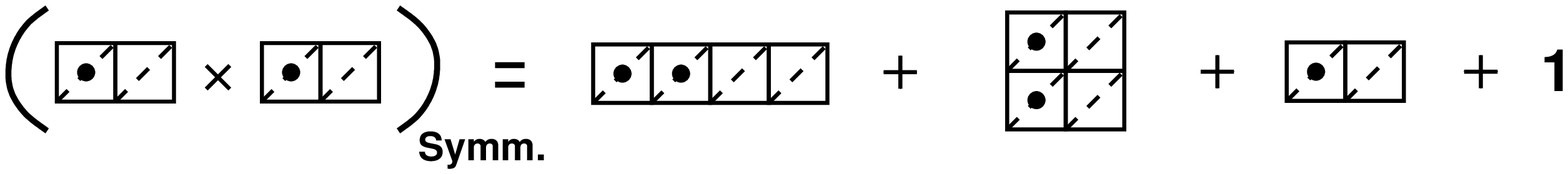}}
	\caption{The symmetric product of two adjoints in $SU(4N |
	4M)$.  The undotted boxes are fundamental representations and
	the dotted boxes are anti-fundamental representations.  The
	result also applies to $SU(N)$ for $N>3$ (with the dashed
	lines removed).}\vspace{-0in} \label{fig:GYT}
\end{figure}

We can now use the graded group theory method of Ref.~\cite{Me} to
determine the number of linearly independent $\CO(a^2p^2)$ 
operators in a general $SU(4N|4M)$ partially quenched theory. 
Because all of the operators in (\ref{eqn:basis1})
transform as bifundamentals under the left-handed chiral
group they all potentially have both adjoint and singlet components.
An operator with two taste spurions therefore comes from the product
of an adjoint plus singlet times an adjoint plus singlet, as shown in
Fig.~\ref{fig:bigGYT} using the graded Young tableaux notation of
Refs.~\cite{BB1, BB2}.  If the two taste spurions are identical (both
$F_L$ or $\Sigma F_R \Sigma^\dagger$), it will only come from the
symmetric part of this product, shown in Fig.~\ref{fig:symGYT}.
The left-handed Lie derivative has vanishing supertrace, however, and
therefore no singlet component.  An operator with two Lie derivatives,
which necessarily have their indices contracted and are therefore
identical, comes from the symmetric product of two adjoints, shown in
Fig.~\ref{fig:GYT}.  Thus the number of linearly-independent operators
is equivalent to the number of singlet representations contained in
the product of Fig.~\ref{fig:GYT}, which comes from the Lie
derivatives, and either Fig.~\ref{fig:bigGYT} or \ref{fig:symGYT},
which comes from the taste spurions.

Consider first operators composed of two Lie Derivatives, 
one $F_L$, and one $\Sigma F_R \Sigma^\dagger$.  
This requires counting singlets in the product of the representations
in Figs.~\ref{fig:bigGYT} and \ref{fig:GYT}. Since
all of the representations in Fig.~\ref{fig:GYT} are self-conjugate,
each can form a singlet with the identical representation in Fig.~\ref{fig:bigGYT}.
This leads to eight linearly-independent operators.
We choose the operator basis shown in Table~\ref{tab:VDDFFP}, in which
operators have been simplified using the anti-Hermiticity of the Lie
derivative and $\Sigma \Sigma^\dagger = 1$.  Each listed operator has
the \emph{same coefficient} with the \emph{same sign} for both V and A
spins because the overall sign and the cross-term sign in
(\ref{eq:OFform}) cancel.  Since group theory only considers
chiral transformation properties, we must impose other symmetries by
hand.  The operators 1 \&\ 3 and 4 \&\ 5 transform into each other under
parity, so we include their sums, each with a single undetermined
coefficient, in the staggered chiral Lagrangian.  Operators 4, 5, and
8 are all proportional to $\Str(F)$, so they vanish unless $F =
\xi_I$.  When this is the case, however, they are proportional to the
LO kinetic term, which we have already included in the
previous sub-section.

\begin{table}\begin{tabular}{lc}  \hline\hline

\multicolumn{1}{c}{\emph{Operator}} & \multicolumn{1}{c}{\emph{Keep?}} \\[0.5mm] \hline

1. $\Str(\partial_\mu \Sigma^\dagger \partial_\mu \Sigma
\Sigma^\dagger F_L \Sigma F_R)$ & Yes -- combined with 3  \\[0.5mm]

2. $\Str(\partial_\mu \Sigma^\dagger F_L \partial_\mu \Sigma F_R)$ &
Yes \\[0.5mm]

3. $\Str(\partial_\mu \Sigma \partial_\mu \Sigma^\dagger \Sigma F_R
\Sigma^\dagger F_L)$ & Yes -- combined with 1 \\[0.5mm]

4. $\Str(\partial_\mu \Sigma \partial_\mu \Sigma^\dagger F_L)
\Str(F_R)$ & No\\[0.5mm]

5. $\Str(\partial_\mu \Sigma^\dagger \partial_\mu \Sigma F_R)
\Str(F_L)$ & No \\[0.5mm]

6. $\Str(\partial_\mu \Sigma^\dagger \partial_\mu \Sigma)\Str(F_L
\Sigma F_R \Sigma^\dagger)$ & Yes \\[0.5mm]

7. $\Str(\Sigma \partial_\mu \Sigma^\dagger F_L)\Str(\Sigma^\dagger
\partial_\mu \Sigma F_R)$ & Yes \\[0.5mm]

8. $\Str(\partial_\mu \Sigma^\dagger \partial_\mu \Sigma) \Str(F_L)
\Str(F_R)$ & No \\[0.5mm]

\hline\hline

\end{tabular}\caption{The eight linearly-independent $\CO(a^2 p^2)$ 
operators with two derivatives and two different taste spurions.  
Note that two of them must be combined and
some can be neglected because of other considerations, as discussed in
the text.}
\label{tab:VDDFFP}\end{table}

Similarly, six linearly-independent operators can be made out of two
Lie Derivatives and two $F_L$'s or $F_R$'s.
These are shown in Table~\ref{tab:VDDFF}.  
Here the minus signs in (\ref{eq:OFform}) do
not cancel, and the coefficients of operators with $V$ and $A$ spins
have the same magnitude, but opposite sign.  
Operator 12 vanishes unless $F = \xi_I$, and 
then is proportional to the LO kinetic term,
as are operators 9, 11, and 14 for all tastes, 
so we can neglect them.  After combining both
lists, we are left with six new linearly independent $\CO(a^2 p^2)$ operators.

\begin{table}\begin{tabular}{llc}  \hline\hline

\multicolumn{2}{c}{\emph{Operator}} & \multicolumn{1}{c}{\emph{Keep?}} \\[0.5mm] \hline

9. & $\pm \Str( \partial_\mu \Sigma \partial_\mu \Sigma^\dagger F_L
F_L)\pm \Str( \partial_\mu \Sigma^\dagger \partial_\mu \Sigma F_R
F_R)$ & No \\[0.5mm]

10. & $\pm \Str(\Sigma \partial_\mu \Sigma^\dagger F_L \Sigma
\partial_\mu \Sigma^\dagger F_L)\pm \Str(\Sigma^\dagger \partial_\mu
\Sigma F_R \Sigma^\dagger \partial_\mu \Sigma F_R)$ & Yes \\[0.5mm]

11. & $\pm \Str(\partial_\mu \Sigma^\dagger \partial_\mu \Sigma)
\Str(F_L F_L)\pm \Str(\partial_\mu \Sigma^\dagger \partial_\mu
\Sigma)\Str(F_R F_R)$ & No \\[0.5mm]

12. & $\pm \Str(\partial_\mu \Sigma \partial_\mu \Sigma^\dagger F_L)
\Str(F_L)\pm \Str(\partial_\mu \Sigma^\dagger \partial_\mu \Sigma F_R)
\Str(F_R)$ & No \\[0.5mm]

13. & $\pm \Str(\Sigma \partial_\mu \Sigma^\dagger F_L)\Str(\Sigma
\partial_\mu \Sigma^\dagger F_L)\pm \Str(\Sigma^\dagger \partial_\mu
\Sigma F_R)\Str(\Sigma^\dagger
\partial_\mu \Sigma F_R)$ & Yes \\[0.5mm]

14. & $\pm \Str(\partial_\mu \Sigma^\dagger \partial_\mu \Sigma)
\Str(F_L) \Str(F_L)\pm \Str(\partial_\mu \Sigma^\dagger \partial_\mu
\Sigma) \Str(F_R) \Str(F_R)$ & No \\[0.5mm]

\hline\hline

\end{tabular}\caption{The six linearly-independent $\CO(a^2 p^2)$ 
operators with two derivatives and two identical taste spurions.  
The upper signs correspond to spin $V$, while
the lower signs correspond to spin $A$.  Some operators can be neglected
because of other considerations, as discussed in the text.}
\label{tab:VDDFF}\end{table}

We note in passing that the
fourteen operators in Tables~\ref{tab:VDDFFP} and \ref{tab:VDDFF}
are those obtained simply by building operators out of the
appropriate elements and using the cyclicity of the supertrace.  
There are no relationships between these operators in the PQ theory,
unlike for chiral $SU(2)$ and $SU(3)$ theories.
Typically, such relationships are found using
Cayley-Hamilton relations, as in Refs.~\cite{FS} and ~\cite{BCE}.
There is, however, no graded analog of Cayley-Hamilton relations,
since the superdeterminant is not a finite polynomial ~\cite{BB2}.

To obtain the actual operators in the chiral Lagrangian we set
$F_L = F_R =F$, a specific taste matrix, here $S$, $P$ or $T$.
Since rotational and $SO(4)$ taste symmetry are unbroken,
the indices on the derivatives must be contracted together,
as must those on the taste matrices.
With these rules, it is easy to see
that the specific $\CO(a^2 p^2)$ operators in the S$\chi\CL$ generated
by operators in $S^{FF(A)}$ with spins $V$ and $A$ are those shown in
Table~\ref{tab:FFA_VA}. We have dropped the $\pm$ signs at this stage
because we do not know the relative size of the coefficients of
the underlying four-fermion operators with spins V and A,
and therefore cannot make use of the fact that their mappings into the
chiral Lagrangian are related.
We have also dropped the operators resulting from taste $S$, since
they are either of the same form as the LO chiral Lagrangian, 
or vanish because $\Str(\Sigma\partial_\mu\Sigma^\dagger)=0$.

\bigskip
We determine the $\CO(a^2 m)$ operators using the same
methodology.  We use the mass spurions introduced above,
which lead to two new left-handed objects:
\begin{equation}
	M \Sigma^\dagger \rightarrow L M \Sigma^\dagger L^\dagger
	\,,\qquad \Sigma M^\dagger \rightarrow L \Sigma M^\dagger
	L^\dagger\,.
\end{equation}
Like the taste spurions, these objects are also bifundamentals under
the left-handed chiral group.  Operators of order $\CO(a^2 m)$ will
contain two taste spurions and one mass spurion, and be chiral
singlets that respect parity.  They will therefore come from the
singlet representations contained in the product of three
bifundamentals.  Such operators may contain either two identical or
two different taste spurions.

If the taste spurions are different, the operators simply come from
the product of two bifundamentals, Figure~\ref{fig:bigGYT}, with a
third, which clearly contains six singlets.  However, because there
are two types of mass spurions, there are twelve corresponding
operators in Table~\ref{tab:VMFFP}.  Each of these operators must be
combined with its parity conjugate to form a parity-invariant
operator, so there are in fact only six operators in the chiral
Lagrangian: (15 + 22), (16 + 21), (17 + 23), (18 + 25), (19 + 24), and
(20 + 26).  Three of these operators are proportional to $\Str(F)$,
and so vanish except for taste $S$, but then reduce to the LO
operator $\Str(\Sigma M^\dagger+M \Sigma^\dagger)$.  
Thus we are left with only three new $\CO(a^2 m)$ operators.

In fact, only two of these three are new if we consider S-matrix elements.
Operator (15 + 22) reduces to the LO operator if $M$ is set to the
mass matrix, since $\CM$ then commutes with ${F}_L$ and ${F}_R$.
The operator does, however, contribute $SU(4)$ taste breaking contributions 
to pseudoscalar matrix elements, for then $M$ is a source and need 
not be proportional to the identity in taste space.

\begin{table}\begin{tabular}{llc}  \hline\hline

\multicolumn{2}{c}{\emph{Operator}} & \multicolumn{1}{c}{\emph{Keep?}} \\[0.5mm] \hline

15. & $\Str( F_L \Sigma F_R M^\dagger)$ & Yes --
combined with 22 \\[0.5mm]

16. & $\Str(F_L \Sigma M^\dagger \Sigma F_R \Sigma^\dagger)$ & Yes --
combined with 21 \\[0.5mm]

17. & $\Str(F_L \Sigma F_R \Sigma^\dagger) \Str(\Sigma M^\dagger)$ &
Yes -- combined with 23 \\[0.5mm]

18. & $\Str(F_L \Sigma M^\dagger) \Str(F_R )$ & No \\[0.5mm]

19. & $\Str(F_L)\Str(F_R M^\dagger \Sigma )$ & No \\[0.5mm]

20. & $\Str(F_L) \Str(F_R ) \Str(\Sigma M^\dagger)$ & No \\[0.5mm]

21. & $\Str(F_R \Sigma^\dagger M \Sigma^\dagger F_L \Sigma )$ & Yes --
combined with 16 \\[0.5mm]

22. & $\Str(F_R \Sigma^\dagger F_L M )$ & Yes -- 
combined with 15 \\[0.5mm]

23. & $\Str(F_L \Sigma F_R \Sigma^\dagger) \Str(M \Sigma^\dagger)$ &
Yes -- combined with 17 \\[0.5mm]

24. & $\Str(F_L M \Sigma^\dagger) \Str(F_R )$ & No \\[0.5mm]

25. & $\Str(F_L)\Str(F_R \Sigma^\dagger M )$ & No \\[0.5mm]

26. & $\Str(F_L) \Str(F_R ) \Str(M \Sigma^\dagger)$ & No \\[0.5mm]

\hline\hline

\end{tabular}\caption{The twelve linearly-independent $\CO(a^2 m)$ 
operators one mass spurion and two different taste spurions.  
Many can be neglected, and the rest must be combined because of symmetry
considerations.}\label{tab:VMFFP}\end{table}

If the operators contain identical taste spurions, two of the three
bifundamentals must be symmetrized, as shown in Figure~\ref{fig:GYT}.
The overall product contains four singlets, and therefore generates
the eight corresponding operators in Table~\ref{tab:VMFF}.  However,
using the fact that $F^2 = 1$ for all tastes as well as $\Str(F) = 0$
except for $\xi_I$, it is easy to see that all of these operators
reduce to $\Str(\Sigma M^\dagger + M \Sigma^\dagger)$ and can be
neglected.

\begin{table}\begin{tabular}{llc}  \hline\hline

\multicolumn{2}{c}{\emph{Operator}} & \multicolumn{1}{c}{\emph{Keep?}} \\[0.5mm] \hline

27. & $\pm \Str(F_L F_L \Sigma M^\dagger) \pm \Str(F_R F_R
\Sigma^\dagger M )$ & No \\[0.5mm]

28. & $\pm \Str(F_L F_L) \Str(\Sigma M^\dagger) \pm \Str(F_R F_R)
\Str(\Sigma^\dagger M)$ & No \\[0.5mm]

29. & $\pm \Str(F_L) \Str(F_L \Sigma M^\dagger) \pm \Str(F_R) \Str(F_R
\Sigma^\dagger M)$ & No \\[0.5mm]

30. & $\pm \Str(F_L) \Str(F_L) \Str(\Sigma M^\dagger) \pm \Str(F_R )
\Str(F_R) \Str(\Sigma^\dagger M)$ & No \\[0.5mm]

31. & $\pm \Str(F_L F_L M \Sigma^\dagger) \pm \Str(F_R F_R M^\dagger
\Sigma)$ & No \\[0.5mm]

32. & $\pm \Str(F_L F_L) \Str(M \Sigma^\dagger) \pm \Str(F_R F_R )
\Str(M^\dagger \Sigma)$ & No \\[0.5mm]

33. & $\pm \Str(F_L) \Str(F_L M \Sigma^\dagger) \pm \Str(F_R) \Str(F_R
M^\dagger \Sigma)$ & No \\[0.5mm]

34. & $\pm \Str(F_L) \Str(F_L) \Str(M \Sigma^\dagger) \pm \Str(F_R)
\Str(F_R ) \Str(M^\dagger \Sigma)$ & No \\[0.5mm]

\hline\hline

\end{tabular}\caption{The eight linearly-independent $\CO(a^2 m)$ 
operators with one mass spurion and two identical taste spurions.  
The upper signs correspond to spin $V$,
while the lower signs correspond to spin $A$.  All of these operators
can be neglected because of additional
considerations.}\label{tab:VMFF}\end{table}

Turning these generic structures into operators in the staggered
chiral Lagrangian is now straightforward: simply let $F_L = F_R =F$.
We do not set $M = M^\dagger = \CM$, however, so as to allow
the derivation of (pseudo)scalar matrix elements. 
The linearly-independent $\CO(a^2 m)$ operators corresponding to
$S_6^{FF(A)}$ operators with spins $V$ and $A$ are shown in
Table~\ref{tab:FFA_VA}.

Finally, we briefly recall the construction of the LO
operators of $\CO(a^2)$, so as to make contact with the discussion
in the main text. These operators contain two taste spurions, and
the only non-trivial $SU(N|M)$ singlet is
\begin{equation}
\Str(F_L \Sigma F_R \Sigma^\dagger).
\label{eq:LOVA}
\end{equation}
This leads to the operators $\CO_1$ and $\CO_6$ in the potential
$\CU$ in (\ref{eq:U}).

\subsubsection{Operators with spin structure $S$ or $P$}
\label{app:FFASP}

Here we consider single insertions of operators
$[S\times F]$ and $[P\times F]$, where the taste can be $F=V$ or $A$.
The chiral structure of these operators is
\begin{equation}
\CO'_F= \left(\bar {Q_L} (1 \otimes \tilde{F}_L) Q_R \pm 
\bar{Q_R} (1 \otimes \tilde{F}_R) Q_L \right)^2 \,,
\label{eq:OFPform}
\end{equation}
where and the upper and lower signs correspond to spin S and P,
respectively, and $\tilde{F}_{L,R}$ are taste matrices.  To
determine the corresponding mesonic operators in the S$\chi\CL$, we
promote the taste matrices to taste spurion fields transforming as
\begin{equation}
\tilde{F}_L\to L \tilde{F}_L R^\dagger, 
\;\;\;\;\; \tilde{F}_R \to R \tilde{F}_R L^\dagger \,.
\end{equation}
Note that these transformations do not maintain the Hermiticity of
the original taste matrices. We can, however, consistently impose
the relation $F_L^\dagger = F_R$, and so do not need to consider
$F_{L,R}^\dagger$ as additional variables.
The left-handed basis for constructing the mesonic operators becomes
\begin{equation}
	\tilde{F}_L \Sigma ^\dagger\,,\ \ 
% & \rightarrow & L (\tilde{F}_L \Sigma ^\dagger) L^\dagger \nonumber 
 \Sigma \tilde{F}_R\,,\ \
% & \rightarrow & L (\Sigma \tilde{F}_R) L^\dagger \nonumber \\
	\Sigma \partial_\mu \Sigma^\dagger \,,\ \ 
% & \rightarrow & L (\Sigma\partial_\mu \Sigma^\dagger) L^\dagger \nonumber \\
  M\Sigma^\dagger \,,\ \ 
% & \rightarrow & L M \Sigma^\dagger L^\dagger	\nonumber \\ 
  \Sigma M^\dagger 
% & \rightarrow & L \SigmaM^\dagger L^\dagger 
\,. \label{eqn:basis2}
\end{equation}

The counting and construction of operators with the correct taste
structure is identical to the $[V,A \times F]$ case, although the
taste spurions are different.  The primary technical difference 
arises when we set $\tilde{F}_L=\tilde{F}_R=\tilde{F}$, because the
relative minus signs within compound operators are different.

As before, there are fourteen $\CO (a^2 p^2)$ operators which
can be constructed from two taste spurions and two Lie derivatives.
These are shown in Table~\ref{tab:SDDFF}.  Notice that the $\pm$ factors
differ from those for $[V,A \times F]$ 
because (\ref{eq:OFPform}) only has an internal minus sign,
not an overall one.  Two pairs of operators, 35 \& 37 and 38 \&
39, transform into each other under parity, so we include their 
sums, each with a single coefficient, in the chiral
Lagrangian. Operators (35 + 37) and 40 both reduce to the LO kinetic operator
 after setting $\tilde{F}_L=\tilde{F}_R=\tilde{F}$.
This leaves ten new linearly independent $\CO (a^2 p^2)$ operators. 

\begin{table}\begin{tabular}{llc}  \hline\hline

\multicolumn{2}{c}{\emph{Operator}} & 
\multicolumn{1}{c}{\emph{Keep?}} \\[0.5mm] \hline \vspace{1.0mm}  

35. & $ \pm \Str(\partial_\mu \Sigma \partial_\mu \Sigma^\dagger
\tilde{F}_L \tilde{F}_R)$ & No \\[0.5mm]

36. & $ \pm \Str(\Sigma \partial_\mu \Sigma^\dagger \tilde{F}_L
\Sigma^\dagger \partial_\mu \Sigma \tilde{F}_R)$ & Yes \\[0.5mm]

37. & $ \pm \Str(\partial_\mu \Sigma^\dagger \partial_\mu \Sigma
\tilde{F}_R \tilde{F}_L )$ & No \\[0.5mm]

38. & $ \pm \Str(\partial_\mu \Sigma^\dagger \partial_\mu \Sigma
\Sigma^\dagger \tilde{F}_L) \Str(\Sigma \tilde{F}_R)$ & Yes --
combined with 39 \\[0.5mm]

39. & $ \pm \Str(\partial_\mu \Sigma \partial_\mu \Sigma^\dagger
\Sigma \tilde{F}_R) \Str(\tilde{F}_L \Sigma ^\dagger)$ & Yes --
combined with 38 \\[0.5mm]

40. & $ \pm \Str(\partial_\mu \Sigma^\dagger \partial_\mu
\Sigma)\Str(\tilde{F}_L \tilde{F}_R)$ & No \\[0.5mm]

41. & $ \pm \Str(\partial_\mu \Sigma^\dagger
\tilde{F}_L)\Str(\partial_\mu \Sigma \tilde{F}_R)$ & Yes \\[0.5mm]

42. & $ \pm \Str(\partial_\mu \Sigma^\dagger \partial_\mu \Sigma)
\Str(\tilde{F}_L \Sigma ^\dagger) \Str(\Sigma \tilde{F}_R)$ & Yes
\\[0.5mm]

43. & $\Str(\partial_\mu \Sigma \partial_\mu \Sigma^\dagger
\tilde{F}_L \Sigma ^\dagger \tilde{F}_L \Sigma
^\dagger)+\Str(\partial_\mu \Sigma^\dagger \partial_\mu \Sigma
\tilde{F}_R \Sigma \tilde{F}_R \Sigma)$ & Yes \\[0.5mm]

44. & $\Str(\partial_\mu \Sigma^\dagger \tilde{F}_L \partial_\mu
\Sigma^\dagger \tilde{F}_L)+\Str(\partial_\mu \Sigma \tilde{F}_R
\partial_\mu \Sigma \tilde{F}_R)$ & Yes
\\[0.5mm]

45. & $\Str(\partial_\mu \Sigma^\dagger \partial_\mu \Sigma)
\Str(\tilde{F}_L \Sigma ^\dagger \tilde{F}_L \Sigma
^\dagger)+\Str(\partial_\mu \Sigma^\dagger \partial_\mu \Sigma)
\Str(\Sigma \tilde{F}_R \Sigma \tilde{F}_R)$ & Yes \\[0.5mm]

46. & $\Str(\partial_\mu \Sigma \partial_\mu \Sigma^\dagger
\tilde{F}_L \Sigma ^\dagger) \Str(\tilde{F}_L \Sigma
^\dagger)+\Str(\partial_\mu \Sigma^\dagger \partial_\mu \Sigma
\tilde{F}_R \Sigma) \Str(\Sigma \tilde{F}_R)$ & Yes \\[0.5mm]

47. & $\Str(\partial_\mu \Sigma^\dagger \tilde{F}_L)\Str(\partial_\mu
\Sigma^\dagger \tilde{F}_L )+\Str(\partial_\mu \Sigma \tilde{F}_R)
\Str(\partial_\mu \Sigma\tilde{F}_R)$ & Yes \\[0.5mm]

48. & $\Str(\partial_\mu \Sigma^\dagger \partial_\mu \Sigma)
\Str(\tilde{F}_L \Sigma ^\dagger) \Str(\tilde{F}_L \Sigma
^\dagger)+\Str(\partial_\mu \Sigma^\dagger \partial_\mu
\Sigma) \Str(\Sigma \tilde{F}_R) \Str(\Sigma \tilde{F}_R)$ & Yes \\[0.5mm]

\hline\hline

\end{tabular}\caption{The fourteen linearly-independent $\CO(a^2 p^2)$ 
operators with two derivatives and two taste spurions.  When present, the upper
and lower signs correspond to spins $S$ and $P$, respectively.}
\label{tab:SDDFF}\end{table}

In addition, there are twenty $\CO(a^2 m)$ operators that contain one
mass and two taste spurions. These are listed in Table~\ref{tab:SMFF}.
Six pairs of operators transform into each other under parity, and
must therefore be included in the Lagrangian with a single
coefficient: 49 \& 56, 50 \& 55, 51 \& 57, 52 \& 59, 53 \& 58, and
54 \& 60. However, operators (49 + 56), (50 + 55), and (51 + 57)
reduce to the LO mass term after setting $\tilde{F}_L = \tilde{F}_R =
\tilde{F}$.  
Thus there are eleven new
linearly-independent, parity-invariant $\CO(a^2 m)$ operators.
Three of these, (52 + 59), 61, and 63,  are pure source terms, however,
since they reduce to LO operators when $M=M^\dagger =\CM$.

\begin{table}\begin{tabular}{llc}  \hline\hline

\multicolumn{2}{c}{\emph{Operator}} & \multicolumn{1}{c}{\emph{Keep?}} \\[0.5mm] \hline

49. & $\pm \Str(\tilde{F}_L \tilde{F}_R \Sigma M^\dagger)$ & No
\\[0.5mm]

50. & $\pm \Str(\tilde{F}_L M^\dagger \Sigma \tilde{F}_R)$ & No
\\[0.5mm]

51. & $\pm \Str(\tilde{F}_L \tilde{F}_R) \Str(\Sigma M^\dagger)$ & No
\\[0.5mm]

52. & $\pm \Str(\tilde{F}_L M^\dagger) \Str(\Sigma \tilde{F}_R)$ & 
Yes -- combined with 59 \\[0.5mm]

53. & $\pm \Str(\tilde{F}_L \Sigma^\dagger)\Str(\Sigma \tilde{F}_R
\Sigma M^\dagger)$ & Yes -- combined with 58 \\[0.5mm]

54. & $\pm \Str(\tilde{F}_L \Sigma^\dagger) \Str(\Sigma \tilde{F}_R)
\Str(\Sigma M^\dagger)$ & Yes -- combined with 60 \\[0.5mm]

55. & $\pm \Str(\tilde{F}_L \tilde{F}_R M \Sigma^\dagger)$ & No
\\[0.5mm]

56. & $\pm \Str(\tilde{F}_L \Sigma^\dagger M \tilde{F}_R)$ & No
\\[0.5mm]

57. & $\pm \Str(\tilde{F}_L \tilde{F}_R) \Str(M \Sigma^\dagger)$ & No
\\[0.5mm]

58. & $\pm \Str(\tilde{F}_L \Sigma^\dagger M \Sigma^\dagger)
\Str(\Sigma \tilde{F}_R)$ & Yes -- combined with 53 \\[0.5mm]

59. & $\pm \Str(\tilde{F}_L \Sigma^\dagger)\Str(\tilde{F}_R M )$ & 
Yes -- combined with 53 \\[0.5mm]

60. & $\pm \Str(\tilde{F}_L \Sigma^\dagger) \Str(\Sigma \tilde{F}_R)
\Str(M \Sigma^\dagger)$ & Yes -- combined with 54 \\[0.5mm]

61. & $\Str(\tilde{F}_L \Sigma^\dagger \tilde{F}_L M^\dagger) +
\Str(\tilde{F}_R \Sigma \tilde{F}_R M)$ & Yes \\[0.5mm]

62. & $\Str(\tilde{F}_L \Sigma^\dagger \tilde{F}_L \Sigma^\dagger)
\Str(\Sigma M^\dagger) + \Str(\tilde{F}_R \Sigma \tilde{F}_R \Sigma)
\Str(\Sigma^\dagger M)$ & Yes \\[0.5mm]

63. & $\Str(\tilde{F}_L \Sigma^\dagger) \Str(\tilde{F}_L M^\dagger) +
\Str(\tilde{F}_R \Sigma) \Str(\tilde{F}_R M)$ & Yes \\[0.5mm]

64. & $\Str(\tilde{F}_L \Sigma^\dagger) \Str(\tilde{F}_L
\Sigma^\dagger) \Str(\Sigma M^\dagger) + \Str(\tilde{F}_R \Sigma)
\Str(\tilde{F}_R \Sigma) \Str(\Sigma^\dagger M)$ & Yes \\[0.5mm]

65. & $\Str(\tilde{F}_L \Sigma^\dagger \tilde{F}_L \Sigma^\dagger M
\Sigma^\dagger) + \Str(\tilde{F}_R \Sigma \tilde{F}_R \Sigma M^\dagger
\Sigma)$ & Yes \\[0.5mm]

66. & $\Str(\tilde{F}_L \Sigma^\dagger \tilde{F}_L \Sigma^\dagger)
\Str(M \Sigma^\dagger) + \Str(\tilde{F}_R \Sigma \tilde{F}_R \Sigma)
\Str(M^\dagger \Sigma)$ & Yes \\[0.5mm]

67. & $\Str(\tilde{F}_L \Sigma^\dagger) \Str(\tilde{F}_L
\Sigma^\dagger M \Sigma^\dagger) + \Str(\tilde{F}_R \Sigma)
\Str(\tilde{F}_R \Sigma M^\dagger \Sigma)$ & Yes \\[0.5mm]

68. & $\Str(\tilde{F}_L \Sigma^\dagger) \Str(\tilde{F}_L
\Sigma^\dagger) \Str(M \Sigma^\dagger) + \Str(\tilde{F}_R \Sigma)
\Str(\tilde{F}_R \Sigma) \Str(M^\dagger \Sigma)$ & Yes \\[0.5mm]

\hline\hline

\end{tabular}\caption{The twenty linearly-independent $\CO(a^2 m)$ 
operators with one mass spurion and two taste spurions.  
When present, the upper signs correspond to spin $S$,
while the lower signs correspond to spin $P$.}
\label{tab:SMFF}\end{table}

To turn these generic taste structures into actual operators in the
S$\chi\CL$, we set $\tilde{F}_L=\tilde{F}_R=\tilde{F}$, with
$\tilde{F}$ being either $V$ or $A$.  However, we choose to leave $M$ and $M^\dagger$ as generic sources which can be set to the quark mass matrix when necessary. 
As before must contract indices on
derivatives and taste matrices separately,
so as not to break rotational or $SO(4)$ taste symmetry.  
The resulting operators are listed in Table~\ref{tab:FFA_SPa2}.

Finally, we recall the construction of the LO
operators of $\CO(a^2)$.
In this case there are three non-trivial $SU(N|M)$ singlets:
\begin{eqnarray}
&& \Str(\tilde F_L \Sigma^\dagger \tilde F_L \Sigma^\dagger) + p.c.  \,,\\
&& \Str(\tilde F_L \Sigma^\dagger) \Str(\tilde F_L \Sigma^\dagger) + p.c. \,, \\
&& \Str(\tilde F_L \Sigma^\dagger) \Str(\Sigma \tilde F_R )\,. 
\label{eq:LOSP}
\end{eqnarray}
The first of these leads to $\CO_3$ and $\CO_4$ in $\CU$, (\ref{eq:U}),
the second to $\CO_{2V}$ and $\CO_{2A}$ in $\CU'$, (\ref{eq:U_prime}),
and the third to $\CO_{5A}$ and $\CO_{5B}$ in $\CU'$.

\subsubsection{Operators with spin structure $T$}
\label{app:FFAT}

The spin $T$ operators, $[T \times V]$ and $[T \times A]$, 
have almost the same chiral structure as those of spins $S$ and $P$:
\begin{equation}
\CO''_F= \sum_{\mu<\nu}
\left(\bar {Q_L} (\gamma_{\mu\nu}\otimes \tilde F_L) Q_R\right)^2 +
\left(\bar{Q_R} (\gamma_{\nu\mu}\otimes \tilde F_R) Q_L \right)^2 \,. \label{eq:OFPPform}
\end{equation}
The only difference is that 
there are no cross terms between $\tilde F_L$ and $\tilde F_R$.  
This means that the corresponding chiral operators are \emph{identical} to
those generated by $[S,P \times V,A]$, except that those coming
from $\tilde F_L$$\tilde F_R$ cross-terms are absent.
Thus the only effect of single insertions of four-fermion
operators with spin structure
$T$ is to change the (unknown) coefficients of 
some of the mesonic operators already listed.

\subsection{Single insertions of operators from $S_6^{FF(B)}$}
\label{app:singleFFB}

The four-fermion operators in $S_6^{FF(B)}$
break both rotational symmetry and the
remaining $SO(4)$ taste symmetry repected by operators in $S_6^{FF(A)}$.
Thus they can map onto mesonic operators that also break
these symmetries.
Indeed they only map onto such operators, because they are
constructed to have no taste singlet component.
In this subsection we construct 
all such operators resulting from a single insertion of $S_6^{FF(B)}$.

We begin with a general comment.
To break the rotational and/or $SO(4)$ taste symmetries requires
that the mesonic operator have more than two repeated indices.
This can be accomplished at NLO either
by having two derivatives and two taste spurions,
or by having four taste spurions.
The former choice leads to the $\CO(a^2 p^2)$ operators 
exemplified by
\begin{equation}
       \sum_\mu \Str(\Sigma \partial_\mu \Sigma^\dagger \xi_\mu
       \Sigma^\dagger \partial_\mu \Sigma \xi_\mu) \,,
\end{equation}
and discussed in this subsection.
The latter choice gives $\CO(a^4)$ operators such as
\begin{equation}
	\sum_{\mu} \sum_{\nu \neq \mu} \Str(\xi_{\mu \nu} \Sigma \xi_\mu \Sigma \xi_{\nu \mu} \Sigma^\dagger \xi_\mu \Sigma^\dagger) \,,
\end{equation}
which will be discussed separately below.
It is not possible, however, to break the symmetries with
two taste spurions and a mass spurion. Thus there are
no new $\CO(a^2 m)$ operators produced by $S_6^{FF(B)}$.

Although the operators in $S_6^{FF(B)}$ appear more complicated than
those in $S_6^{FF(A)}$, because of the
coupled spin and taste matrices, we can actually reuse the bulk of our
work from the previous subsection.  This is because we can still use
spurion analysis and group theory to determine the \emph{chiral}
structure of the mesonic operators.  We just need to be more careful
about the index structure and the specific taste matrices.  
We illustrate the procedure by working through the mapping
of $[V_\mu \times T_\mu]$ in detail; the method then
can be generalized straighforwardly to the remaining 
operators in $S_6^{FF(B)}$.

We begin by recalling the definition of this operator:
\begin{equation}
[V_\mu\times T_\mu] \equiv \sum_{\mu}\sum_{\nu\ne\mu}
\Big\{ \bar Q \sfno{\mu}{\mu\nu} Q
\,\bar Q \sfno{\mu}{\nu\mu} Q -
\bar Q \sfno{\mu}{\mu\nu5} Q
\,\bar Q \sfno{\mu}{5\nu\mu} Q \Big\} \,.
\label{eq:VTdef}
\end{equation} 
The second term in this expression removes the taste singlet component.
It is, however, cumbersome and unnecessary to keep both terms,
so we first simplify using
\begin{equation}
\frac12 \left( [V_\mu\times T_\mu] + [V\times T]\right)\equiv \sum_{\mu}\sum_{\nu\ne\mu}
\bar Q \sfno{\mu}{\mu\nu} Q
\,\bar Q \sfno{\mu}{\nu\mu} Q  \,.
\label{eq:VTdefsimp}
\end{equation} 
Since $[V\times T]$ is from $S_6^{FF(A)}$ this addition has
no impact on matching to operators which break rotation and/or taste
$SO(4)$ symmetries.  The chiral structure of the operator in (\ref{eq:VTdefsimp}) is given by
\begin{equation}
\CO(\mu)_{F} = \pm \left[\bar {Q_R} (\gamma_\mu\otimes F(\mu)_R) Q_R \pm
\bar{Q_L} (\gamma_\mu\otimes F(\mu)_L) Q_L \right]^2 \,,
\label{eq:OFformB}
\end{equation}
which has the same form as for $[V\times T]$,
(\ref{eq:OFform}), except that the taste spurions depend on $\mu$,
and the sum over $\mu$ has been removed for now.
The mapping then proceeds as follows:
first, determine the mapping of $\CO(\mu)_F$ onto mesonic operators,
 for fixed $\mu$;
next, set the taste spurions to their appropriate values;
and, finally, sum over $\mu$ and any other remaining indices.

The group theory required to determine the
independent mappings of $\CO(\mu)_{F}$ onto mesonic operators is 
identical to that for $[V\times T]$, since the
spurions $F(\mu)_{R,L}$ must transform like $F_{R,L}$.
Thus one obtains the types enumerated in
Tables~\ref{tab:VDDFFP} and \ref{tab:VDDFF}, with the provisos that
$F_{R,L}\longrightarrow F(\mu)_{R,L}$ and that we must be
careful with Lorentz (or, more precisely,
hypercubic group) indices. In particular, we do not yet know
how the indices $\mu$ in the operators in the tables are
connected with the index $\mu$ in $\CO(\mu)_F$.
To work this out we need to determine how
$\CO(\mu)_F$ transforms under rotations \emph{while keeping
the spurions fixed}. This may seem confusing because of
the index $\mu$ in $F(\mu)_{R,L}$, but once, for a given $\mu$, we
convert the taste matrices into spurion fields, they are 
to be treated as rotationally-invariant (pseudo-)scalar fields. 
Then, since $\CO(\mu)_F$ has
two vector indices, it has a component which is a singlet under
hypercubic rotations
(corresponding to summing over $\mu$), and a component which is part
of a 3-d hypercubic irrep (the diagonal
part of the two-index symmetric tensor representation of the 
Euclidean rotation group). The former is not of interest here,
since the sum over $\mu$ decouples the spin and taste indices.
It is the latter which we need to match onto mesonic operators.

It is straightforward to do this matching. For each of the
operators in Tables~\ref{tab:VDDFFP} and \ref{tab:VDDFF}, we
need to replace ``$\partial_\mu \partial_\mu$'' with
``$\partial_\mu \partial_\mu - (1/4)\sum_\rho \partial_\rho \partial_\rho$''.
This projects out the singlet component.
The singlet component has decoupled spin and taste indices, however,
so we can drop it if we wish, and keep only ``$\partial_\mu \partial_\mu$''.
Doing so gives the following simple prescription: keep
the operators in Tables~\ref{tab:VDDFFP} and \ref{tab:VDDFF} as is,
but substitute $F_{R,L}\longrightarrow F(\mu)_{R,L}$.
In this way the spin and taste indices become correlated.  Now that the operator mapping is done for fixed $\mu$, we can set $F(\mu)$ to the appropriate taste matrix (in this case $\xi_{\mu \nu}$) and then sum over the remaining indices (in this case both $\mu$ and $\nu$ with the constraint that $\mu \neq \nu$).  For example, after completing the three steps of
the mapping procedure outlined above for operator 1
(Table~\ref{tab:VDDFFP}) we obtain
\begin{equation}
\sum_{\mu}\sum_{\nu\ne\mu} \Str(\partial_\mu \Sigma^\dagger \partial_\mu \Sigma
\Sigma^\dagger \xi_{\mu\nu} \Sigma \xi_{\nu\mu})
\,.
\end{equation}
This operator, like the $S_6^{FF(B)}$ operator it came from,
breaks the symmetries down to the lattice spin-taste group
$\Gamma_4 \semitimes SW_4$.

Carrying out the same procedure for all the types
in Tables~\ref{tab:VDDFFP} and \ref{tab:VDDFF} leads to
the mesonic operators collected in 
the upper panel of Table~\ref{tab:FFB}.
Note that the mapping of the four-fermion operator
$[A_\mu \times T_\mu]$ leads to the same set of 
mesonic operators as those from $[V_\mu\times T_\mu]$.

The mapping of operators $[T_\mu \times V_\mu]$ and
$[T_\mu \times A_\mu]$ can be done similarly.
The chiral structure is given by
\begin{equation}
\CO''(\mu)_F= \sum_{\nu\ne \mu}
\bar {Q_L} (\gamma_{\mu\nu}\otimes \tilde F(\mu)_L) Q_R \; \bar{Q_R} (\gamma_{\nu\mu}\otimes \tilde F(\mu)_R) Q_L \,,
\label{eq:OFPPformB}
\end{equation}
where the taste spurions depend on $\mu$,
and $\mu$ is not (yet) summed.\footnote{Note that the chiral structure of $[T_\mu \times V_\mu]$ is not the same as that of $[T \times V]$.}$^,$\footnote{We thank Claude Bernard for pointing out an
error in our discussion of this operator
in earlier version of this paper. }
On the other hand, $\nu$ is summed, so the rotational properties
of the operator are the same as those of $\CO(\mu)_F$ discussed above.
Thus when it is mapped onto mesonic operators with two derivatives
their indices can both be set to $\mu$.
The types of mesonic operators that arise are as for
$[S\times V,A]$, except that here we need keep only those with one $\tilde F_L$ and one $\tilde F_R$, namely operator types 35-42 in Table~\ref{tab:SDDFF}.  Furthermore, here we must substitute $\tilde F_{R,L} \longrightarrow \tilde F(\mu)_{R,L}$
(with $F(\mu)_{R,L}\to \xi_\mu$ or $\xi_{\mu5}$ in the end).
The resulting specific mesonic operators are collected
in the lower panel of Table~\ref{tab:FFB}.

In Ref.~\cite{LS} it was claimed that one of the operators
generated by $[T_\mu \times V_\mu]$ at $\CO(a^2p^2)$ was
\begin{equation}
\sum_{\mu\ne\nu\ne\rho\ne\mu}
\Str\left[(\partial_\mu - \partial_\nu) \Sigma \xi_\rho\right]
\Str\left[(\partial_\mu - \partial_\nu) \Sigma \xi_\rho\right]
\,.
\end{equation}
This operator does not, however, appear in our lists.
In fact, it is not invariant under the 
``spin-taste locked'' rotation
\begin{eqnarray}
	\partial_\nu \rightarrow \partial_\mu, & \;\;\;\; &
	\partial_\mu \rightarrow -\partial_\nu \nonumber \\ \gamma_\nu
	\rightarrow \gamma_\mu, & \;\;\;\; & \gamma_\mu \rightarrow
	-\gamma_\nu \nonumber \\ \xi_\nu \rightarrow \xi_\mu, &
	\;\;\;\; & \xi_\mu \rightarrow -\xi_\nu \,.
\end{eqnarray}
which corresponds to a lattice $SW_4$ transformation~\cite{Verstegen,LS}.
Thus it does not arise.

\subsection{Double insertions of four-fermion operators}
\label{app:double}

Mesonic operators of $\CO(a^4)$ can arise from the combination of two four-fermion
operators, each of which brings a power of $a^2$,
or by mapping a single $\CO(a^4)$ quark-level operator. 
However, as noted in Sec.~\ref{sec:Review}, the $\CO(a^4)$ operators
do not break any additional symmetries and do not contain any more
correlated taste indices, so they do not lead to additional mesonic operators.

Thus we consider only the mesonic operators which result from two
insertions of four-fermion operators. These two insertions need not have
same spin or taste, and may come from either $S_6^{FF(A)}$ or
$S_6^{FF(B)}$.  We first consider pairs of $FF(A)$ operators,
separating the various possible spin combinations.  We then generalize
these results to the case in which one or both of the operators is in
$S_6^{FF(B)}$.

\subsubsection{Spin $V$ or $A$ four-fermion operators
 with spin $V$ or $A$ four-fermion operators}

Operators of $\CO(a^4)$ must contain four taste spurions -- two from
each four-fermion operator.  The taste spurions corresponding to
operators with spins $V$ and $A$ are $F_L$ and $F_R$, which in the end
can be set to $S$, $P$ or $T$. In fact, as we have seen above, the
taste singlet always leads to operators which either vanish or
are of the same form as LO operators. We have checked that this 
remains true for $\CO(a^4)$ operators, and so drop taste $S$ from
the beginning. Thus we can set $\Str(F_L)=\Str(F_R)=0$, which
simplifies the operator enumeration.  

There are five parity-invariant ways to combine the four
taste spurions:
\begin{eqnarray}
	&&(F_L F_R)(F_L' F_R') \nonumber \\ &&(F_L F_L)(F_L' F_R') +
	p.c. \nonumber \\ &&(F_L F_R)(F_L' F_L') + p.c. \nonumber \\
	&&(F_L F_L)(F_L' F_L') + p.c. \nonumber \\ &&(F_L F_L)(F_R'
	F_R') + p.c.\,.
\label{eqn:taste_VA_VA}\end{eqnarray}
Here we use primes to distinguish between spurions from the separate
four-fermion operators because they may have different tastes.  To
determine the number of linearly-independent chiral operators, we use
graded group theory as before.  For simplicity, we build operators out
of the left-handed basis defined in (\ref{eqn:basis1}).  Both $F_L$
and $\Sigma F_R \Sigma^\dagger$ are now supertraceless, so they have no
singlet component and transform as adjoints under the chiral symmetry
group.  The corresponding $\CO(a^4)$ operators are therefore chiral
singlets contained in the product of four adjoints, with various
symmetries imposed based on the handedness of the four spurions.

\begin{figure}
	\epsfysize=0.27in \vspace{0in}\hspace{-.0in}\rotatebox{0}
	{\leavevmode\epsffile{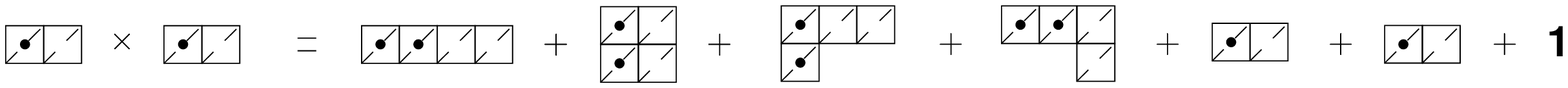}} \caption{The
	product of two adjoints in $SU(4N | 4M)$.  The result
	also applies to $SU(N)$ for $N > 3$ (with the dashed lines
	removed).}  \vspace{-0in} \label{fig:Adj_YT}
\end{figure}

Operators of the first type, $(F_L F_R)(F_L' F_R')$, contain four
different taste spurions, and therefore possess no additional
symmetry.  The number of linearly-independent operators of this sort
is simply the number of singlets in the product of four adjoints.
(The product of two adjoints is shown in Fig.~\ref{fig:Adj_YT}.)
There are nine such operators, and they are given in
Table~\ref{tab:a4_VA_VA_s1}.  Most, however, reduce to LO
operators or field-independent constants after setting $F$
and $F'$ to specific taste matrices; only three are new.

\begin{table}\begin{tabular}{llc}  \hline\hline

\multicolumn{2}{c}{\emph{Operator}} & \multicolumn{1}{c}{\emph{Keep?}} \\[0.5mm] \hline

69. & $\Str(F_L \Sigma F_R \Sigma^\dagger F_L' \Sigma F_R'
\Sigma^\dagger)$ & Yes -- combined with 74 \\[0.5mm]

70. & $\Str(F_L \Sigma F_R F_R' \Sigma^\dagger F_L')$ & No \\[0.5mm]

71. & $\Str(F_L F_L' \Sigma F_R F_R' \Sigma^\dagger)$ & No \\[0.5mm]

72. & $\Str(F_L F_L' \Sigma F_R' F_R \Sigma^\dagger)$ & No \\[0.5mm]

73. & $\Str(F_L \Sigma F_R' F_R \Sigma^\dagger F_L')$ & No \\[0.5mm]

74. & $\Str(F_L \Sigma F_R' \Sigma^\dagger F_L' \Sigma F_R
\Sigma^\dagger)$ & Yes -- combined with 69 \\[0.5mm]

75. & $\Str(F_L \Sigma F_R \Sigma^\dagger)\Str(F_L' \Sigma F_R'
\Sigma^\dagger)$ & Yes \\[0.5mm]

76. & $\Str(F_L F_L')\Str( F_R F_R' )$ & No \\[0.5mm]

77. & $\Str(F_L \Sigma F_R' \Sigma^\dagger)\Str(F_L' \Sigma F_R
\Sigma^\dagger )$ & Yes \\[0.5mm]

\hline\hline

\end{tabular}\caption{The nine linearly-independent $\CO(a^4)$ 
operators composed of taste spurions $F_L$, $F_R$, $F'_L$ and $F'_R$.}
\label{tab:a4_VA_VA_s1}\end{table}

Operators of the last type, $(F_L F_L)(F_R' F_R') + p.c.$, contain two
pairs of identical taste spurions.  They therefore come from the
singlet irreps in the product of one symmetric product of two adjoints
with another symmetric product of two adjoints, i.e. the product of
Fig.~\ref{fig:GYT} with itself.  There are four such singlets; they
correspond to the operators in Table~\ref{tab:a4_VA_VA_s2}.  However,
only two of these operators are new and nontrivial.

\begin{table}\begin{tabular}{llc}  \hline\hline

\multicolumn{2}{c}{\emph{Operator}} & \multicolumn{1}{c}{\emph{Keep?}} \\[0.5mm] \hline

78. & $\Str(F_L F_L \Sigma F_R' F_R' \Sigma^\dagger) + p.c.$ & No
\\[0.5mm]

79. & $\Str(F_L \Sigma F_R' \Sigma^\dagger F_L \Sigma F_R'
\Sigma^\dagger) + p.c.$ & Yes \\[0.5mm]

80. & $\Str(F_L F_L)\Str( F_R' F_R' ) + p.c.$ & No \\[0.5mm]

81. & $\Str(F_L \Sigma F_R' \Sigma^\dagger)\Str(F_L \Sigma F_R'
\Sigma^\dagger) + p.c.$ & Yes \\[0.5mm]

\hline\hline

\end{tabular}\caption{The four linearly-independent $\CO(a^4)$ 
operators composed of two $F_L$ and two $F'_R$ taste spurions.}
\label{tab:a4_VA_VA_s2}\end{table}

It is straightforward to show that the other three spurion
combinations in (\ref{eqn:taste_VA_VA}) do not generate any additional
$\CO(a^4)$ operators, so we choose to omit their discussion here.  All
that remains is the somewhat tedious task of replacing the taste
spurions with specific taste matrices. 
There are three non-trivial pairings of tastes,
leading to three different taste matrix substitutions:
\begin{eqnarray}
	\,[V,A \times P] \textrm{ with } [V,A \times P] & \Rightarrow
	& F = F' = \xi_5 \nonumber \\ \,[V,A \times P] \textrm{ with }
	[V,A \times T] & \Rightarrow & F = \xi_5, \;\; F' = \xi_{\mu
	\nu} \nonumber \\ \,[V,A \times T] \textrm{ with } [V,A \times
	T] & \Rightarrow & F = \xi_{\mu \nu}, \;\; F' = \xi_{\rho
	\sigma}\,.
\end{eqnarray} 
Note that, in the case of taste $T$ with $T$, the two pairs of taste
indices must be contracted separately because $S_6^{FF(A)}$ respects
both $SO(4)$-taste and Lorentz symmetry.  The three corresponding sets
of chiral operators are given in Table~\ref{tab:a4_VA_VA}.
  
\subsubsection{Spin $S$, $P$, or $T$ four-fermion operators with 
spin $S$, $P$, or $T$ four-fermion operators}

The taste spurions corresponding to operators with spins
$S$, $P$, and $T$ are $\tilde{F_L}$ and $\tilde{F_R}$, and can
be combined in five parity-invariant ways:
\begin{eqnarray}
	&&(\tilde{F_L} \tilde{F_R})(\tilde{F_L}' \tilde{F_R}')
	\nonumber \\ &&(\tilde{F_L} \tilde{F_L})(\tilde{F_L}'
	\tilde{F_R}') + p.c. \nonumber \\ &&(\tilde{F_L}
	\tilde{F_R})(\tilde{F_L}' \tilde{F_L}') + p.c. \nonumber \\
	&&(\tilde{F_L} \tilde{F_L})(\tilde{F_L}' \tilde{F_L}') +
	p.c. \nonumber \\ &&(\tilde{F_L} \tilde{F_L})(\tilde{F_R}'
	\tilde{F_R}') + p.c.\,.
\label{eqn:taste_SP_SP}\end{eqnarray}
In this case we build operators out of the left-handed basis defined
in (\ref{eqn:basis2}).  Neither $F_L \Sigma^\dagger$ nor $\Sigma F_R$
is supertraceless, so each transforms as a bifundamental under the
chiral symmetry group.  Thus their product contains slightly more
singlet irreps, and therefore generates more operators, than in the
spin $V$ case.

Operators of type $(\tilde{F_L} \tilde{F_R})(\tilde{F_L}'
\tilde{F_R}')$ correspond to singlet irreps in the product of four
bifundamentals, i.e. the product of Fig.~\ref{fig:bigGYT} with
itself.  There are twenty-four singlet irreps in this product, but we
only show, in Table~\ref{tab:a4_SP_SP_s1},
the three that lead to new $\CO(a^4)$ operators.

\begin{table}\begin{tabular}{llc}  \hline\hline

\multicolumn{2}{c}{\emph{Operator}} & \multicolumn{1}{c}{\emph{Keep?}} \\[0.5mm] \hline

82. & $\Str(\tilde{F_L} \Sigma^\dagger \tilde{F_L}' \Sigma^\dagger
)\Str(\Sigma \tilde{F_R} \Sigma \tilde{F_R}')$ & Yes \\[0.5mm]

83. & $\Str(\tilde{F_L} \Sigma^\dagger \tilde{F_L}' \Sigma^\dagger
)\Str(\Sigma \tilde{F_R})\Str(\Sigma \tilde{F_R}') + p.c.$ & Yes
\\[0.5mm]

84. & $\Str(\tilde{F}_L \Sigma ^\dagger) \Str(\tilde{F}_L' \Sigma
^\dagger) \Str(\Sigma \tilde{F}_R) \Str(\Sigma \tilde{F}_R')$ & Yes
\\[0.5mm]

\hline\hline

\end{tabular}\caption{The three non-trivial $\CO(a^4)$ 
operators composed of the four taste spurions
$\tilde F_L$, $\tilde F_R$, $\tilde F'_L$ and $\tilde F'_R$.}
\label{tab:a4_SP_SP_s1}\end{table}

In operators of type $\big((\tilde{F_L} \tilde{F_L})(\tilde{F_L}'
\tilde{F_R}') + p.c.\big)$ and $\big((\tilde{F_L}
\tilde{F_R})(\tilde{F_L}' \tilde{F_L}') + p.c.\big)$, one pair of
taste spurions is identical.  Thus the number of such operators equals
the number of singlet irreps in the product of one symmetric product
of two bifundamentals with another two bifundamentals, i.e. the
product of Fig.~\ref{fig:symGYT} with Fig.~\ref{fig:bigGYT}.  This
product contains fourteen singlets, so there are twenty-eight such
operators.  However, only eight are new, and these are listed in
Table~\ref{tab:a4_SP_SP_s2}.

\begin{table}\begin{tabular}{llc}  \hline\hline

\multicolumn{2}{c}{\emph{Operator}} & \multicolumn{1}{c}{\emph{Keep?}} \\[0.5mm] \hline

85. & $\Str(\tilde{F}_L \Sigma ^\dagger \tilde{F}_L \Sigma ^\dagger
\tilde{F}_L' \Sigma ^\dagger) \Str(\Sigma \tilde{F}_R') + p.c.$ & Yes
\\[0.5mm]

86. & $\Str(\tilde{F}_L' \Sigma ^\dagger \tilde{F}_L' \Sigma ^\dagger
\tilde{F}_L \Sigma ^\dagger) \Str(\Sigma \tilde{F}_R) + p.c.$ & Yes
\\[0.5mm]

87. & $\Str(\tilde{F}_L \Sigma ^\dagger \tilde{F}_L \Sigma ^\dagger)
\Str(\tilde{F}_L' \Sigma ^\dagger) \Str(\Sigma \tilde{F}_R') + p.c.$ &
Yes \\[0.5mm]

88. & $\Str(\tilde{F}_L' \Sigma ^\dagger \tilde{F}_L' \Sigma ^\dagger)
\Str(\tilde{F}_L \Sigma ^\dagger) \Str(\Sigma \tilde{F}_R) + p.c.$ &
Yes \\[0.5mm]

89. & $\Str(\tilde{F}_L \Sigma ^\dagger \tilde{F}_L' \Sigma ^\dagger)
\Str(\tilde{F}_L \Sigma ^\dagger) \Str(\Sigma \tilde{F}_R') + p.c.$ &
Yes \\[0.5mm]

90. & $\Str(\tilde{F}_L' \Sigma ^\dagger \tilde{F}_L \Sigma ^\dagger)
\Str(\tilde{F}_L' \Sigma ^\dagger) \Str(\Sigma \tilde{F}_R) + p.c.$ &
Yes \\[0.5mm]

91. & $\Str(\tilde{F}_L \Sigma ^\dagger) \Str(\tilde{F}_L \Sigma
^\dagger) \Str(\tilde{F}_L' \Sigma ^\dagger) \Str(\Sigma \tilde{F}_R')
+ p.c.$ & Yes \\[0.5mm]

92. & $\Str(\tilde{F}_L' \Sigma ^\dagger) \Str(\tilde{F}_L' \Sigma
^\dagger) \Str(\tilde{F}_L \Sigma ^\dagger) \Str(\Sigma \tilde{F}_R) +
p.c.$ & Yes \\[0.5mm]

\hline\hline

\end{tabular}\caption{The eight non-trivial $\CO(a^4)$ 
operators composed of two
$\tilde F_L$, one $\tilde F'_L$ and one $\tilde F'_R$
taste spurions.}
\label{tab:a4_SP_SP_s2}\end{table}

Finally, in operators of type $\big((\tilde{F_L}
\tilde{F_L})(\tilde{F_L}' \tilde{F_L}') + p.c.\big)$ and
$\big((\tilde{F_L} \tilde{F_L})(\tilde{F_R}' \tilde{F_R}') +
p.c.\big)$, both pairs of taste spurions are identical.  Therefore
these operators correspond to chiral singlets in the product of one
symmetric product of bifundamentals with another symmetric product of
bifundamentals, or of Fig.~\ref{fig:symGYT} with itself.  There are
ten such singlets, and therefore twenty such operators, but only
fourteen are new. We list these in Table~\ref{tab:a4_SP_SP_s3}.

\begin{table}\begin{tabular}{llc}  \hline\hline

\multicolumn{2}{c}{\emph{Operator}} & \multicolumn{1}{c}{\emph{Keep?}} \\[0.5mm] \hline

93. & $\Str(\tilde{F}_L \Sigma ^\dagger \tilde{F}_L \Sigma ^\dagger
\tilde{F}_L' \Sigma ^\dagger \tilde{F}_L' \Sigma ^\dagger) + p.c.$ &
Yes \\[0.5mm]

94. & $\Str(\tilde{F}_L \Sigma ^\dagger \tilde{F}_L' \Sigma ^\dagger
\tilde{F}_L \Sigma ^\dagger \tilde{F}_L' \Sigma ^\dagger) + p.c.$ &
Yes \\[0.5mm]

95. & $\Str(\tilde{F}_L \Sigma ^\dagger \tilde{F}_L \Sigma ^\dagger
\tilde{F}_L' \Sigma ^\dagger) \Str(\tilde{F}_L' \Sigma ^\dagger)+
p.c.$ & Yes \\[0.5mm]

96. & $\Str(\tilde{F}_L' \Sigma ^\dagger \tilde{F}_L' \Sigma ^\dagger
\tilde{F}_L \Sigma ^\dagger) \Str(\tilde{F}_L \Sigma ^\dagger)+ p.c.$
& Yes \\[0.5mm]

97. & $\Str(\tilde{F}_L \Sigma ^\dagger \tilde{F}_L \Sigma
^\dagger)\Str(\tilde{F}_L' \Sigma ^\dagger \tilde{F}_L' \Sigma
^\dagger)+ p.c.$ & Yes \\[0.5mm]

98. & $\Str(\tilde{F}_L \Sigma ^\dagger \tilde{F}_L' \Sigma
^\dagger)\Str(\tilde{F}_L \Sigma ^\dagger \tilde{F}_L' \Sigma
^\dagger)+ p.c.$ & Yes \\[0.5mm]

99. & $\Str(\tilde{F}_L \Sigma ^\dagger \tilde{F}_L \Sigma
^\dagger)\Str(\tilde{F}_L' \Sigma ^\dagger)\Str(\tilde{F}_L' \Sigma
^\dagger) + p.c.$ & Yes \\[0.5mm]

100. & $\Str(\tilde{F}_L \Sigma ^\dagger \tilde{F}_L' \Sigma
^\dagger)\Str(\tilde{F}_L \Sigma ^\dagger)\Str(\tilde{F}_L' \Sigma
^\dagger) + p.c.$ & Yes \\[0.5mm]

101. & $\Str(\tilde{F}_L' \Sigma ^\dagger \tilde{F}_L' \Sigma
^\dagger)\Str(\tilde{F}_L \Sigma ^\dagger)\Str(\tilde{F}_L \Sigma
^\dagger) + p.c.$ & Yes \\[0.5mm]

102. & $\Str(\tilde{F}_L \Sigma ^\dagger) \Str(\tilde{F}_L \Sigma
^\dagger) \Str(\tilde{F}_L' \Sigma^\dagger) \Str(\tilde{F}_L' \Sigma
^\dagger)+ p.c.$ & Yes \\[0.5mm]

103. & $\Str(\tilde{F}_L \Sigma ^\dagger \tilde{F}_L \Sigma ^\dagger)
\Str(\Sigma \tilde{F}_R' \Sigma \tilde{F}_R') + p.c.$ & Yes \\[0.5mm]

104. & $\Str(\tilde{F}_L \Sigma ^\dagger \tilde{F}_L \Sigma ^\dagger)
\Str(\Sigma \tilde{F}_R') \Str(\Sigma \tilde{F}_R') + p.c.$ & Yes
\\[0.5mm]

105. & $\Str(\tilde{F}_L \Sigma^\dagger) \Str(\tilde{F}_L\Sigma^\dagger)
\Str(\Sigma \tilde{F}_R' \Sigma \tilde{F}_R')  + p.c.$ &
Yes \\[0.5mm]

106. & $\Str(\tilde{F}_L \Sigma ^\dagger) \Str(\tilde{F}_L \Sigma
^\dagger) \Str(\Sigma \tilde{F}_R') \Str(\Sigma \tilde{F}_R') + p.c.$
& Yes \\[0.5mm]

\hline\hline

\end{tabular}\caption{The fourteen non-trivial $\CO(a^4)$ 
operators composed of two $\tilde F_L$ and either two $\tilde F'_L$ or two $\tilde F'_R$ 
taste spurions.}
\label{tab:a4_SP_SP_s3}\end{table}

As we know from the previous section, spin $T$ four-fermion operators
generate a subset of those chiral operators generated by spins $S$ and
$P$; we can therefore ignore them.  Thus, when pairing four-fermion
operators, there are three combinations requiring different taste
matrix substitutions:
\begin{eqnarray}
	\,[S,P \times V] \textrm{ with } [S,P \times V] & \Rightarrow
	& \tilde{F} = \xi_\mu, \;\; \tilde{F}' = \xi_\nu \nonumber \\
	\,[S,P \times V] \textrm{ with } [S,P \times A] & \Rightarrow
	& \tilde{F} = \xi_\mu, \;\; \tilde{F}' = \xi_{\nu 5} \nonumber
	\\ \,[S,P \times A] \textrm{ with } [S,P \times A] &
	\Rightarrow & \tilde{F} = \xi_{\mu 5}, \;\; \tilde{F}' =
	\xi_{\nu 5}\,.
\end{eqnarray} 
In all of these operators $\mu$ and $\nu$ must be separately
contracted to maintain $SO(4)$-taste and Lorentz symmetry.
Table~\ref{tab:a4_SP_SP} contains all of the corresponding operators
in the chiral Lagrangian.

\subsubsection{Spin $V$ or $A$ four-fermion operators with spin $S$, $P$, or $T$ four-fermion operators}

Operators that arise from the insertion of one operator with spin $V$
or $A$ and one with spin $S$, $P$, or $T$ necessarily have mixed
tastes.  Each four-fermion operator contributes
two taste spurions, so the corresponding chiral operators contain two
$F$'s and two $\tilde{F}$'s.  The five parity-invariant ways of
combining these spurions are
\begin{eqnarray}
	&&(F_L F_R)(\tilde{F_L} \tilde{F_R}) \nonumber \\ &&(F_L
	F_L)(\tilde{F_L} \tilde{F_R}) + p.c. \nonumber \\ &&(F_L
	F_R)(\tilde{F_L} \tilde{F_L}) + p.c. \nonumber \\ &&(F_L
	F_L)(\tilde{F_L} \tilde{F_L}) + p.c. \nonumber \\ &&(F_L
	F_L)(\tilde{F_R} \tilde{F_R}) + p.c.\,.
\label{eqn:taste_VS_SP}\end{eqnarray}
We do not need to use primes, since the spurions from the two four-fermion
operators are already distinguished by the presence or absence of the tilde.
As before, we build our operators using a left-handed basis, and in
this case we must use both taste spurion bases defined in
(\ref{eqn:basis1}) and (\ref{eqn:basis2}).  Recall that under chiral
symmetry transformations, $F_L$ and $\Sigma F_R \Sigma^\dagger$
transform as adjoints, while $\tilde{F_L} \Sigma^\dagger$ and $\Sigma
\tilde{F_R}$ transform as bifundamentals.  Thus the group theory
counting will be slightly different than in the two previous sections,
and each of the five spurion combinations in
(\ref{eqn:taste_VS_SP}) must be considered separately.

First consider operators of type $(F_L F_R)(\tilde{F_L} \tilde{F_R})$.
These come from the singlet irreps in the product of two adjoints with
two bifundamentals.  There are fourteen, but only four correspond to
new NLO operators.  Next consider the \emph{third} type, $(F_L
F_R)(\tilde{F_L} \tilde{F_L}) + p.c.$.  Because there are two
$\tilde{F_L}$'s, chiral operators correspond to chiral singlets in the
product of two adjoints with the symmetric product of two
bifundamentals.  This product contains eight singlets, but only leads
to four new NLO operators.  Finally, the other three combinations in
(\ref{eqn:taste_VS_SP}) can be shown not to produce any new operators,
so we do not discuss them further.  Table~\ref{tab:a4_VA_SP_s1} shows
the eight new operators that can be built out of two $F$'s and two
$\tilde{F}$'s.

\begin{table}\begin{tabular}{llc}  \hline\hline

\multicolumn{2}{c}{\emph{Operator}} & \multicolumn{1}{c}{\emph{Keep?}} \\[0.5mm] \hline

107. & $\Str(F_L \Sigma \tilde{F_R} \Sigma F_R \Sigma^\dagger
\tilde{F_L} \Sigma^\dagger)$ & Yes \\[0.5mm]

108. & $\Str(F_L \Sigma \tilde{F_R} \Sigma F_R
\Sigma^\dagger)\Str(\tilde{F_L} \Sigma^\dagger)$ & Yes -- combined
with 109 \\[0.5mm]

109. & $\Str(F_L \Sigma F_R \Sigma^\dagger \tilde{F_L}
\Sigma^\dagger)\Str(\Sigma \tilde{F_R})$ & Yes -- combined with 108
\\[0.5mm]

110. & $\Str(F_L \Sigma F_R \Sigma^\dagger)\Str(\tilde{F_L}
\Sigma^\dagger)\Str(\Sigma \tilde{F_R})$ & Yes \\[0.5mm]

111. & $\Str(\tilde{F_L} \Sigma^\dagger \tilde{F_L} \Sigma^\dagger F_L
\Sigma F_R \Sigma^\dagger) + p.c.$ & Yes\\[0.5mm]

112. & $\Str(\tilde{F_L} \Sigma^\dagger \tilde{F_L}
\Sigma^\dagger)\Str(F_L \Sigma F_R \Sigma^\dagger) + p.c.$ & Yes
\\[0.5mm]

113. & $\Str(\tilde{F_L} \Sigma^\dagger F_L \Sigma F_R
\Sigma^\dagger)\Str(\tilde{F_L} \Sigma^\dagger) + p.c.$ & Yes
\\[0.5mm]

114. & $\Str(\tilde{F_L} \Sigma^\dagger)\Str(\tilde{F_L}
\Sigma^\dagger)\Str(F_L \Sigma F_R \Sigma^\dagger) + p.c.$ & Yes
\\[0.5mm]

\hline\hline

\end{tabular}\caption{The eight non-trivial $\CO(a^4)$ 
operators composed of spurions
$F_L$, $F_R$  and two $\tilde F's$ .}
\label{tab:a4_VA_SP_s1}\end{table}

In this mixed-taste and mixed-spurion case, there are four possible
four-fermion operator combinations and corresponding taste matrix
substitutions:
\begin{eqnarray}
	\,[V,A \times P] \textrm{ with } [S,P \times V] & \Rightarrow
	& F = \xi_5, \;\; \tilde{F} = \xi_\mu \nonumber \\ \,[V,A
	\times P] \textrm{ with } [S,P \times A] & \Rightarrow & F =
	\xi_5, \;\; \tilde{F} = \xi_{\mu 5} \nonumber \\ \,[V,A \times
	T] \textrm{ with } [S,P \times V] & \Rightarrow & F = \xi_{\mu
	\nu}, \;\; \tilde{F} = \xi_{\rho} \nonumber \\ \,[V,A \times
	T] \textrm{ with } [S,P \times A] & \Rightarrow & F = \xi_{\mu
	\nu}, \;\; \tilde{F} = \xi_{\rho 5}\,.
\end{eqnarray} 
Once again, all index pairs are contracted separately, thereby maintaining $SO(4)$-taste and Lorentz symmetry.
The chiral operators of $\CO(a^4)$ arising from the insertion of a
spin $V$ or $A$ four-fermion operator and the insertion of a spin $S$,
$P$, or $T$ four-fermion operator are given in
Table~\ref{tab:a4_VA_SP}.
 
\subsubsection{Double insertions involving four-fermion operators in $S_6^{FF(B)}$}

In this section we determine mesonic operators of $\CO(a^4)$ which arise from
the double insertion of four-fermion operators, at least one of which
is from $S_6^{FF(B)}$.  
As in the single four-fermion operator insertion case, the chiral structures
are identical to those for insertions of the corresponding 
operators in $S_6^{FF(A)}$,
but new index structures arise which break the $SO(4)$ taste
 symmetry.  Also like before, we are not interested in
operators which \emph{do not} break this symmetry, since
these have already been completely determined.  Thus we first consider
insertions of two four-fermion operators from $S_6^{FF(B)}$.  After
this discussion, it will be easy to see that combinations of one
$FF(A)$ operator with one $FF(B)$ operator cannot generate anything
additional, and in fact only generate operators that respect
$SO(4)$-taste and Lorentz symmetries.

Although the spin order is reversed as compared to previous sections,
we begin with double insertions of $[T_\mu \times V_\mu,A_\mu]$.  This
is because it is easy to guess the correct $SO(4)$-taste symmetry
breaking chiral operators.  They have the same chiral structure as a subset of
operators that come from double insertions of $[S,P \times V,A]$, namely those which arise from $L-R$ cross-terms (operators 82-84).  A quick glance at Table~\ref{tab:a4_SP_SP}
reveals that there is only one possible symmetry-breaking index
contraction: $\mu = \nu$.  As this contraction does not break any of
the discrete lattice symmetries, it must be present at some order
in our power counting.  We can show, however, that it appears at
$\CO(a^4)$ using an argument similar to those in
Section~\ref{app:singleFFB}.  For fixed $\mu$ and $\mu'$, the general
spin-taste index structure of these operators (ignoring factors of $\xi_5$ in $A_\mu$) is the following:
\begin{equation}
	\sum_{\nu \neq \mu} \sum_{\nu' \neq \mu'} (\mu \nu \otimes
	\mu)(\nu \mu \otimes \mu)(\mu' \nu' \otimes \mu')(\nu' \mu'
	\otimes \mu')
\end{equation}
Because mesonic operators of $\CO(a^4)$ have no derivatives, invariance
under the lattice symmetries implies that they must be \emph{singlets under
hypercubic rotations}.  As was
discussed previously, both $\mu \mu$ and $\mu'\mu'$ contain a singlet
and a 3-d irrep.  Thus their product contains two overall
singlets -- one from the product of the singlets and one from the 
product of the 3-d irreps. Only the latter gives new operators,
for it contains a part in which $\mu = \mu'$,
confirming our initial guess.  
The $SO(4)$-taste symmetry breaking operators arising from
double insertions of $[T_\mu \times V_\mu,A_\mu]$ are given in
the first two panels of Table~\ref{tab:a4_FFB}.

Next we proceed to mixed-spin operators, i.e. the result of an
insertion of $[V_\mu,A_\mu \times T_\mu]$ with an insertion of $[T_\mu
\times V_\mu,A_\mu]$.  These operators have identical chiral
structures to the ones in the top portion of
Table~\ref{tab:a4_VA_SP}.\footnote{%
Note that we do not need the
$``-\mu\nu5\,"$ parts of $[V_\mu,A_\mu \times T_\mu]$, since they can
be removed by adding an appropriate amount of $[V,A \times T]$,
as discussed in Sec.~\ref{app:singleFFB}.  This
simplifies the final taste matrix substitution.}
Once again, there is
only one possible $SO(4)$-taste breaking index contraction, $\mu =
\rho$, but we must still show that this is the correct mapping of the
four-fermion operator product.\footnote{Note that $\nu = \rho$ does
not produce different operators since these are all just dummy
indices.}  In this case, the operators have the following index
structure at fixed $\mu$ and $\mu'$:
\begin{equation}
	\sum_{\nu \neq \mu} \sum_{\nu' \neq \mu'} (\mu \otimes \mu
	\nu)(\mu \otimes \nu \mu)(\mu' \nu' \otimes \mu')(\nu' \mu'
	\otimes \mu')\,.
\end{equation}
Just as before, one hypercubic singlet in this product comes from the two
3-d irreps, and contains a part with $\mu = \mu'$.  This generates the
$\CO(a^4)$ mixed-spin operators in the third panel of
Table~\ref{tab:a4_FFB}.

Finally we address double insertions of $[V_\mu,A_\mu \times T_\mu]$.
This will produce operators with the same chiral structure as double
insertions of $[V,A \times T]$.  Such operators are given in the
bottom portion of Table~\ref{tab:a4_VA_VA}.  In this case, multiple
combinations of index contractions are possible: $\mu = \rho$ and $\mu
= \rho$, $\nu = \sigma$.  Both are allowed by the lattice
symmetries, but we must determine whether either arises at $\CO(a^4)$.
After fixing $\mu$ and $\mu'$, the index structure is
\begin{equation}
	\sum_{\nu \neq \mu} \sum_{\nu' \neq \mu'} (\mu \otimes \mu
	\nu)(\mu \otimes \nu \mu)(\mu' \otimes \mu' \nu')(\mu' \otimes
	\nu' \mu')\,.
\end{equation}
Now we can see that the only nontrivial way of producing a 
hypercubic singlet is $\mu = \mu'$, just like in the two previous
examples.  The indices $\nu$ and $\nu'$ are independently summed
from the beginning. We conclude that only the first of the two possible index contractions arises
from the double insertion of four fermion operators.
The final panel of Table~\ref{tab:a4_FFB} shows the $\CO(a^4)$ chiral operators
generated by double insertions of $[V_\mu,A_\mu \times T_\mu]$.

After these examples, it is easy to see that the combination of an
$FF(A)$ four-fermion operator with an $FF(B)$ four-fermion operator
cannot generate any additional operators.  This is because, in the
$FF(A)$ operator, spin and taste indices are completely
uncorrelated. Thus these operators contain only a singlet under
hypercubic rotations, and can only form a overall singlet by
combining with the singlet component of the $FF(B)$ operator.
But doing so removes the correlations between taste indices.
Such mixed-symmetry double insertions only modify the coefficients 
of the $\CO(a^4)$ $FF(A)$ operators.

\bigskip

A natural question that arises is what happened to the second possible
set of index contractions in the $[V_\mu, A_\mu \times T_\mu]^2$ case,
in which both $\mu = \mu'$ and $\nu = \nu'$.  This would lead to
mesonic operators such as
\begin{equation}
\sum_{\mu \neq \nu} \left\{\Str(\xi_{\mu \nu} \Sigma \xi_{\nu \mu} \Sigma^\dagger \xi_{\mu \nu} \Sigma \xi_{\nu \mu} \Sigma^\dagger) + p.c.\right\}
\,,
\label{eq:a8}
\end{equation}
i.e. with both $\mu$ and $\nu$ appearing four times.
Such operators are 
consistent with all of the lattice symmetries, so they must be present
in the chiral Lagrangian at some order.
The answer is that these operators
\emph{are} present, but they are of $\CO(a^8)$.  
It is clear from the
previous examples that contracting the spin indices to make hypercube
singlets can never contract all of the taste indices, some of
which are independently summed.  
Thus chiral operators such as (\ref{eq:a8}) can only
arise if all of the tensor indices are coupled 
\emph{in the quark-level operator itself}.  
This requires a local eight-quark operator, e.g.
\begin{equation}
\sum_\mu \sum_{\nu \neq \mu}
[\bar Q(\gamma_\mu\otimes \xi_{\mu\nu}) Q ]
[\bar Q(\gamma_\mu\otimes \xi_{\nu\mu}) Q ]
[\bar Q(\gamma_\mu\otimes \xi_{\mu\nu}) Q ]
[\bar Q(\gamma_\mu\otimes \xi_{\nu\mu}) Q ]
\,.
\end{equation}
which enters only at $\CO(a^8)$.
   
A more physical way of looking at this is in terms of gluon exchange.
Taste-changing interactions are due to the exchange of hard gluons
with momentum $k \sim \pi/a$. To obtain a complete correlation of
taste indices requires a complete correlation of the hard gluon momenta.
This can arise directly, for example, with four bilinears connected by four gluons
joining at a four-gluon vertex, as in the above operator. The presence of four hard gluon propagators
gives the factor $(1/k^2)^4\sim a^8$.
In contrast, the $\CO(a^4)$ mesonic operators that we are considering arise
from two separate insertions of $\CO(a^2)$ four-fermion operators tied
together by gluons from the $\CO(a^0)$ Lagrangian.\footnote{%
These lead to a local operator in the chiral Lagrangian because the gluons
are confined to a small region by confinement. In effect, this
replaces four powers of $a$ in the eight fermion operator with
$R^4$, where $R\sim 1/\Lambda_{\rm QCD}$ is the confinement length scale.}
Thus these gluons do not violate taste symmetry.
The correlation between taste indices in the two four-fermion operators can only come \emph{indirectly}, through the Lorentz structure.

We can check the conclusion that operators such as (\ref{eq:a8})
do not appear at $\CO(a^4)$ by examining the counterterms
generated by other $SO(4)$-taste symmetry breaking operators.  
Consider four-pion vertices generated  using the 
$\CO(a^2 p^2)$ operators from the top portion of Table~\ref{tab:FFB}.  
Form a single diagram with only four external pions 
by joining the four pion legs with derivatives to form a loop.  
The loop integral couples the indices on the derivatives and
gives rise to a quartically divergent counterterm with the following
four tastes of external pions: $\mu \nu$, $\nu \mu$, $\mu \sigma$, and
$\sigma \mu$.  This is of the same form as resulting from
the $\CO(a^4)$ operator that we derived above.

The general conclusion we draw from this discussion is that one
\emph{cannot} determine the staggered chiral Lagrangian simply by
testing the invariance of all possible operators under all of the lattice
symmetries.  While this procedure does filter out incorrect operators,
it does not lead to the correct power-counting.  
The only foolproof method is therefore to map each 
quark-level operator directly onto chiral operators.

\subsection{Additional source terms for vector and axial currents}
\label{app:source}

Up to this stage we have not explicitly included source terms for
left- and right-handed (or equivalently vector and axial) currents.
The standard method for doing so
is to promote derivatives to covariant derivatives,
\begin{equation}
\partial_\mu \Sigma \longrightarrow
D_\mu \Sigma = \partial_\mu \Sigma - i \ell_\mu \Sigma + i \Sigma r_\mu  \,,
\qquad
\partial_\mu \Sigma^\dagger \longrightarrow
D_\mu \Sigma^\dagger = \partial_\mu \Sigma^\dagger 
- i r_\mu \Sigma^\dagger + i \Sigma^\dagger \ell_\mu \,,
\end{equation}
and enforce local chiral symmetry~\cite{GassLeut1, GassLeut2}.
Here $\ell_\mu$ and $r_\mu$ are, respectively, matrix sources for the 
left and right-handed currents.
One also needs the field strengths associated with these sources, from
which one can build contact terms at NLO. These do not contribute to
S-matrix elements or decay constants, however, so we do not list them here.

In continuum $\chi$PT, the NLO Lagrangian takes the same form as that without
the sources except for the change of regular to covariant derivatives, 
aside from aforementioned contact terms. Possible terms involving covariant
derivatives acting on $M$ or $M^\dagger$, for example, can be removed using
the equations of motion (EoM). 
This turns out not to be true for S$\chi$PT. There are additional terms
involving covariant derivatives acting on the taste spurions which cannot
be removed, and which play an important role in the breaking of
$SO(4)$ for axial decay constants.
Of course, in addition to these terms one must change regular
to covariant derivatives in all the $\CO(a^2p^2)$ terms given previously.

The extra ``source terms'' are built from a covariant derivative
acting on a taste spurion, a second taste spurion (since they always come
in pairs), and a Lie derivative (to match the Lorentz indices).
Thus they are of $\CO(a^2 p^2)$ and are produced by single insertions of
four-fermion operators. We run through the possibilities in turn.

\subsubsection{Single insertion of $FF(A)$ operators}

We will go through the group theory and operator enumeration rather quickly, as the process is the same as in all previous sections.  We will concentrate instead on eliminating operators through use of the EoM, 
as this is where there is a difference between staggered and continuum $\chi$PT.

\medskip

Operators arising from a single insertion of $[V,A \times F]$ can be built out of the left-handed basis in (\ref{eqn:basis1}) along with two additional elements:
\begin{equation}
	D_\mu F_L\,, \qquad \Sigma D_\mu F_R \Sigma^\dagger \,.
\end{equation}
Because $F_L$ and $F_R$ transform differently than $\Sigma$ and $\Sigma^\dagger$, the covariant derivative acts in the following way on the taste spurions: 
\begin{equation}
	D_\mu F_L = \partial_\mu F_L - i [\ell_\mu, F_L]\,, \qquad D_\mu F_R = \partial_\mu F_L - i [r_\mu, F_R]\,. 
\label{eqn:CovDer1}\end{equation}
Once $F_{L,R}$ are set to fixed taste matrices, the partial derivative pieces will vanish.  However, the commutator terms can still contribute to the left- and right-handed currents, and therefore to PGB decay constants. 

It is clear from (\ref{eqn:CovDer1}) that taste $S$ cannot contribute, since then the commutator vanishes.  Thus we need only consider tastes $P$ and $T$, in which case $F_L$ and $F_R$ are supertraceless, and transform as adjoints under the chiral symmetry group.  Moreover, $D_\mu F_L$ and $D_\mu F_R$ are also supertraceless because of the cyclicity of the supertrace, so they too are adjoints.  Thus operators with one taste spurion, one covariant derivative acting on a taste spurion, and one Lie derivative will correspond to the singlet irreps in the product of three adjoints.  There are two singlets in this product, and therefore four operators.  These operators are given in Table~\ref{tab:source1}.     

\begin{table}\begin{tabular}{llc}  \hline\hline

\multicolumn{2}{c}{\emph{Operator}} & \multicolumn{1}{c}{\emph{Keep?}} \\[0.5mm] \hline

115. & $\Str(D_\mu F_L  F_L \Sigma D_\mu \Sigma^\dagger) + p.c.$ & Yes \\[0.5mm]

116. & $\Str(D_\mu F_L  \Sigma D_\mu \Sigma^\dagger F_L ) + p.c.$ & Sometimes \\[0.5mm]

117. & $\Str(D_\mu F_L \Sigma F_R D_\mu \Sigma^\dagger) + p.c.$ & Yes \\[0.5mm]

118. & $\Str(D_\mu F_L  \Sigma D_\mu \Sigma^\dagger \Sigma F_R \Sigma^\dagger) + p.c.$ 
& Sometimes \\[0.5mm]

\hline\hline

\end{tabular}\caption{The four $a^2$ source operators with one $D_\mu F$ and one Lie derivative.  Note that the covariant derivative only acts on the object directly following. The operators which are to be kept "sometimes" can be removed using the equations of motion if the underlying operator is of type $FF(A)$.}\label{tab:source1}\end{table}

We can remove some of these operators using the EoM.  First recall (qualitatively) how this is done in the case of covariant derivatives acting on mass spurions.  One such operator is $\Str(D_\mu M D_\mu \Sigma^\dagger)$.  Integrating by parts, the covariant derivative can be moved onto the field, resulting in $-\Str(M D_\mu D_\mu \Sigma^\dagger)$.  Then  $D_\mu D_\mu \Sigma^\dagger$ can be replaced with operators involving either
one or no derivatives acting on each of $\Sigma$ or $\Sigma^\dagger$ using the EoM,
all of which are already included in the standard NLO effective Lagrangian.
By trying to follow the same procedure for operators involving taste spurions, we will see where it breaks down.

For example, first consider operator 115.  We must be careful when integrating by parts because there are more than just two terms in this operator -- we cannot simply move $D_\mu$ from one piece to the other and add a minus sign.  Derivatives can be moved around using integration by parts because we are integrating over all spacetime in the action, so total derivative terms become surface terms and can be neglected.  We therefore
 consider a total derivative term which gives rise to operator 115:
\begin{equation}
	D_\mu(F_L  F_L \Sigma D_\mu \Sigma^\dagger) = (D_\mu F_L)  F_L \Sigma D_\mu \Sigma^\dagger + F_L (D_\mu F_L) \Sigma D_\mu \Sigma^\dagger + F_L  F_L D_\mu \Sigma  D_\mu \Sigma^\dagger + F_L  F_L \Sigma (D_\mu D_\mu \Sigma^\dagger)\,.
\end{equation}           
After integrating over all spacetime and taking the supertrace, the l.h.s. vanishes, 
so the four operators on the r.h.s. are all related.  The first is operator 115, while the second is operator 116.  The third operator is from the LO Lagrangian, while the fourth can be related to other standard operators using the EoM.  We learn two lessons from this exercise:  first, that operators 115 and 116 are redundant -- they are related by integration by parts;  second, that we cannot remove both using the EoM---and we choose
to keep operator 115. The difference from the example given above of the covariant
derivative acting on the mass matrix is the presence of  
 \emph{two} taste spurions in each operator.  Integration by parts can never remove the covariant derivative from \emph{both} taste matrices and produce something such as $D_\mu D_\mu \Sigma$, which can be removed using EoM.  The same  holds for operators 117 and 118;  we can remove only one of these two using the EoM, and we choose to keep operator 117.  

\medskip

Operators arising from a single insertion of $[S,P \times F]$ can be built similarly
 out of the left-handed basis in (\ref{eqn:basis2}) plus the elements
\begin{equation}
	D_\mu \tilde{F_L} \Sigma^\dagger \,, \qquad \Sigma D_\mu \tilde{F_R}\,.
\end{equation}
In this case $\tilde{F_L}$ and $\tilde{F_R}$ transform in the same way as $\Sigma$ and $\Sigma^\dagger$, so
\begin{equation}
D_\mu \tilde{F_L} = \partial_\mu \tilde{F_L} - i \ell_\mu \tilde{F_L} + i \tilde{F_L} r_\mu  \,,
\qquad
D_\mu \tilde{F_R} = \partial_\mu \tilde{F_R} 
- i r_\mu \tilde{F_R} + i \tilde{F_R} \ell_\mu \,.
\end{equation} 
Neither $\tilde{F_{L}}\Sigma^\dagger$ nor $D_\mu \tilde{F_{L}}\Sigma^\dagger$ 
(or their parity conjugates) are supertraceless, so they transform as bifundamentals.  Thus all operators with one $\tilde{F}$, one $D_\mu \tilde{F}$, and a Lie derivative come from singlet irreps in the product of two bifundamentals with one adjoint.  There are four such singlets, and eight such operators, given in Table~\ref{tab:source2}.

\begin{table}\begin{tabular}{llc}  \hline\hline

\multicolumn{2}{c}{\emph{Operator}} & \multicolumn{1}{c}{\emph{Keep?}} \\[0.5mm] \hline

119. & $\Str(D_\mu \tilde{F_L} \Sigma^\dagger \tilde{F_L}  D_\mu \Sigma^\dagger) + p.c.$ & Yes \\[0.5mm]

120. & $\Str(D_\mu \tilde{F_L}  D_\mu \Sigma^\dagger \tilde{F_L} \Sigma^\dagger ) + p.c.$ & Sometimes \\[0.5mm]

121. & $\Str(D_\mu \tilde{F_L} \Sigma^\dagger)  \Str(\tilde{F_L}  D_\mu \Sigma^\dagger) + p.c.$ & Yes \\[0.5mm]

122. & $\Str(D_\mu \tilde{F_L}  D_\mu \Sigma^\dagger) \Str( \tilde{F_L} \Sigma^\dagger ) + p.c.$ & Sometimes \\[0.5mm]

123. & $\Str(D_\mu \tilde{F_L}  \tilde{F_R} \Sigma D_\mu \Sigma^\dagger) + p.c.$ & Yes \\[0.5mm]

124. & $\Str(D_\mu \tilde{F_L}  D_\mu \Sigma^\dagger \Sigma \tilde{F_R} ) + p.c.$ & No \\[0.5mm]

125. & $\Str(D_\mu \tilde{F_L} \Sigma^\dagger)  \Str(\Sigma \tilde{F_R} \Sigma D_\mu \Sigma^\dagger) + p.c.$ & Yes \\[0.5mm]

126. & $\Str(D_\mu \tilde{F_L}  D_\mu \Sigma^\dagger) \Str( \Sigma \tilde{F_R} ) + p.c.$ & Sometimes \\[0.5mm]

\hline\hline

\end{tabular}\caption{The eight $a^2$ source operators with one $D_\mu \tilde{F}$ and one Lie derivative.  Notation as in Table~\ref{tab:source1}.}\label{tab:source2}\end{table}

We can now eliminate operators using integration by parts and EoM.  Operator pairs 119 \& 120, 121 \& 122, 123 \& 124, and 124 \& 125 are each related in this way, 
and we choose to keep
only the first operator in each pair.  In fact, operator 124 becomes \emph{equal to} operator 123 once one uses the result that $D_\mu (\tilde{F_L} \tilde{F_R}) = 0$, i.e. that the spurions are momentum-independent and their product has no taste.  Thus we are left with four additional operators. 

\medskip

As in the previous sections, four fermion operators with spin $T$ do not lead to any additional mesonic operators, and simply modify some of the unknown coefficients of the spin $S$ and $P$ operators.   

\medskip

Inserting the appropriate taste matrices is now straightforward.  We simplify expressions using the fact that $\partial_\mu F = 0$.  We also write operators in terms of $\ell_\mu$ and $r_\mu$ because the covariant derivative acts differently on the two types of taste spurions.  The final ``source term" operators are listed in Table~\ref{tab:source_FFA}.  

\subsubsection{Single insertion of $FF(B)$ operators} 

In Section~\ref{app:singleFFB} we discussed in detail how to map onto $\CO(a^2 p^2)$ mesonic operators arising from a single insertion of a four fermion operator in $S_6^{FF(B)}$, so we just summarize the result here:  the \emph{chiral} structures are the same as those of single $FF(A)$ insertions, but the \emph{index} structures are different, breaking $SO(4)$-taste and Lorentz symmetry.  Simply put, each $FF(B)$ operator is identical to the corresponding $FF(A)$ operator, but with the derivative indices contracted with a pair of taste indices.  However, there is one important difference:  because the indices on the covariant derivative and the taste matrices are \emph{correlated}, the number of operators can \emph{no longer be reduced} using the EoM.  Thus there are more $a^2 source$ operators corresponding to a single insertion of $[V_\mu \times T_\mu]$ or $[A_\mu \times T_\mu]$ than those those from a single insertion of $[V,A \times T]$.\footnote{Note that the operators resulting from generic forms 115 and 116 are
the same up to a sign, so the doubling of operators does not apply in this case.}  There are, however, the \emph{same} number of $a^2 source$ operators from $[T_\mu \times V_\mu, A_\mu]$ as from $[S \times V,A]$ because the doubling from the inability to use the EoM is counteracted by a halving because there are only $L-R$ cross-terms.  Table~\ref{tab:source_FFB} lists all of the resulting $FF(B)$ ``source term'' operators.   

Note that $[V_\mu \times T_\mu]$ only generates single supertrace operators, so there are no ``source term'' hairpins with tensor taste.  This has interesting consequences for $SO(4)$-taste symmetry breaking in PGB decay constants.  

\section{$\CO(a^2 p^2)$, $\CO(a^2 m)$ and $\CO(a^4)$ operators in the S$\chi$$\CL$}
\label{app:tables}

\begin{table}[b]\begin{tabular}{l|ccc}  \hline\hline

\multicolumn{1}{l|}{\emph{Generic}} & 
\multicolumn{3}{c}{\emph{Specific $\CO(a^2 p^2)$ Operator}} \\[0.5mm] 

\emph{Operator}& $[V,A \times P]$ 
&& $[V,A \times T]$  \\ \hline 

1 + 3 & 
$\Str(\partial_\mu \Sigma^\dagger \partial_\mu \Sigma \Sigma^\dagger
\xi_5 \Sigma \xi_5) + p.c.$
&& $ \Str(\partial_\mu \Sigma^\dagger \partial_\mu \Sigma
\Sigma^\dagger \xi_{\nu\rho} \Sigma \xi_{\rho\nu}) + p.c.$
\\[0.5mm]

2 & 
$\Str(\partial_\mu \Sigma^\dagger \xi_5 \partial_\mu \Sigma \xi_5)$
&& $\Str(\partial_\mu \Sigma^\dagger \xi_{\nu\rho} \partial_\mu
\Sigma \xi_{\rho\nu})$ 
\\[0.5mm]

6 & 
$\Str(\partial_\mu \Sigma^\dagger \partial_\mu \Sigma)\Str(\xi_5
\Sigma \xi_5 \Sigma^\dagger)$ 
&& $\Str(\partial_\mu \Sigma^\dagger \partial_\mu
\Sigma)\Str(\xi_{\nu\rho} \Sigma \xi_{\rho\nu} \Sigma^\dagger)$
\\[0.5mm]

7 & $\Str(\Sigma \partial_\mu \Sigma^\dagger
\xi_5)\Str(\Sigma^\dagger \partial_\mu\Sigma \xi_5)$ 
&& $\Str(\Sigma \partial_\mu \Sigma^\dagger
\xi_{\nu\rho})\Str(\Sigma^\dagger \partial_\mu \Sigma \xi_{\rho\nu})$
\\[0.5mm]

10 &
$\Str(\Sigma\partial_\mu \Sigma^\dagger \xi_5 \Sigma \partial_\mu
\Sigma^\dagger \xi_5) + p.c.$ 
&& $\Str(\Sigma \partial_\mu \Sigma^\dagger \xi_{\nu\rho} \Sigma
\partial_\mu \Sigma^\dagger \xi_{\rho\nu}) + p.c.$
\\[0.5mm]

13 &
$\Str(\Sigma \partial_\mu \Sigma^\dagger \xi_5)\Str(\Sigma \partial_\mu
\Sigma^\dagger \xi_5) + p.c. $
&& $ \Str(\Sigma \partial_\mu \Sigma^\dagger
\xi_{\nu\rho})\Str(\Sigma \partial_\mu \Sigma^\dagger \xi_{\rho\nu}) + p.c.$
\\[0.5mm]

\hline\hline

\end{tabular}

\vspace{3mm}

\begin{tabular}{l|ccc}  \hline\hline

\multicolumn{1}{l|}{\emph{Generic}} & 
\multicolumn{3}{c}{\emph{Specific $\CO(a^2 m)$ Operator}} \\[0.5mm] 

\emph{Operator}  & $[V,A \times P]$  && $[V,A \times T]$ \\ \hline 

15 + 22 &
$\Str(\xi_5 \Sigma \xi_5 M^\dagger) + p.c.$
&&
$\Str(\xi_{\mu\nu} \Sigma \xi_{\nu\mu} M^\dagger) + p.c.$
\\[0.5mm]

16 + 21 & 
$\Str(\xi_5 \Sigma M^\dagger \Sigma \xi_5 \Sigma^\dagger) + p.c.$ 
&& $\Str(\xi_{\mu \nu} \Sigma M^\dagger \Sigma \xi_{\nu \mu}
\Sigma^\dagger) + p.c.$  
\\[0.5mm]

17 + 23  & 
$\Str(\xi_5 \Sigma \xi_5 \Sigma^\dagger)\Str(\Sigma M^\dagger) + p.c.$
&& $\Str(\xi_{\mu \nu} \Sigma \xi_{\nu \mu} \Sigma^\dagger)
\Str(\Sigma M^\dagger) + p.c.$ 
\\[0.5mm]

\hline\hline

\end{tabular}\caption{
Operators in the staggered chiral Lagrangian arising
from a single insertion of an $S_6^{FF(A)}$ operator
with spin $V$ or $A$.
Implicit sums over repeated indices follow the
summation convention defined in (\ref{eqn:LSnotation}),
and $p.c.$ indicates parity conjugate.  
Each chiral operator has an independent undetermined coefficient
of $\CO(a^2)$. Partial derivatives act by convention only on the
object immediately to their right.
To include left- and right-handed sources, derivatives
should be replaced with covariant derivatives.
$M$ and $M^\dagger$ serve as (pseudo)scalar sources; if such sources are
absent then $M=M^\dagger=\CM$.}
\label{tab:FFA_VA}\end{table}

\begin{table}[t]\begin{tabular}{l|ccc}  \hline\hline

\multicolumn{1}{l|}{\emph{Generic}} & 
\multicolumn{3}{c}{\emph{Specific $\CO(a^2 p^2)$ Operator}} \\[0.5mm] 

\emph{Operator} & $[S,P \times V]$ &  & $[S,P \times A]$ \\ \hline 

36 & $ \Str(\Sigma \partial_\mu \Sigma^\dagger \xi_\nu
\Sigma^\dagger \partial_\mu \Sigma \xi_\nu)$ & & 
$ \Str(\Sigma\partial_\mu \Sigma^\dagger \xi_{\nu 5}
\Sigma^\dagger \partial_\mu \Sigma \xi_{5 \nu})$ \\[0.5mm]

38 + 39 & $ \Str(\partial_\mu \Sigma^\dagger \partial_\mu
\Sigma \Sigma^\dagger \xi_\nu ) \Str(\Sigma \xi_\nu ) + p.c. $ & & 
$ \Str(\partial_\mu \Sigma^\dagger
\partial_\mu \Sigma \Sigma^\dagger \xi_{\nu 5}) \Str(\Sigma \xi_{5 \nu}) +p.c.$ 
\\[0.5mm]

41 & $ \Str(\partial_\mu \Sigma^\dagger \xi_\nu )\Str(\partial_\mu
\Sigma \xi_\nu )$ & & $ \Str(\partial_\mu \Sigma^\dagger \xi_{\nu
5})\Str(\partial_\mu \Sigma \xi_{5
\nu})$ \\[0.5mm]

42 & $ \Str(\partial_\mu \Sigma^\dagger \partial_\mu \Sigma)
\Str(\xi_\nu \Sigma ^\dagger) \Str(\Sigma \xi_\nu )$ & & 
$\Str(\partial_\mu \Sigma^\dagger \partial_\mu
\Sigma) \Str(\xi_{\nu 5} \Sigma ^\dagger) \Str(\Sigma \xi_{5 \nu})$ \\[0.5mm]

43 & $\Str(\partial_\mu \Sigma \partial_\mu \Sigma^\dagger \xi_\nu
\Sigma ^\dagger \xi_\nu \Sigma ^\dagger)+p.c.$
 &  & $\Str(\partial_\mu \Sigma \partial_\mu \Sigma^\dagger \xi_{\nu 5} 
\Sigma ^\dagger \xi_{5 \nu} \Sigma ^\dagger)+p.c.$
\\[0.5mm]

44 & $\Str(\partial_\mu \Sigma^\dagger \xi_\nu \partial_\mu
\Sigma^\dagger \xi_\nu )+p.c.$ &&
$\Str(\partial_\mu\Sigma^\dagger \xi_{\nu 5}  
\partial_\mu \Sigma^\dagger \xi_{5 \nu})+p.c.$
 \\[0.5mm]

45 & $\Str(\partial_\mu \Sigma^\dagger \partial_\mu \Sigma)
\Str(\xi_\nu \Sigma ^\dagger \xi_\nu \Sigma ^\dagger)+p.c.$ & &
$\Str(\partial_\mu \Sigma^\dagger\partial_\mu \Sigma) 
\Str(\xi_{\nu 5} \Sigma ^\dagger \xi_{5 \nu} \Sigma ^\dagger) + p.c.$
\\[0.5mm]

46 & $\Str(\partial_\mu \Sigma \partial_\mu \Sigma^\dagger
\xi_\nu \Sigma ^\dagger) \Str(\xi_\nu \Sigma ^\dagger) + p.c.$ & &
$\Str(\partial_\mu \Sigma \partial_\mu \Sigma^\dagger \xi_{\nu 5} 
\Sigma ^\dagger) \Str(\xi_{5 \nu} \Sigma ^\dagger) + p.c.$
\\[0.5mm]

47 & $\Str(\partial_\mu \Sigma^\dagger \xi_\nu )^2+p.c.$ & & 
$\Str(\partial_\mu \Sigma^\dagger \xi_{\nu5})
\Str(\partial_\mu \Sigma^\dagger \xi_{5 \nu})+p.c.$
\\[0.5mm]

48 & $\Str(\partial_\mu \Sigma^\dagger \partial_\mu \Sigma)
\Str(\xi_\nu \Sigma ^\dagger)^2+p.c.$ &&
$\Str(\partial_\mu \Sigma^\dagger \partial_\mu \Sigma) \Str(\xi_{\nu 5}
\Sigma ^\dagger)\Str(\xi_{5 \nu} \Sigma ^\dagger) + p.c.$ 
\\[0.5mm]

\hline\hline
\end{tabular}

\vspace{3mm}

\begin{tabular}{l|ccc}  \hline\hline

\multicolumn{1}{l|}{\emph{Generic}} & 
\multicolumn{3}{c}{\emph{Specific $\CO(a^2 m)$ Operator}} \\[0.5mm] 

\emph{Operator} & $[S,P \times V]$ &  & $[S,P \times A]$ \\ \hline 

52 + 59 & $\Str(\xi_\mu M^\dagger) \Str(\Sigma \xi_\mu)$ && 
$\Str(\xi_{\mu5} M^\dagger) \Str(\Sigma \xi_{5\mu})$ 
\\[0.5mm]

53 + 58 & $\Str(\xi_\mu \Sigma^\dagger)\Str(\Sigma \xi_\mu \Sigma M^\dagger) +p.c.$ &&
$\Str(\xi_{\mu 5} \Sigma^\dagger)\Str(\Sigma \xi_{5\mu} \Sigma M^\dagger) +p.c.$ 
\\[0.5mm]

54 + 60 & 
$\Str(\xi_\mu \Sigma^\dagger) \Str(\Sigma \xi_\mu)\Str(\Sigma M^\dagger) +p.c.$ & & 
$\Str(\xi_{\mu 5} \Sigma^\dagger)\Str(\Sigma \xi_{5 \mu}) \Str(\Sigma M^\dagger) + p.c.$
  \\[0.5mm]

61 & $\Str(\xi_\mu \Sigma^\dagger \xi_\mu M^\dagger) + p.c.$ &&
$\Str(\xi_{\mu 5} \Sigma^\dagger\xi_{5 \mu} M^\dagger) + p.c.$
\\[0.5mm]

62 & $\Str(\xi_\mu \Sigma^\dagger \xi_\mu \Sigma^\dagger) \Str(\Sigma M^\dagger) + p.c.$ &&
$\Str(\xi_{\mu 5} \Sigma^\dagger\xi_{5 \mu} \Sigma^\dagger) \Str(\Sigma M^\dagger) +p.c.$
\\[0.5mm]

63 & $\Str(\xi_\mu \Sigma^\dagger) \Str(\xi_\mu M^\dagger) + p.c.$ &&
$\Str(\xi_{\mu 5} \Sigma^\dagger) \Str(\xi_{5\mu} M^\dagger) + p.c.$
\\[0.5mm]

64 & $\Str(\xi_\mu \Sigma^\dagger) 
\Str(\xi_\mu \Sigma^\dagger)\Str(\Sigma M^\dagger) +p.c. $ 
& & $\Str(\xi_{\mu 5} \Sigma^\dagger)
\Str(\xi_{5 \mu} \Sigma^\dagger) \Str(\Sigma M^\dagger) +p.c.$ 
\\[0.5mm]

65 & $\Str(\xi_\mu \Sigma^\dagger \xi_\mu \Sigma^\dagger M
\Sigma^\dagger) + p.c.$&&
$\Str(\xi_{\mu 5} \Sigma^\dagger \xi_{5 \mu}
\Sigma^\dagger M \Sigma^\dagger) + p.c.$
\\[0.5mm]

66 & $\Str(\xi_\mu \Sigma^\dagger \xi_\mu \Sigma^\dagger) \Str(M
\Sigma^\dagger) + p.c.$ &&
$\Str(\xi_{\mu 5} \Sigma^\dagger
\xi_{5 \mu} \Sigma^\dagger) \Str(M \Sigma^\dagger) + p.c.$ 
\\[0.5mm]

67 & $\Str(\xi_\mu \Sigma^\dagger) \Str(\xi_\mu \Sigma^\dagger M
\Sigma^\dagger) + p.c.$ &&
$\Str(\xi_{\mu 5}\Sigma^\dagger) \Str(\xi_{5 \mu} 
\Sigma^\dagger M \Sigma^\dagger) +  p.c.$
\\[0.5mm]

68 & $\Str(\xi_\mu \Sigma^\dagger) \Str(\xi_\mu \Sigma^\dagger)
\Str(M \Sigma^\dagger) + p.c.$ & & 
$\Str(\xi_{\mu 5} \Sigma^\dagger)\Str(\xi_{5 \mu} \Sigma^\dagger)
\Str(M \Sigma^\dagger)+p.c.$ 
\\[0.5mm]

\hline\hline

\end{tabular} 
\caption{Operators in the staggered chiral Lagrangian arising
from a single insertion of an $S_6^{FF(A)}$ operator
with spin $S$ or $P$.
Notation as in table~\ref{tab:FFA_VA}.
}
\label{tab:FFA_SPa2} \end{table}

\begin{table}\begin{tabular}{l|c}  \hline\hline

\multicolumn{1}{l|}{\emph{Generic}} & 
\multicolumn{1}{c}{\emph{Specific $\CO(a^2 p^2)$ Operator}} \\[0.5mm] 

\emph{Operator} & $[V_\mu \times T_\mu]$ and $[A_\mu \times T_\mu]$  \\ \hline 

1 + 3 & $\Str(\partial_\mu \Sigma^\dagger \partial_\mu \Sigma
\Sigma^\dagger \xi_{\mu\nu} \Sigma \xi_{\nu\mu}) + p.c.$
\\[0.5mm]

2 & $\Str(\partial_\mu \Sigma^\dagger \xi_{\mu\nu} \partial_\mu \Sigma
\xi_{\nu\mu})$
\\[0.5mm]

6 & $\Str(\partial_\mu \Sigma^\dagger \partial_\mu
\Sigma)\Str(\xi_{\mu\nu} \Sigma \xi_{\nu\mu} \Sigma^\dagger)$
\\[0.5mm]

7 & $\Str(\Sigma \partial_\mu \Sigma^\dagger
\xi_{\mu\nu})\Str(\Sigma^\dagger \partial_\mu \Sigma \xi_{\nu\mu})$
\\[0.5mm]

10 & $\Str(\Sigma \partial_\mu \Sigma^\dagger \xi_{\mu\nu}
\Sigma \partial_\mu \Sigma^\dagger \xi_{\nu\mu})+p.c.$
\\[0.5mm]

13 & $\Str(\Sigma \partial_\mu \Sigma^\dagger
\xi_{\mu\nu})\Str(\Sigma \partial_\mu \Sigma^\dagger
\xi_{\nu\mu}) + p.c.$
\\[0.5mm]

\hline\hline
\end{tabular}

\vspace{3mm}

\begin{tabular}{l|ccc}  \hline\hline

\multicolumn{1}{l|}{\emph{Generic}} & 
\multicolumn{3}{c}{\emph{Specific $\CO(a^2 p^2)$ Operator}} \\[0.5mm] 

\emph{Operator} & $[T_\mu \times V_\mu]$ &  & $[T_\mu \times A_\mu]$ \\ \hline 

36 & $ \Str(\Sigma \partial_\mu \Sigma^\dagger \xi_\mu
\Sigma^\dagger \partial_\mu \Sigma \xi_\mu)$ & & 
$ \Str(\Sigma\partial_\mu \Sigma^\dagger \xi_{\mu 5}
\Sigma^\dagger \partial_\mu \Sigma \xi_{5 \mu})$ \\[0.5mm]

38 + 39 & $ \Str(\partial_\mu \Sigma^\dagger \partial_\mu
\Sigma \Sigma^\dagger \xi_\mu ) \Str(\Sigma \xi_\mu ) + p.c. $ & & 
$ \Str(\partial_\mu \Sigma^\dagger
\partial_\mu \Sigma \Sigma^\dagger \xi_{\mu 5}) \Str(\Sigma \xi_{5 \mu}) +p.c.$ 
\\[0.5mm]

41 & $ \Str(\partial_\mu \Sigma^\dagger \xi_\mu )\Str(\partial_\mu
\Sigma \xi_\mu )$ & & $ \Str(\partial_\mu \Sigma^\dagger \xi_{\mu
5})\Str(\partial_\mu \Sigma \xi_{5
\mu})$ \\[0.5mm]

42 & $ \Str(\partial_\mu \Sigma^\dagger \partial_\mu \Sigma)
\Str(\xi_\mu \Sigma ^\dagger) \Str(\Sigma \xi_\mu )$ & & 
$\Str(\partial_\mu \Sigma^\dagger \partial_\mu
\Sigma) \Str(\xi_{\mu 5} \Sigma ^\dagger) \Str(\Sigma \xi_{5 \mu})$ \\[0.5mm]

\hline\hline
\end{tabular}

\caption{
Rotationally non-invariant operators in the staggered chiral Lagrangian arising
from a single insertion of $S_6^{FF(B)}$ operators.
There is an implicit summation over both $\mu$ and $\nu$ with the
constraint $\nu\ne\mu$.}
\label{tab:FFB} \end{table}

%BREAK!!!

\begin{table}\begin{tabular}{l|ccc}  \hline\hline

\multicolumn{1}{l|}{\emph{Generic Op.}} & 
\multicolumn{1}{c}{$[V,A \times P]$ with $[V,A \times P]$} \\ \hline 

69 + 74 & $\Str(\xi_5 \Sigma \xi_5 \Sigma^\dagger \xi_5 \Sigma \xi_5 \Sigma^\dagger)$ 
\\[0.5mm]

75 & $\Str(\xi_5 \Sigma \xi_5 \Sigma^\dagger)\Str(\xi_5 \Sigma \xi_5 \Sigma^\dagger)$ 
 \\[0.5mm]

\hline\hline

\end{tabular}

\vspace{3mm}

\begin{tabular}{l|ccc}  \hline\hline

\multicolumn{1}{l|}{\emph{Generic Op.}} & 
\multicolumn{1}{c}{$[V,A \times P]$ with $[V,A \times T]$} \\ \hline  

69 + 74 & $\Str(\xi_5 \Sigma \xi_5 \Sigma^\dagger \xi_{\mu \nu} 
		\Sigma \xi_{\nu \mu} \Sigma^\dagger) + p.c.$ \\[0.5mm]

75 & $\Str(\xi_5 \Sigma \xi_5 \Sigma^\dagger)\Str(\xi_{\mu \nu} 
			\Sigma \xi_{\nu \mu} \Sigma^\dagger)$  \\[0.5mm]

77 & $\Str(\xi_5 \Sigma \xi_{\mu \nu} \Sigma^\dagger)\Str(\xi_{\nu \mu} 
			\Sigma \xi_5 \Sigma^\dagger )$  \\[0.5mm]

79 & $\Str(\xi_5 \Sigma \xi_{\mu \nu} \Sigma^\dagger \xi_5 
			\Sigma \xi_{\nu \mu} \Sigma^\dagger) + p.c.$  \\[0.5mm]

81 & $\Str(\xi_5 \Sigma \xi_{\mu \nu} \Sigma^\dagger)\Str(\xi_5 
			\Sigma \xi_{\nu \mu} \Sigma^\dagger) + p.c.$  \\[0.5mm]

\hline\hline

\end{tabular}

\vspace{3mm}

\begin{tabular}{l|ccc}  \hline\hline

\multicolumn{1}{l|}{\emph{Generic Op.}} & 
\multicolumn{1}{c}{$[V,A \times T]$ with $[V,A \times T]$} \\ \hline

69 + 74 & $\Str(\xi_{\mu \nu} \Sigma \xi_{\nu \mu} \Sigma^\dagger \xi_{\rho \sigma} 
\Sigma \xi_{\sigma \rho} \Sigma^\dagger) + p.c.$ \\[0.5mm]

75 & $\Str(\xi_{\mu \nu} \Sigma \xi_{\nu \mu} \Sigma^\dagger)\Str(\xi_{\rho \sigma} 
\Sigma \xi_{\sigma \rho} \Sigma^\dagger)$  \\[0.5mm]

77 & $\Str(\xi_{\mu \nu} \Sigma \xi_{\rho \sigma} \Sigma^\dagger)\Str(\xi_{\sigma \rho} 
\Sigma \xi_{\nu \mu} \Sigma^\dagger )$  \\[0.5mm]

79 & $\Str(\xi_{\mu \nu} \Sigma \xi_{\rho \sigma} \Sigma^\dagger \xi_{\nu \mu} 
\Sigma \xi_{\sigma \rho} \Sigma^\dagger) + p.c.$  \\[0.5mm]

81 & $\Str(\xi_{\mu \nu} \Sigma \xi_{\rho \sigma} \Sigma^\dagger)\Str(\xi_{\nu \mu} 
\Sigma \xi_{\sigma \rho} \Sigma^\dagger) + p.c.$  \\[0.5mm]

\hline\hline

\end{tabular}
\caption{$\CO(a^4)$ operators in the staggered chiral Lagrangian arising
from two insertions of an $S_6^{FF(A)}$ operator
with spin $V$ or $A$.
Notation as in table~\ref{tab:FFA_VA}.
}
\label{tab:a4_VA_VA}\end{table}

\begin{table}\begin{tabular}{l|cc}  \hline\hline

{\emph{Generic Op.}} & 
$[S,P \times V]$ with $[S,P \times V]$ & $[S,P \times A]$ with $[S,P \times A]$ \\ \hline

82 & $\Str(\xi_\mu \Sigma^\dagger \xi_\nu \Sigma^\dagger)
\Str(\Sigma \xi_\mu \Sigma \xi_\nu)$ & $\;\;
\Str(\xi_{\mu 5} \Sigma^\dagger \xi_{\nu 5} \Sigma^\dagger)
\Str(\Sigma \xi_{5 \mu} \Sigma \xi_{5 \nu})$  \\[0.5mm]

83 & $\Str(\xi_\mu \Sigma^\dagger \xi_\nu \Sigma^\dagger)
\Str(\Sigma \xi_\mu)\Str(\Sigma \xi_\nu) + p.c.$ & $\;\;
\Str(\xi_{\mu 5} \Sigma^\dagger \xi_{\nu 5} \Sigma^\dagger)
\Str(\Sigma \xi_{5 \mu})\Str(\Sigma \xi_{5 \nu}) + p.c.$  \\[0.5mm]

84 & $\Str(\xi_\mu \Sigma ^\dagger) \Str(\xi_\nu \Sigma ^\dagger) 
\Str(\Sigma \xi_\mu) \Str(\Sigma \xi_\nu)$ & $\;\;
\Str(\xi_{\mu 5} \Sigma ^\dagger) \Str(\xi_{\nu 5} \Sigma ^\dagger) 
\Str(\Sigma \xi_{5 \mu}) \Str(\Sigma \xi_{5 \nu})$  \\[0.5mm]

85 & $\Str(\xi_\mu \Sigma ^\dagger \xi_\mu \Sigma ^\dagger \xi_\nu \Sigma ^\dagger) 
\Str(\Sigma \xi_\nu) + p.c.$ & $\;\;
\Str(\xi_{\mu 5} \Sigma ^\dagger \xi_{5 \mu} \Sigma ^\dagger \xi_{\nu 5} \Sigma ^\dagger) 
\Str(\Sigma \xi_{5 \nu}) + p.c.$ \\[0.5mm]

87 & $\Str(\xi_\mu \Sigma ^\dagger \xi_\mu \Sigma ^\dagger) 
\Str(\xi_\nu \Sigma ^\dagger) \Str(\Sigma \xi_\nu) + p.c.$ & $\;\;
\Str(\xi_{\mu 5} \Sigma ^\dagger \xi_{5 \mu} \Sigma ^\dagger) 
\Str(\xi_{\nu 5} \Sigma ^\dagger) \Str(\Sigma \xi_{5 \nu}) + p.c.$ \\[0.5mm]

89 & $\Str(\xi_\mu \Sigma ^\dagger \xi_\nu \Sigma ^\dagger) 
\Str(\xi_\mu \Sigma ^\dagger) \Str(\Sigma \xi_\nu) + p.c.$ & $\;\;
\Str(\xi_{\mu 5} \Sigma ^\dagger \xi_{\nu 5} \Sigma ^\dagger) 
\Str(\xi_{5 \mu} \Sigma ^\dagger) \Str(\Sigma \xi_{5 \nu}) + p.c.$ \\[0.5mm]

91 & $\Str(\xi_\mu \Sigma ^\dagger) \Str(\xi_\mu \Sigma ^\dagger) 
\Str(\xi_\nu \Sigma ^\dagger) \Str(\Sigma \xi_\nu) + p.c.$ & $\;\;
\Str(\xi_{\mu 5} \Sigma ^\dagger) \Str(\xi_{5 \mu} \Sigma ^\dagger) 
\Str(\xi_{\nu 5} \Sigma ^\dagger) \Str(\Sigma \xi_{5 \nu}) + p.c.$ \\[0.5mm]

93 & $\Str(\xi_\mu \Sigma ^\dagger \xi_\mu \Sigma ^\dagger \xi_\nu 
\Sigma ^\dagger \xi_\nu \Sigma ^\dagger) + p.c.$ & $\;\;
\Str(\xi_{\mu 5} \Sigma ^\dagger \xi_{5 \mu} \Sigma ^\dagger \xi_{\nu 5} 
\Sigma ^\dagger \xi_{5 \nu} \Sigma ^\dagger) + p.c.$ \\[0.5mm]

94 & $\Str(\xi_\mu \Sigma ^\dagger \xi_\nu \Sigma ^\dagger \xi_\mu 
\Sigma ^\dagger \xi_\nu \Sigma ^\dagger) + p.c.$  & $\;\;\Str(\xi_{\mu 5} 
\Sigma ^\dagger \xi_{\nu 5} \Sigma ^\dagger \xi_{5 \mu} 
\Sigma ^\dagger \xi_{5 \nu} \Sigma ^\dagger) + p.c.$ \\[0.5mm]

95 & $\Str(\xi_\mu \Sigma ^\dagger \xi_\mu \Sigma ^\dagger \xi_\nu 
\Sigma ^\dagger) \Str(\xi_\nu \Sigma ^\dagger)+ p.c.$ & $\;\;
\Str(\xi_{\mu 5} \Sigma ^\dagger \xi_{5 \mu} \Sigma ^\dagger \xi_{\nu 5} 
\Sigma ^\dagger) \Str(\xi_{5 \nu} \Sigma ^\dagger)+ p.c.$ \\[0.5mm]

97 & $\Str(\xi_\mu \Sigma ^\dagger \xi_\mu \Sigma ^\dagger)
\Str(\xi_\nu \Sigma ^\dagger \xi_\nu \Sigma ^\dagger)+ p.c.$ 
& $\;\;\Str(\xi_{\mu 5} \Sigma ^\dagger \xi_{5 \mu} \Sigma ^\dagger)\Str(\xi_{\nu 5} \Sigma ^\dagger \xi_{5 \nu} \Sigma ^\dagger)+ p.c.$ \\[0.5mm]

98 & $\Str(\xi_\mu \Sigma ^\dagger \xi_\nu \Sigma ^\dagger)
\Str(\xi_\mu \Sigma ^\dagger \xi_\nu \Sigma ^\dagger)+ p.c.$ 
& $\;\;\Str(\xi_{\mu 5} \Sigma ^\dagger \xi_{\nu 5} \Sigma ^\dagger)
\Str(\xi_{5 \mu} \Sigma ^\dagger \xi_{5 \nu} \Sigma ^\dagger)+ p.c.$ \\[0.5mm]

99 & $\Str(\xi_\mu \Sigma ^\dagger \xi_\mu \Sigma ^\dagger)\Str(\xi_\nu 
\Sigma ^\dagger)\Str(\xi_\nu \Sigma ^\dagger) + p.c.$ 
& $\;\;\Str(\xi_{\mu 5} \Sigma ^\dagger \xi_{5 \mu} 
\Sigma ^\dagger)\Str(\xi_{\nu 5} \Sigma ^\dagger)
\Str(\xi_{5 \nu} \Sigma ^\dagger) + p.c.$ \\[0.5mm]

100 & $\Str(\xi_\mu \Sigma ^\dagger \xi_\nu \Sigma ^\dagger)
\Str(\xi_\mu \Sigma ^\dagger)\Str(\xi_\nu \Sigma ^\dagger) + p.c.$  
& $\;\;\Str(\xi_{\mu 5} \Sigma ^\dagger \xi_{\nu 5} \Sigma ^\dagger)
\Str(\xi_{5 \mu} \Sigma ^\dagger)\Str(\xi_{5 \nu} \Sigma ^\dagger) + p.c.$ \\[0.5mm]

102 & $\Str(\xi_\mu \Sigma ^\dagger) \Str(\xi_\mu \Sigma ^\dagger) 
\Str(\xi_\nu \Sigma^\dagger) \Str(\xi_\nu \Sigma ^\dagger)+ p.c.$  
& $\;\;\Str(\xi_{\mu 5} \Sigma ^\dagger) \Str(\xi_{5 \mu} \Sigma ^\dagger) 
\Str(\xi_{\nu 5} \Sigma^\dagger) \Str(\xi_{5 \nu} \Sigma ^\dagger)+ p.c.$ \\[0.5mm]

103 & $\Str(\xi_\mu \Sigma ^\dagger \xi_\mu \Sigma ^\dagger) 
\Str(\Sigma \xi_\nu \Sigma \xi_\nu)$ 
& $\;\;\Str(\xi_{\mu 5} \Sigma ^\dagger \xi_{5 \mu} \Sigma ^\dagger) 
\Str(\Sigma \xi_{\nu 5} \Sigma \xi_{5 \nu})$  \\[0.5mm]

104 & $\Str(\xi_\mu \Sigma ^\dagger \xi_\mu \Sigma ^\dagger) 
\Str(\Sigma \xi_\nu) \Str(\Sigma \xi_\nu) + p.c.$  
& $\;\;\Str(\xi_{\mu 5} \Sigma ^\dagger \xi_{5 \mu} \Sigma ^\dagger) 
\Str(\Sigma \xi_{\nu 5}) \Str(\Sigma \xi_{5 \nu}) + p.c.$ \\[0.5mm]

106 & $\Str(\xi_\mu \Sigma ^\dagger) \Str(\xi_\mu \Sigma ^\dagger) 
\Str(\Sigma \xi_\nu) \Str(\Sigma \xi_\nu)$  
& $\;\;\Str(\xi_{\mu 5} \Sigma ^\dagger) \Str(\xi_{5 \mu} \Sigma ^\dagger) 
\Str(\Sigma \xi_{\nu 5}) \Str(\Sigma \xi_{5 \nu})$ \\[0.5mm]

\hline\hline

\end{tabular}

\vspace{3mm}

\begin{tabular}{l|ccc}  \hline\hline

\multicolumn{1}{l|}{\emph{Generic Op.}} 
& \multicolumn{1}{c}{$[S,P \times V]$ with $[S,P \times A]$} \\\hline

82 & $\Str(\xi_\mu \Sigma^\dagger \xi_{\nu 5} \Sigma^\dagger )
\Str(\Sigma \xi_\mu \Sigma \xi_{5 \nu})$  \\[0.5mm]

83 & $\Str(\xi_\mu \Sigma^\dagger \xi_{\nu 5} \Sigma^\dagger )
\Str(\Sigma \xi_\mu)\Str(\Sigma \xi_{5 \nu}) + p.c.$  \\[0.5mm]

84 & $\Str(\xi_\mu \Sigma ^\dagger) \Str(\xi_{\nu 5} \Sigma ^\dagger) 
\Str(\Sigma \xi_\mu) \Str(\Sigma \xi_{5 \nu})$  \\[0.5mm]

85 & $\Str(\xi_\mu \Sigma ^\dagger \xi_\mu \Sigma ^\dagger \xi_{\nu 5} \Sigma ^\dagger) 
\Str(\Sigma \xi_{5 \nu}) + p.c.$  \\[0.5mm]

86 & $\Str(\xi_{\nu 5} \Sigma ^\dagger \xi_{5 \nu} \Sigma ^\dagger \xi_\mu \Sigma ^\dagger) 
\Str(\Sigma \xi_\mu) + p.c.$  \\[0.5mm]

87 & $\Str(\xi_\mu \Sigma ^\dagger \xi_\mu \Sigma ^\dagger) 
\Str(\xi_{\nu 5} \Sigma ^\dagger)  \Str(\Sigma \xi_{5 \nu}) + p.c.$  \\[0.5mm]

88 & $\Str(\xi_{\nu 5} \Sigma ^\dagger \xi_{5 \nu} \Sigma ^\dagger) 
\Str(\xi_\mu \Sigma ^\dagger) \Str(\Sigma \xi_\mu) + p.c.$  \\[0.5mm]

89 & $\Str(\xi_\mu \Sigma ^\dagger \xi_{\nu 5} \Sigma ^\dagger) 
\Str(\xi_\mu \Sigma ^\dagger) \Str(\Sigma \xi_{5 \nu}) + p.c.$  \\[0.5mm]

90 & $\Str(\xi_{\nu 5} \Sigma ^\dagger \xi_\mu \Sigma ^\dagger) 
\Str(\xi_{5 \nu} \Sigma ^\dagger) \Str(\Sigma \xi_\mu) + p.c.$  \\[0.5mm]

91 & $\Str(\xi_\mu \Sigma ^\dagger) \Str(\xi_\mu \Sigma ^\dagger) 
\Str(\xi_{\nu 5} \Sigma ^\dagger) \Str(\Sigma \xi_{5 \nu}) + p.c.$  \\[0.5mm]

92 & $\Str(\xi_{\nu 5} \Sigma ^\dagger) \Str(\xi_{5 \nu} \Sigma ^\dagger) 
\Str(\xi_\mu \Sigma ^\dagger) \Str(\Sigma \xi_\mu) + p.c.$  \\[0.5mm]

93 & $\Str(\xi_\mu \Sigma ^\dagger \xi_\mu \Sigma ^\dagger \xi_{\nu 5} 
\Sigma ^\dagger \xi_{5 \nu} \Sigma ^\dagger) + p.c.$  \\[0.5mm]

94 & $\Str(\xi_\mu \Sigma ^\dagger \xi_{\nu 5} \Sigma ^\dagger \xi_\mu 
\Sigma ^\dagger \xi_{5 \nu} \Sigma ^\dagger) + p.c.$  \\[0.5mm]

95 & $\Str(\xi_\mu \Sigma ^\dagger \xi_\mu \Sigma ^\dagger \xi_{\nu 5} \Sigma ^\dagger) 
\Str(\xi_{5 \nu} \Sigma ^\dagger)+ p.c.$  \\[0.5mm]

96 & $\Str(\xi_{\nu 5} \Sigma ^\dagger \xi_{5 \nu} \Sigma ^\dagger \xi_\mu \Sigma ^\dagger) 
\Str(\xi_\mu \Sigma ^\dagger)+ p.c.$  \\[0.5mm]

97 & $\Str(\xi_\mu \Sigma ^\dagger \xi_\mu \Sigma ^\dagger)
\Str(\xi_{\nu 5} \Sigma ^\dagger \xi_{5 \nu} \Sigma ^\dagger)+ p.c.$  \\[0.5mm]

98 & $\Str(\xi_\mu \Sigma ^\dagger \xi_{\nu 5} \Sigma ^\dagger)
\Str(\xi_\mu \Sigma ^\dagger \xi_{5 \nu} \Sigma ^\dagger)+ p.c.$  \\[0.5mm]

99 & $\Str(\xi_\mu \Sigma ^\dagger \xi_\mu \Sigma ^\dagger)
\Str(\xi_{\nu 5} \Sigma ^\dagger)\Str(\xi_{5 \nu} \Sigma ^\dagger) + p.c.$  \\[0.5mm]

100 & $\Str(\xi_\mu \Sigma ^\dagger \xi_{\nu 5} \Sigma^\dagger)
\Str(\xi_\mu \Sigma ^\dagger)\Str(\xi_{5 \nu} \Sigma ^\dagger) + p.c.$  \\[0.5mm]

101 & $\Str(\xi_{\nu 5} \Sigma ^\dagger \xi_{5 \nu} \Sigma^\dagger)\
Str(\xi_\mu \Sigma ^\dagger)\Str(\xi_\mu \Sigma ^\dagger) + p.c.$  \\[0.5mm]

102 & $\Str(\xi_\mu \Sigma ^\dagger) \Str(\xi_\mu \Sigma ^\dagger) 
\Str(\xi_{\nu 5} \Sigma^\dagger) \Str(\xi_{5 \nu} \Sigma ^\dagger)+ p.c.$  \\[0.5mm]

103 & $\Str(\xi_\mu \Sigma ^\dagger \xi_\mu \Sigma ^\dagger) 
\Str(\Sigma \xi_{\nu 5} \Sigma \xi_{5 \nu}) + p.c.$  \\[0.5mm]

104 & $\Str(\xi_\mu \Sigma ^\dagger \xi_\mu \Sigma ^\dagger) 
\Str(\Sigma \xi_{\nu 5}) \Str(\Sigma \xi_{5 \nu}) + p.c.$  \\[0.5mm]

105 & $\Str(\xi_{\nu 5} \Sigma ^\dagger \xi_{5 \nu} \Sigma ^\dagger) 
\Str(\Sigma \xi_\mu) \Str(\Sigma \xi_\mu) + p.c.$  \\[0.5mm]

106 & $\Str(\xi_\mu \Sigma ^\dagger) \Str(\xi_\mu \Sigma ^\dagger) 
\Str(\Sigma \xi_{\nu 5}) \Str(\Sigma \xi_{5 \nu}) + p.c.$  \\[0.5mm]

\hline\hline

\end{tabular}
\caption{$\CO(a^4)$ operators in the staggered chiral Lagrangian arising
from two insertions of an $S_6^{FF(A)}$ operator
with spin $S$ or $P$.
Notation as in table~\ref{tab:FFA_VA}.
}
\label{tab:a4_SP_SP}\end{table}

\begin{table}\begin{tabular}{l|ccc}  \hline\hline

\emph{Generic Op.} 
& $[V,A \times P]$ with $[S,P \times V]$ & & $[V,A \times P]$ with $[S,P \times A]$ 
\\ \hline 

107 &  $\Str(\xi_5 \Sigma \xi_\mu \Sigma \xi_5 \Sigma^\dagger \xi_\mu \Sigma^\dagger)$ 
& &  $\Str(\xi_5 \Sigma \xi_{\mu 5} \Sigma \xi_5 \Sigma^\dagger \xi_{5 \mu} \Sigma^\dagger)$
  \\[0.5mm]

108 + 109 & $\Str(\xi_5 \Sigma \xi_\mu \Sigma \xi_5 \Sigma^\dagger)\Str(\xi_\mu \Sigma^\dagger) + p.c.$ & & $\Str(\xi_5 \Sigma \xi_{\mu 5} \Sigma \xi_5 \Sigma^\dagger)\Str(\xi_{5 \mu} \Sigma^\dagger) + p.c.$ \\[0.5mm]

110 & $\Str(\xi_5 \Sigma \xi_5 \Sigma^\dagger)\Str(\xi_\mu \Sigma^\dagger)\Str(\Sigma \xi_\mu)$ & & $\Str(\xi_5 \Sigma \xi_5 \Sigma^\dagger)\Str(\xi_{\mu 5} \Sigma^\dagger)\Str(\Sigma \xi_{5 \mu})$  \\[0.5mm]

111 & $\Str(\xi_\mu \Sigma^\dagger \xi_\mu \Sigma^\dagger \xi_5 \Sigma \xi_5 \Sigma^\dagger) + p.c.$ & & $\Str(\xi_{\mu 5} \Sigma^\dagger \xi_{5 \mu} \Sigma^\dagger \xi_5 \Sigma \xi_5 \Sigma^\dagger) + p.c.$  \\[0.5mm]

112 & $\Str(\xi_\mu \Sigma^\dagger \xi_\mu \Sigma^\dagger)\Str(\xi_5 \Sigma \xi_5 \Sigma^\dagger) + p.c.$ & & $\Str(\xi_{\mu 5} \Sigma^\dagger \xi_{5 \mu} \Sigma^\dagger)\Str(\xi_5 \Sigma \xi_5 \Sigma^\dagger) + p.c.$ \\[0.5mm]

113 & $\Str(\xi_\mu \Sigma^\dagger \xi_5 \Sigma \xi_5 \Sigma^\dagger)\Str(\xi_\mu \Sigma^\dagger) + p.c.$ & & $\Str(\xi_{\mu 5} \Sigma^\dagger \xi_5 \Sigma \xi_5 \Sigma^\dagger)\Str(\xi_{5 \mu} \Sigma^\dagger) + p.c.$  \\[0.5mm]

114 & $\Str(\xi_\mu \Sigma^\dagger)\Str(\xi_\mu \Sigma^\dagger)\Str(\xi_5 \Sigma \xi_5 \Sigma^\dagger) + p.c.$ & & $\Str(\xi_{\mu 5} \Sigma^\dagger)\Str(\xi_{5 \mu} \Sigma^\dagger)\Str(\xi_5 \Sigma \xi_5 \Sigma^\dagger) + p.c.$  \\[0.5mm]

\hline\hline

\end{tabular}

\vspace{3mm}

\begin{tabular}{l|ccc}  \hline\hline

\emph{Generic Op.} 
& $[V,A \times T]$ with $[S,P \times V]$ & & $[V,A \times T]$ with $[S,P \times A]$ 
\\ \hline 

107 &  $\Str(\xi_{\mu \nu} \Sigma \xi_\rho \Sigma \xi_{\nu \mu} \Sigma^\dagger \xi_\rho \Sigma^\dagger)$ & &  $\Str(\xi_{\mu \nu} \Sigma \xi_{\rho 5} \Sigma \xi_{\nu \mu} \Sigma^\dagger \xi_{5 \rho} \Sigma^\dagger)$  \\[0.5mm]

108 + 109 & $\Str(\xi_{\mu \nu} \Sigma \xi_\rho \Sigma \xi_{\nu \mu} \Sigma^\dagger)\Str(\xi_\rho \Sigma^\dagger) + p.c.$ & & $\Str(\xi_{\mu \nu} \Sigma \xi_{\rho 5} \Sigma \xi_{\nu \mu} \Sigma^\dagger)\Str(\xi_{5 \rho} \Sigma^\dagger) + p.c.$ \\[0.5mm]

110 & $\Str(\xi_{\mu \nu} \Sigma \xi_{\nu \mu} \Sigma^\dagger)\Str(\xi_\rho \Sigma^\dagger)\Str(\Sigma \xi_\rho)$ & & $\Str(\xi_{\mu \nu} \Sigma \xi_{\nu \mu} \Sigma^\dagger)\Str(\xi_{\rho 5} \Sigma^\dagger)\Str(\Sigma \xi_{5 \rho})$  \\[0.5mm]

111 & $\Str(\xi_\rho \Sigma^\dagger \xi_\rho \Sigma^\dagger \xi_{\mu \nu} \Sigma \xi_{\nu \mu} \Sigma^\dagger) + p.c.$ & & $\Str(\xi_{\rho 5} \Sigma^\dagger \xi_{5 \rho} \Sigma^\dagger \xi_{\mu \nu} \Sigma \xi_{\nu \mu} \Sigma^\dagger) + p.c.$  \\[0.5mm]

112 & $\Str(\xi_\rho \Sigma^\dagger \xi_\rho \Sigma^\dagger)\Str(\xi_{\mu \nu}\Sigma \xi_{\nu \mu} \Sigma^\dagger) + p.c.$ & & $\Str(\xi_{\rho 5} \Sigma^\dagger \xi_{5 \rho} \Sigma^\dagger)\Str(\xi_{\mu \nu} \Sigma \xi_{\nu \mu} \Sigma^\dagger) + p.c.$ \\[0.5mm]

113 & $\Str(\xi_\rho \Sigma^\dagger \xi_{\mu \nu} \Sigma \xi_{\nu \mu} \Sigma^\dagger)\Str(\xi_\rho \Sigma^\dagger) + p.c.$ & & $\Str(\xi_{\rho 5} \Sigma^\dagger \xi_{\mu \nu} \Sigma \xi_{\nu \mu} \Sigma^\dagger)\Str(\xi_{5 \rho} \Sigma^\dagger) + p.c.$  \\[0.5mm]

114 & $\Str(\xi_\rho \Sigma^\dagger)\Str(\xi_\rho \Sigma^\dagger)\Str(\xi_{\mu \nu} \Sigma \xi_{\nu \mu} \Sigma^\dagger) + p.c.$ & & $\Str(\xi_{\rho 5} \Sigma^\dagger)\Str(\xi_{5 \rho} \Sigma^\dagger)\Str(\xi_{\mu \nu} \Sigma \xi_{\nu \mu} \Sigma^\dagger) + p.c.$  \\[0.5mm]

\hline\hline

\end{tabular}
\caption{$\CO(a^4)$ operators in the staggered chiral Lagrangian arising
from two insertions of $S_6^{FF(A)}$ operators, one 
with spin $V$ or $A$, the other with spin $S$ or $P$.
Notation as in table~\ref{tab:FFA_VA}.}
\label{tab:a4_VA_SP}\end{table}

\begin{table}\begin{tabular}{l|cc}  \hline\hline

{\emph{Generic Op.}} 
& $[T_\mu \times V_\mu]$ with $[T_\mu \times V_\mu]$ 
& $[T_\mu \times A_\mu]$ with $[T_\mu \times A_\mu]$ \\ \hline

82 & $\Str(\xi_\mu \Sigma^\dagger \xi_\mu \Sigma^\dagger)
\Str(\Sigma \xi_\mu \Sigma \xi_\mu)$ & $\;\;
\Str(\xi_{\mu 5} \Sigma^\dagger \xi_{\mu 5} \Sigma^\dagger)
\Str(\Sigma \xi_{5 \mu} \Sigma \xi_{5 \mu})$  \\[0.5mm]

83 & $\Str(\xi_\mu \Sigma^\dagger \xi_\mu \Sigma^\dagger)
\Str(\Sigma \xi_\mu)\Str(\Sigma \xi_\mu) + p.c.$ & $\;\;
\Str(\xi_{\mu 5} \Sigma^\dagger \xi_{\mu 5} \Sigma^\dagger)
\Str(\Sigma \xi_{5 \mu})\Str(\Sigma \xi_{5 \mu}) + p.c.$  \\[0.5mm]

84 & $\Str(\xi_\mu \Sigma ^\dagger) \Str(\xi_\mu \Sigma ^\dagger) 
\Str(\Sigma \xi_\mu) \Str(\Sigma \xi_\mu)$ & $\;\;
\Str(\xi_{\mu 5} \Sigma ^\dagger) \Str(\xi_{\mu 5} \Sigma ^\dagger) 
\Str(\Sigma \xi_{5 \mu}) \Str(\Sigma \xi_{5 \mu})$  \\[0.5mm]

\hline\hline

\end{tabular}

\vspace{3mm}

\begin{tabular}{l|ccc}  \hline\hline

\multicolumn{1}{l|}{\emph{Generic Op.}} 
& \multicolumn{1}{c}{$[T_\mu \times V_\mu]$ with $[T_\mu \times A_\mu]$} \\ \hline 

82 & $\Str(\xi_\mu \Sigma^\dagger \xi_{\mu 5} \Sigma^\dagger )
\Str(\Sigma \xi_\mu \Sigma \xi_{5 \mu})$  \\[0.5mm]

83 & $\Str(\xi_\mu \Sigma^\dagger \xi_{\mu 5} \Sigma^\dagger )
\Str(\Sigma \xi_\mu)\Str(\Sigma \xi_{5 \mu}) + p.c.$  \\[0.5mm]

84 & $\Str(\xi_\mu \Sigma ^\dagger) \Str(\xi_{\mu 5} \Sigma ^\dagger) 
\Str(\Sigma \xi_\mu) \Str(\Sigma \xi_{5 \mu})$  \\[0.5mm]

\hline\hline

\end{tabular}

\vspace{3mm}

\begin{tabular}{l|ccc}  \hline\hline

\emph{Generic Op.} 
& $[V_\mu,A_\mu \times T_\mu]$ with $[T_\mu \times V_\mu]$ 
& & $[V_\mu,A_\mu \times T_\mu]$ with $[T_\mu \times A_\mu]$ \\\hline

107 &  $\Str(\xi_{\mu \nu} \Sigma \xi_\mu \Sigma \xi_{\nu \mu} \Sigma^\dagger \xi_\mu \Sigma^\dagger)$ & &  $\Str(\xi_{\mu \nu} \Sigma \xi_{\mu 5} \Sigma \xi_{\nu \mu} \Sigma^\dagger \xi_{5 \mu} \Sigma^\dagger)$  \\[0.5mm]

108 + 109 & $\Str(\xi_{\mu \nu} \Sigma \xi_\mu \Sigma \xi_{\nu \mu} \Sigma^\dagger)\Str(\xi_\mu \Sigma^\dagger) + p.c.$ & & $\Str(\xi_{\mu \nu} \Sigma \xi_{\mu 5} \Sigma \xi_{\nu \mu} \Sigma^\dagger)\Str(\xi_{5 \mu} \Sigma^\dagger) + p.c.$ \\[0.5mm]

110 & $\Str(\xi_{\mu \nu} \Sigma \xi_{\nu \mu} \Sigma^\dagger)\Str(\xi_\mu \Sigma^\dagger)\Str(\Sigma \xi_\mu)$ & & $\Str(\xi_{\mu \nu} \Sigma \xi_{\nu \mu} \Sigma^\dagger)\Str(\xi_{\mu 5} \Sigma^\dagger)\Str(\Sigma \xi_{5 \mu})$  \\[0.5mm]

\hline\hline
\end{tabular}

\vspace{3mm}

\begin{tabular}{l|ccc}  \hline\hline

\multicolumn{1}{l|}{\emph{Generic Op.}} 
& \multicolumn{1}{c}{$[V_\mu,A_\mu \times T_\mu]$ with $[V_\mu,A_\mu \times T_\mu]$}
 \\ \hline

69 + 74 & $\Str(\xi_{\mu \nu} \Sigma \xi_{\nu \mu} \Sigma^\dagger \xi_{\mu \sigma} \Sigma \xi_{\sigma \mu} \Sigma^\dagger) + p.c.$ \\[0.5mm]

75 & $\Str(\xi_{\mu \nu} \Sigma \xi_{\nu \mu} \Sigma^\dagger)\Str(\xi_{\mu \sigma} \Sigma \xi_{\sigma \mu} \Sigma^\dagger)$  \\[0.5mm]

77 & $\Str(\xi_{\mu \nu} \Sigma \xi_{\mu \sigma} \Sigma^\dagger)\Str(\xi_{\sigma \mu} \Sigma \xi_{\nu \mu} \Sigma^\dagger )$  \\[0.5mm]

79 & $\Str(\xi_{\mu \nu} \Sigma \xi_{\mu \sigma} \Sigma^\dagger \xi_{\nu \mu} \Sigma \xi_{\sigma \mu} \Sigma^\dagger) + p.c.$  \\[0.5mm]

81 & $\Str(\xi_{\mu \nu} \Sigma \xi_{\mu \sigma} \Sigma^\dagger)\Str(\xi_{\nu \mu} \Sigma \xi_{\sigma \mu} \Sigma^\dagger) + p.c.$  \\[0.5mm]

\hline\hline
\end{tabular}

\caption{$\CO(a^4)$ operators in the staggered chiral Lagrangian arising
from two insertions of $S_6^{FF(B)}$ operators.
The indices $\mu$ and $\nu$ are separately summed, with the constraint
that $\mu\ne\nu$.}
\label{tab:a4_FFB}\end{table}

\begin{table}[t]
\begin{tabular}{l|ccc}  \hline\hline

\emph{Generic Op.} 
& $[V,A \times P]$ && $[V,A \times T]$ \\\hline

115 &  $i\;\Str([\ell_\mu, \xi_5]\xi_5 \Sigma D_\mu \Sigma^\dagger) + p.c.$
& & $i\;\Str([\ell_\mu, \xi_{\nu\rho}] \xi_{\rho\nu}
 \Sigma D_\mu \Sigma^\dagger) + p.c.$ \\[0.5mm]

117 &  $i\;\Str([\ell_\mu, \xi_5] \Sigma \xi_5 D_\mu \Sigma^\dagger)  + p.c.$
& & $i\;\Str([\ell_\mu, \xi_{\nu\rho}] \Sigma \xi_{\rho\nu} D_\mu \Sigma^\dagger) + p.c.$
\\[0.5mm]

\hline\hline
\end{tabular}

\vspace{3mm}

\begin{tabular}{l|ccc}  \hline\hline

\emph{Generic Op.} 
& $[S,P \times V]$ && $[S,P \times A]$ \\\hline

119 &  $i\;\Str([r_\mu \xi_\nu - \xi_\nu \ell_\mu] \Sigma \xi_\nu D_\mu \Sigma) + p.c.$
& & $i\;\Str([r_\mu \xi_{\nu5}-\xi_{\nu5}\ell_\mu] \Sigma \xi_{5\nu} D_\mu \Sigma) + p.c.$
\\[0.5mm]

121 &  $i\;\Str([r_\mu \xi_\nu- \xi_\nu \ell_\mu] \Sigma) \Str(\xi_\nu D_\mu \Sigma) + p.c.$
& & $i\;\Str([r_\mu \xi_{\nu5} - \xi_{\nu5} \ell_\mu] \Sigma) 
\Str(\xi_{5\nu} D_\mu \Sigma) + p.c.$
\\[0.5mm]

123 & $i\;\Str([r_\mu \xi_\nu - \xi_\nu \ell_\mu] \xi_\nu \Sigma^\dagger D_\mu \Sigma) + p.c.$
& & $i\;\Str([r_\mu \xi_{\nu 5} - \xi_{\nu5} \ell_\mu] \xi_{5\nu}
 \Sigma^\dagger D_\mu \Sigma) + p.c.$ \\[0.5mm]

125 &  $i\;\Str([r_\mu \xi_\nu - \xi_\nu \ell_\mu]\Sigma) 
		\Str(\xi_\nu D_\mu \Sigma^\dagger) + p.c.$ 
& & $i\;\Str([r_\mu \xi_{\nu5}-\xi_{\nu5}\ell_\mu] \Sigma) 
	\Str(\xi_{5\nu} D_\mu \Sigma^\dagger) + p.c.$ \\[0.5mm]

\hline\hline
\end{tabular}

\caption{$\CO(a^2 p^2)$ operators proportional to sources,
 arising from single insertions of $S_6^{FF(A)}$ operators.}
\label{tab:source_FFA}\end{table}

\begin{table}
\begin{tabular}{l|c}  \hline\hline

\emph{Generic Op.} 
& $[V_\mu,A_\mu \times T_\mu]$ \\\hline

115 & $i\;\Str([\ell_\mu, \xi_{\mu\nu}]  \xi_{\nu\mu} \Sigma D_\mu \Sigma^\dagger) + p.c.$  \\[0.5mm]

117 & $i\;\Str([\ell_\mu, \xi_{\mu\nu}] \Sigma \xi_{\nu\mu} D_\mu \Sigma^\dagger) + p.c.$  \\[0.5mm]

118 & $i\;\Str([\ell_\mu, \xi_{\mu\nu}]  \Sigma D_\mu \Sigma^\dagger \Sigma \xi_{\nu\mu} \Sigma^\dagger) + p.c.$  \\[0.5mm]

\hline\hline
\end{tabular}

\vspace{3mm}

\begin{tabular}{l|ccc}  \hline\hline

\emph{Generic Op.} 
& $[T_\mu \times V_\mu]$ && $[T_\mu \times A_\mu]$ \\\hline

123 & $i\;\Str([r_\mu \xi_\mu - \xi_\mu \ell_\mu] \xi_\mu \Sigma^\dagger D_\mu \Sigma) + p.c.$
& & $i\;\Str([r_\mu \xi_{\mu 5} - \xi_{\mu5} \ell_\mu] \xi_{5\mu} \Sigma^\dagger D_\mu \Sigma) + p.c.$ \\[0.5mm]

125 &  $i\;\Str([r_\mu \xi_\mu - \xi_\mu \ell_\mu]\Sigma) \Str(\xi_\mu D_\mu \Sigma^\dagger) + p.c.$ 
& & $i\;\Str([r_\mu \xi_{\mu5}-\xi_{\mu5}\ell_\mu] \Sigma) \Str(\xi_{5\mu} D_\mu \Sigma^\dagger) + p.c.$ \\[0.5mm]

126 &  $i\;\Str([r_\mu \xi_\mu - \xi_\mu \ell_\mu] D_\mu \Sigma) \Str( \Sigma^\dagger \xi_\mu ) + p.c$. 
& & $i\;\Str([r_\mu \xi_{\mu 5} - \xi_{\mu 5} \ell_\mu] D_\mu \Sigma) \Str( \Sigma^\dagger \xi_{5 \mu} ) + p.c.$ \\[0.5mm]

\hline\hline
\end{tabular}

\caption{$\CO(a^2 p^2)$ operators proportional to sources,
 arising from single insertions of $S_6^{FF(B)}$ operators.
Indices are independently summed, with the constraint $\mu\ne\nu$. }
\label{tab:source_FFB}\end{table}

\end{document}